\documentclass[letterpaper,11pt]{article}
\pdfoutput=1

\usepackage{jheppub}
\usepackage{hyperref} 
\usepackage{epsfig}
\usepackage[normalem]{ulem}
\usepackage{bm}
\usepackage{bbm}
\usepackage{slashed}
\usepackage{xspace}  
\usepackage[export]{adjustbox}
\usepackage{subfig}
\usepackage[countmax]{subfloat}
\usepackage{multirow}

\newcommand{\setcaptionskip}{\setlength\baselineskip{14pt}}
\newcommand{\setmainskip}{\setlength\baselineskip{18pt}}
 
\usepackage{amsmath,latexsym}
\usepackage{pstricks}
\usepackage{color} 
 
\usepackage{calc}
\usepackage{accents}
\newcommand{\dbtilde}[1]{\accentset{\approx}{#1}}

\def\cB{\mathcal{B}}
\def\cC{\mathcal{C}}

\def\cN{\mathcal{N}}
\def\cS{\mathcal{S}}
\def\cT{\mathcal{T}}

\newcommand{\sdt}{\!\cdot\!}

\newcommand{\eq}[1]{Eq.~\eqref{eq:#1}}
\newcommand{\eqs}[2]{Eqs.~\eqref{eq:#1} and \eqref{eq:#2}} 
\renewcommand{\sec}[1]{Sec.~\ref{sec:#1}}

\newcommand{\fig}[1]{Fig.~\ref{fig:#1}}

\newcommand{\app}[1]{App.~\ref{app:#1}}

\newcommand{\abs}[1]{\lvert#1\rvert}

\newcommand{\plus}{\!+\!}

\newcommand{\la}{\lambda}

\newcommand{\gs}{g}

\newcommand{\hard}{\mathrm{hard}}
\newcommand{\dyn}{\mathrm{dyn}}

\newcommand{\df}{\mathrm{d}}
\newcommand{\img}{\mathrm{i}}

\newcommand{\w}{\omega}

\newcommand{\cL}{\mathcal{L}}
\newcommand{\cM}{\mathcal{M}}
\newcommand{\cO}{\mathcal{O}}

\newcommand{\ddslash}{{d\!\!{}^-}}
\newcommand{\deltaslash}{{\delta\!\!\!{}^-}}

\newcommand{\fd}[2]{\parbox{#1}{\includegraphics[width=#1]{#2}}}

\newcommand{\as}{\alpha_s}

\newcommand{\nn}{\nonumber}
\newcommand{\mcdot}{\!\cdot\!}
\newcommand{\beq}{\begin{equation}}
\newcommand{\eeq}{\end{equation}}
\newcommand{\bea}{\begin{eqnarray}}
\newcommand{\eea}{\end{eqnarray}}

\newcommand{\MSbar}{\mbox{${\overline {\rm MS}}$}}

\newcommand{\Eq}[1]{Equation~\eqref{#1}}
\DeclareRobustCommand{\Sec}[1]{Sec.~\ref{#1}}

\DeclareRobustCommand{\Fig}[1]{Fig.~\ref{#1}}

\DeclareRobustCommand{\Eq}[1]{Eq.~(\ref{#1})}
\DeclareRobustCommand{\Eqs}[2]{Eqs.~(\ref{#1}) and (\ref{#2})}

\newcommand\prp[1]{\vec #1_\perp}
\newcommand\prpsq[1]{\vec{#1}^{2}_\perp}
\newcommand\prpsqm[2]{\bigl( \prp{#1} \!-\! \prp{#2} \bigr)^{\!2}}
\newcommand\prpsqp[2]{\bigl( \prp{#1} \!+\! \prp{#2} \bigr)^{\!2}}

\newcommand\prnth[1]{\left(#1\right)}
\newcommand\reg[2]{\left| \frac{2 #1^z}{\nu} \right|^{-#2\eta}}

\newcommand{\IInt}{\int\!\!\!\!\!\int}

\newcommand{\bn}{{\bar n}}
\newcommand{\bT}{{\bar T}}
\newcommand{\bnP}{\bar {\cal P}}

\newcommand{\cP}{{\cal P}}
\def\bnslash{\bar n\!\!\!\slash}
\def\nslash{n\!\!\!\slash}

\newcommand{\dbar}{d\hspace*{-0.08em}\bar{}\hspace*{0.1em}}
\newcommand{\Sl}[1]{\slashed{#1}}

\usepackage{marginnote}

\renewcommand{\tabcolsep}{1ex}

\allowdisplaybreaks[1]

\begin{document}



\preprint{\vbox{\hbox{MIT--CTP 5628}}}

\title{\boldmath A Collinear Perspective on the Regge Limit}

\author[1]{Anjie Gao,}
\author[2]{Ian Moult,}
\author[1,3]{Sanjay Raman,}
\author[1]{Gregory Ridgway,}
\author[1]{and Iain W. Stewart}

\affiliation[1]{Center for Theoretical Physics, Massachusetts Institute of Technology, Cambridge, MA 02139, USA}
\affiliation[2]{Department of Physics, Yale University, New Haven, CT 06511}
\affiliation[3]{Department of Physics, Harvard University, Cambridge, MA 02138, USA}

\emailAdd{anjiegao@mit.edu}
\emailAdd{ian.moult@yale.edu}
\emailAdd{sanjayraman@fas.harvard.edu}
\emailAdd{gridgway@mit.edu}
\emailAdd{iains@mit.edu}

\abstract{The high energy (Regge) limit provides a playground for understanding all loop structures of scattering amplitudes, and plays an important role in the description of many phenomenologically relevant cross-sections.
While well understood in the planar limit, the structure of non-planar corrections introduces many fascinating complexities, for which a general organizing principle is still lacking.
We study the structure of multi-reggeon exchanges in the context of the effective field theory for forward scattering, and derive their factorization into collinear operators (impact factors) and soft operators. 
We derive the structure of the renormalization group consistency equations in the effective theory, showing how the anomalous dimensions of the soft operators are related to those of the collinear operators,  allowing us to derive renormalization group equations in the Regge limit purely from a collinear perspective.
The rigidity of the consistency equations provides considerable insight into the all orders organization of Regge amplitudes in the effective theory, as well as its relation to other approaches. Along the way we derive a number of technical results that improve the understanding of the effective theory.
We illustrate this collinear perspective by re-deriving all the standard BFKL equations for two-Glauber exchange from purely collinear calculations, and we show that this perspective provides a number of conceptual and computational advantages as compared to the standard view from soft or Glauber physics. 
We anticipate that this formulation in terms of collinear operators will enable a better understanding of the relation between BFKL and DGLAP in gauge theories, and facilitate the analysis of renormalization group evolution equations describing Reggeization beyond next-to-leading order.
}

\maketitle



\section{Introduction}
\label{sec:intro}

Simplifications of scattering amplitudes in the high energy (Regge) limit, due to the separation of scales $|t| \ll s$, provide valuable insight into their all loop behavior. This limit has therefore been the subject of significant interest \cite{Gell-Mann:1964aya,Mandelstam:1965zz,McCoy:1976ff,Grisaru:1974cf,Grisaru:1973ku,Grisaru:1973vw,Fadin:1975cb,Kuraev:1976ge,Lipatov:1976zz,Kuraev:1977fs,Balitsky:1978ic,Lipatov:1985uk,Lipatov:1995pn}. Take, for concreteness, a $2\to 2$ scattering amplitude. The Regge limit is most often studied by considering the dynamics of the partons exchanged between the two high-energy projectiles. In particular, these dynamics are often reformulated in terms of emergent degrees of freedom called Reggeons. This has been extensively studied since the program of Gribov \cite{Gribov:1968fc,Abarbanel:1975me,Baker:1976cv} from a number of perspectives including the classic work of Balitsky, Fadin, Kuraev and Lipatov \cite{Fadin:1975cb,Lipatov:1976zz,Kuraev:1976ge,Kuraev:1977fs,Balitsky:1978ic,Lipatov:1985uk}, Lipatov's effective action \cite{Lipatov:1995pn}, and more modern approaches in terms of Wilson lines \cite{Balitsky:1995ub,Balitsky:2001gj,Caron-Huot:2013fea,Caron-Huot:2017fxr,Caron-Huot:2017zfo}. 
This has led to a large amount of progress; however, there are still areas where the relevant calculations are complicated. 
For example, a derivation of BFKL-type evolution equations that separate the color octet Regge cut and pole contributions required significant effort~\cite{Fadin:2016wso,Fadin:2017nka,Caron-Huot:2017fxr,Fadin:2019tdt,Fadin:2019qfo,Fadin:2019lnb} before the appearance of more attractive recent results in Refs.~\cite{Caola:2021izf,Falcioni:2021dgr}. An all-orders organization for the amplitude structure predicted by the Reggeon field theory has also not yet emerged.

In this paper we present a collinear approach to Reggeization derived in a top-down effective field theory (EFT) of forward scattering \cite{Rothstein:2016bsq}, which is formulated in the Soft-Collinear Effective Theory (SCET) \cite{Bauer:2000ew, Bauer:2000yr, Bauer:2001ct, Bauer:2001yt}.  We show that rapidity renormalization group \cite{Chiu:2011qc,Chiu:2012ir} consistency of evolution equations in the Regge limit enable the derivation of classic evolution equations for gluon Reggeization~\cite{Fadin:1993wh}, the BFKL pomeron~\cite{Fadin:1993wh}, and other color channels, from purely collinear calculations of SCET impact factors. 
Such calculations are simplified as compared to their soft counterparts since one only needs to consider a single scattering object. We will show that such calculations in the collinear sector also make manifest the connection between planarity/non-planarity and Regge poles/cuts. For example, in the calculation of the BFKL pomeron, the pole contributions come only from planar EFT graphs, while the non-trivial convolution (Regge cut) structure arises only from the non-planar graphs. This is quite distinct from the picture from the soft sector, and we believe that this clean separation between pole and cut contributions will be particularly convenient when moving to the study of Regge cuts in non-planar theories. It also allows us to give a straightforward proof of Reggeization into a single pole in a planar gauge theory using the instantaneous nature of Glauber exchanges. This is expected from general proofs \cite{Mandelstam:1963cw,Caron-Huot:2013fea}, explicit examples of all loop planar amplitudes  \cite{Drummond:2007aua,Korchemsky:2018hnb}, and is even expected to continue to higher multiplicity \cite{DelDuca:2019tur}. However, the difficulty is in understanding corrections beyond this limit, which we believe that our organization can achieve.

There are several motivations for this work. First, we would like to provide a general organizing principle for studying the Regge limit of $2\to 2$ scattering amplitudes. The RG consistency equations that we derive in this paper are a key step towards that goal, since they elucidate the mixing structure in the EFT to all loop orders, as well as providing gauge invariant definitions of Regge anomalous dimensions. Additionally, they allow us to understand how the effective theory organizes Regge amplitudes, as compared with other approaches to the Regge limit, such as the Reggeon Field Theory.
Second, it has long been known that there are interesting connections between anomalous dimensions in the Regge limit, and the analytic continuation of twist-2 anomalous dimensions \cite{Jaroszewicz:1980mq,Jaroszewicz:1982gr,Kotikov:2007cy,Kotikov:2004er,Kotikov:2002ab,Kotikov:2000pm,Brower:2006ea}. In the context of conformal field theories, this has recently been put on a rigorous footing using the recent developments of analyticity in spin \cite{Caron-Huot:2017vep} and lightray operators \cite{Kravchuk:2018htv,Kologlu:2019mfz,Caron-Huot:2022eqs}. We believe that developing a collinear perspective on the Regge limit, may allow for a connection with these recent developments, as well as a better understanding of the connection between BFKL and DGLAP.
Third, the effective field theory for forward scattering can be applied to a number of phenomenological problems, ranging from better understanding jet substructure in the quark-gluon plasma \cite{Vaidya:2021mjl,Vaidya:2020lih,Vaidya:2021vxu,Vaidya:2021mly}, saturation \cite{Stewart:2023lwz}, evolution equations for deep inelastic scattering \cite{Neill:2023jcd}, to proofs of factorization for hadron collider observables, and to relations between rapidity anomalous dimensions and Regge exponents~\cite{DelDuca:2017pmn,Moult:2022lfy,Rothstein:2023dgb}. By studying the simplest case of $2\to 2$ forward scattering at the amplitude level, we are able to better understand the general structure of the effective field theory, which will then help in its application to these other phenomenological problems.

An outline of the paper is as follows. In \Sec{sec:reviewRandG} we briefly review Reggeons and Glaubers as the degrees of freedom for forward scattering, as well as relevant ingredients of the effective field theory of forward scattering.
In \Sec{sec:multi_glauber} we derive the factorization of the $2\to 2$ forward scattering amplitude into multi-Glauber operators, and study in detail their renormalization group structure. A key result of this paper is to derive the renormalization group consistency equations relating the renormalization group evolution of the soft/Glauber degrees of freedom, and the collinear degrees of freedom. This will allow us to derive renormalization group equations in the Regge limit purely from a collinear perspective. We also show that graphs with a single Glauber exchange attached to one impact factor and $j>1$ Glaubers attached to the other impact factor, vanish in the EFT. This differs from the alternate Reggeon exchange picture, where the analogous graphs do not vanish for odd $j$.
In \Sec{sec:BFKL} we derive the known BFKL equations for two Glauber exchange from the collinear perspective, highlighting the simplicity of this approach, and formalizing several issues, such as the treatment of color mixing, that will be useful in going to higher number of Glauber exchanges.
In \Sec{sec:planar}, we use our collinear perspective combined with the instantaneous nature of Glauber exchange to give a simple all orders proof of the pole structure of Reggeization in planar gauge theories. This proof is constructive, makes clear the connection between planarity and Reggeization, and also makes clear how to incorporate non-planar corrections. 
We conclude and discuss a number of future goals of our approach in \Sec{sec:conc}.

\section{Review of Reggeon and Glauber Exchange}
\label{sec:reviewRandG}

Throughout this paper we will be considering the simplest scattering process, namely two-to-two scattering. Denoting the incoming momenta as $p_1, p_2$ and the outgoing momenta as $p_3,p_4$,
\begin{align}
\fd{4.5cm}{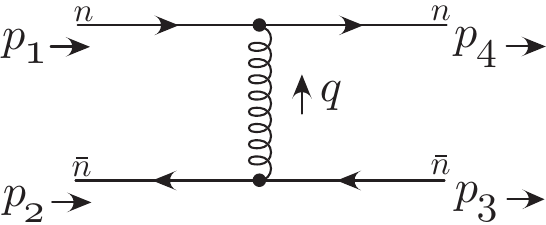}\,,
\end{align}
the forward limit is characterized by a small momentum transfer, $|q^2| \ll s=(p_1+p_2)^2$, and is also referred to as the Regge limit. One of the complexities of Regge physics is that there are a variety of closely related approaches, which have evolved significantly over time, ultimately into forms unrecognizable from what they were named after. Furthermore, these approaches often use different languages, some of which predate quantum field theory. While our goal here is not to give a comprehensive review of the different approaches, we wish to provide some context so as to be able to understand the differences between the approaches. In particular, while the language of Reggeons is most likely familiar to those studying the Regge limit, the language of Glauber gluons that we will use is perhaps less familiar. We will therefore attempt to put the distinction into context. After briefly reviewing the history, we will discuss the effective field theory approach to forward scattering, formulated in terms of Glauber operators, in some detail in \Sec{sec:review}. In \Sec{sec:neqglaub} we will then highlight a number of distinctions between Glaubers and Reggeons. These distinctions will be further developed throughout the paper, but it will be important to clearly have in mind that there is a distinction.

Building on the work of Regge in non-relativistic quantum mechanics \cite{Regge:1959mz}, the program initiated by Gribov \cite{Gribov:1968fc,Abarbanel:1975me,Baker:1976cv} was to develop a two dimensional field theory in impact parameter space, the Reggeon field theory \cite{Gribov:1968fc,Abarbanel:1975me,Baker:1976cv}, in which rapidity played the role of an external parameter.\footnote{For an interesting historical account of the Regge limit, see \cite{DelDuca:2018nsu}. } In this field theory the Reggeons are the emergent degrees of freedom which provide the appropriate description of QCD in the high energy limit. In the case of a fixed number of Reggeons, the standard BFKL and BK equations can be viewed as equations of a two dimensional quantum mechanical system. These equations exhibit a number of remarkable symmetries \cite{Lipatov:1993yb,Faddeev:1994zg} and have therefore been extensively studied. The development of a complete Reggeon field theory requires vertices describing reggeon transitions, which would promote it from a quantum mechanical problem to a true interacting field theory. While there has been much progress on deriving such vertices \cite{Bartels:1995kf,Bartels:1999aw,Bartels:2002au,Bartels:2004hb,Bartels:2007dm}, the complete Reggeon field theory remains elusive. Another approach to deriving the Reggeon field theory is through Lipatov's effective action \cite{Lipatov:1995pn} (for a review, see e.g. Refs.~\cite{Lipatov:1996ts,Hentschinski:2009zz}). Significant recent progress has also been made by Caron-Huot and collaborators \cite{Caron-Huot:2013fea,Caron-Huot:2017fxr,Caron-Huot:2017zfo}. Here they use Wilson lines and modern amplitude techniques, however the degrees of freedom should still be thought of as Reggeons, and in particular obey the standard properties of Reggeons (such as signature).

The study of Glauber gluons, on the other hand, arose primarily in the study of factorization and factorization breaking, and in particular, played an important role in the factorization proofs of Refs.~\cite{Collins:1985ue,Collins:1989gx,Collins:1988ig}. The particular name ``Glauber'' seems to have arisen in the paper \cite{Bodwin:1981fv}. The relation between these seemingly different physical situations is that Glauber gluons can mediate forward scattering subprocesses in hard scattering processes, which must be proven to cancel to arrive at standard factorization formulas. A framework for using Glauber exchange simultaneously as a tool to study forward scattering processes and to investigate factorization violation was put forward in Ref.~\cite{Rothstein:2016bsq}.

The fact that the momentum of Glauber gluons scales as $p^2 \simeq \cP_\perp^2$, and that they are integrated into a potential that is non-local in the transverse plane provides a relation to the two dimensional Reggeon field theory. However, it will be important that the Glaubers correspond to particular momentum region of diagrams of a four dimensional field theory. While this may seem like a technical detail, we will see that  it leads to quite significant differences in the interpretation of diagrams. Therefore, while we will adopt much of the language of Reggeon Field Theory to describe the behavior of amplitudes, we will be careful to refer to the objects passed in the t-channel in the EFT as Glauber gluons, or Glauber potential operators, and not Reggeons.  We will denote Glauber operators by dashed red lines, such as
\begin{align}
  \fd{3cm}{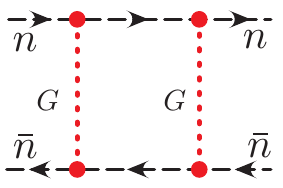}\,,
\end{align} 
while we will denote Reggeons by blue doubled gluons, such as
\begin{align}
  \fd{3.7cm}{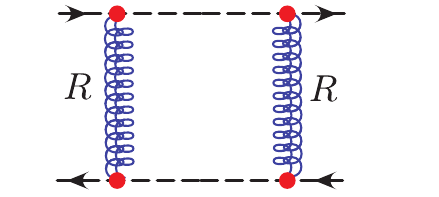}\,.
\end{align}  
In this paper we will take a considerable step towards understanding  the reduction of the four-dimensional Glauber EFT to the two-dimensional Reggeon field theory. In particular, in \Sec{sec:consistent}, we will argue that the Reggeon Field Theory is in fact the rapidity renormalization group equations of the Glauber operators in the EFT.

\subsection{Review of Glauber SCET}
\label{sec:review}

We now describe in more technical detail the effective field theory approach to forward scattering.
SCET is an effective field theory describing soft and collinear degrees of freedom in the presence of a hard or forward scattering process \cite{Bauer:2000ew, Bauer:2000yr, Bauer:2001ct, Bauer:2001yt}. Soft and collinear degrees of freedom are described by distinct fields in the EFT. Glauber interactions are incorporated in the theory using non-local potentials that provide interactions between collinear particles scattering in different directions, and between soft and collinear particles~\cite{Rothstein:2016bsq}. The form of the leading power interactions involving Glauber gluon potentials were derived in Ref.~\cite{Rothstein:2016bsq}, and the leading interactions (at subleading power) for Glauber quark induced potentials were derived in Ref.~\cite{Moult:2017xpp}.

For forward scattering processes we are interested in the dynamics of energetic projectiles traveling near the lightcone, so it is convenient to use lightcone coordinates. We can define two lightlike reference vectors $n^\mu$ and $\bn^\mu$ such that $n^2 = \bn^2 = 0$ and $n\sdt\bn = 2$. Any momentum, $p$, can then be written as
\begin{equation} \label{eq:lightcone_dec}
p^\mu = \bn\sdt p\,\frac{n^\mu}{2} + n\sdt p\,\frac{\bn^\mu}{2} + p^\mu_{\perp}\
\,.\end{equation}
For notational convenience, we will often use the shorthand $p^-=\bn\cdot p$ and $p^+=n\cdot p$.

We will consider two-to-two scattering, denoting the incoming momenta as $p_1, p_2$ and the outgoing momenta as $p_3,p_4$:
\begin{align}
\fd{4.5cm}{figsCO/Full_Glaub_tree}\,.
\end{align}
The forward limit is characterized by a small momentum transfer, $|q^2| \ll s=(p_1+p_2)^2$. In this case, the high energy particles being scattered have a large momentum along the $n$ or $\bar n$ direction, with small momentum fluctuations about this motion.  These are referred to as collinear particles. A particle is said to be $n$-collinear if its momentum $p$ scales as $(n\cdot p, \bn \cdot p, p_{\perp}) \sim (\la^2,1,\la)$, where $\lambda\ll 1$ is a formal power counting parameter. The $\bar n$ collinear particles are defined in an analogous manner with $(n\cdot p, \bn \cdot p, p_{\perp}) \sim (1,\la^2,\la)$. For a collinear particle, the forward scattering condition enforces that the large component of its momentum is preserved. In the forward limit, gluons that can be exchanged between the two collinear sectors without knocking particles offshell (giving them parametrically larger $p^2$) have a scaling $p^\mu \sim Q(\lambda^2, \lambda^2, \lambda)$. These are referred to as $n$-$\bn$ Glauber gluons, and to leading power have $p^2=p_\perp^2$, therefore behaving as off-shell potentials. They are analogous to potential Coulombic gluons in non-relativistic scattering, but are now instantaneous in both light-cone times, or equivalently in time and a longitudinal direction.

Since the Glauber gluons are off-shell, they should be integrated out, and are described by potentials in the EFT. These potentials are non-local in the plane transverse to the scattering. Due to this non-locality, we will draw a Glauber potential interaction as an extended dashed red line
\begin{align}
\fd{3cm}{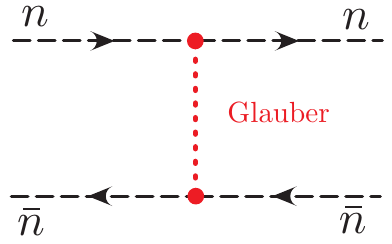}
 \,.
\end{align}
Although we will be studying the forward scattering of collinear quarks and gluons as above, soft particles will also play an important role. Particles are said to be soft if they have momentum scaling as $(\lambda,\lambda, \lambda)$.\footnote{Technically this scaling for the soft momentum defines a SCET$_\text{II}$~\cite{Bauer:2002aj} theory, although this distinction is not important for us here.} For the forward scattering of a soft gluon with a $n$-collinear quark, 
\begin{align}\label{eq:ns_scattering}
\fd{3cm}{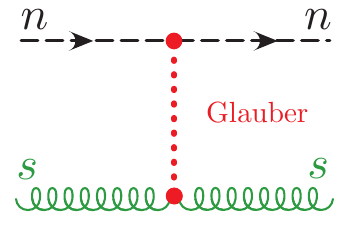}\,,
\end{align}
the forward scattering condition on the soft gluon is that its $n\cdot p$ momentum is conserved. The $n$-$s$ Glauber gluons that can be exchanged in $n$-soft scattering have the scaling  $p^\mu \sim Q(\lambda^2, \lambda, \lambda)$. An identical line of reasoning applies for $\bar n$-s forward scattering, with exchanged $\bn$-$s$ Glauber gluons scaling as $p^\mu \sim Q(\lambda, \lambda^2, \lambda)$.  Glauber exchanges always have $n\cdot p\, \bn\cdot p \ll \vec p_\perp^{\,2}$.

In this paper the primary role of $n$-$s$ and $\bar n$-$s$ Glauber potentials will be as time ordered products (T-products) in loops appearing in $n$-$\bar n$ collinear scattering, such as
\begin{align}\label{eq:Tproduct_picture}
 \fd{3cm}{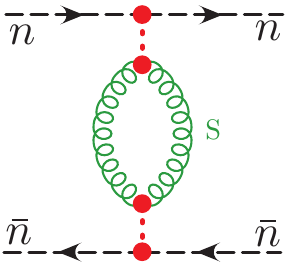}
  \qquad \text{or}\qquad 
 \fd{5cm}{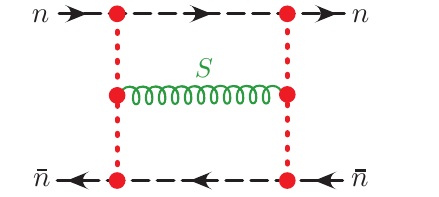}\,.
\end{align}
These T-products are required to correctly describe a number of interesting effects, such as the running of Glauber potentials and Glauber transitions. They will play an important role in our discussion.

The full SCET Lagrangian including Glauber interactions is given by 
\begin{align} \label{eq:SCETLagExpand}
\cL_{\text{SCET}}=\cL_\hard+\cL_\dyn= \cL^{(0)} + \cL_G^{(0)} +\sum_{i\geq0} \cL_\hard^{(i)}+\sum_{i\geq1} \cL^{(i)}\,,
\end{align}
Here  $\cL^{(0)}$ is the standard SCET Lagrangian describing the dynamics of soft and collinear interactions without any Glauber potential insertions. As is well known, the interactions between soft and collinear particles can be decoupled in  $\cL^{(0)}$~\cite{Bauer:2002nz}. However, the interactions between soft and collinear particles can no longer be decoupled once Glauber potentials are included, due to diagrams such as \Eq{eq:ns_scattering}. These interactions are contained in the leading power Lagrangian $ \cL_G^{(0)}$, whose form we now discuss in more detail.

The $ \cL_G^{(0)}$ Lagrangian describes the potentials for both $n$-$\bar n$ and $n$-$s$ scattering. 
For $n$-$\bar n$ collinear scattering, the leading power Glauber operators are
\begin{align}\label{eq:nnops}
\cO^{q\bar q}_{sn\bar n} &= \cO_n^{qB} \frac{1}{\cP_\perp^2} \cO_s^{BC} \frac{1}{\cP_\perp^2} \cO_\bn^{qC}
 \,, \qquad    
\cO^{g q}_{sn\bar n} = \cO_n^{gB} \frac{1}{\cP_\perp^2} \cO_s^{BC} \frac{1}{\cP_\perp^2} \cO_\bn^{qC}
  \,, \nn \\
\cO^{qg}_{sn\bar n} &= \cO_n^{qB} \frac{1}{\cP_\perp^2} \cO_s^{BC} \frac{1}{\cP_\perp^2} \cO_\bn^{gC}
 \,, \qquad  
 \cO^{gg}_{sn\bar n} = \cO_n^{gB} \frac{1}{\cP_\perp^2} \cO_s^{BC} \frac{1}{\cP_\perp^2} \cO_\bn^{gC}
 \,,
\end{align}
where 
\begin{align} \label{eq:Onops}
\cO_n^{qB} = \bar \chi_n T^B \frac{\Sl{\bar n}}{2} \chi_n \,, \qquad \cO_n^{gB} = \frac{\img}{2} f^{BCD} \cB^C_{n\perp \mu} \frac{\bar n}{2} \cdot (\cP+\cP^\dagger ) \cB^{D\mu}_{n\perp}\,,
\end{align}
with the analogous definitions for $\cO_\bn^{qB}$ and $\cO_\bn^{gB}$. 
These operators are written in terms of the gauge invariant quark and gluon fields \cite{Bauer:2000yr,Bauer:2001ct}
\begin{align} \label{eq:chiB}
\chi_{{n}}(x) &= \Bigl[W_{n}^\dagger(x)\, \xi_{n}(x) \Bigr]
\,,\qquad 
\cB_{{n}\perp}^\mu(x)
= \frac{1}{g}\Bigl[ W_{n}^\dagger(x)\,\img  D_{n\perp}^\mu W_{n}(x)\Bigr]
 \,,
\end{align}
where
\begin{equation}
\img  D_{{n_i}\perp}^\mu = \cP^\mu_{\perp} + g A^\mu_{{n_i}\perp}\,,
\end{equation}
is a collinear covariant perpendicular derivative and
\begin{equation} \label{eq:Wn}
W_{n_i}(x) = \biggl[~\sum_\text{perms} \exp\Bigl(-\frac{g}{\bnP_{n_i}}\,\bn\sdt A_{n_i}(x)\Bigr)~\biggr]\,,
\end{equation}
is a collinear Wilson line. Here $\bnP_{n_i} = \bar n_i \cdot \cP$ and $\cP_\perp$ are collinear derivatives that give ${\cal O}(\lambda^0)$ $\bar n_i\cdot p$, and ${\cal O}(\lambda)$ perpendicular momenta, respectively.

The most non-trivial component of the Glauber operators is the soft operator which sits in between the collinear sectors, given by
\begin{align}\label{eq:iain_ira_soft}
\cO_s^{BC}=8\pi \alpha_s &\left\{  \cP^\mu_\perp S_n^T S_{\bar n} \cP_{\perp \mu} -\cP^\perp_\mu g\tilde \cB_{S\perp}^{n\mu} S_n^T S_{\bar n} - S_n^T S_{\bar n} g \tilde \cB_{S\perp}^{\bar n\mu} \cP^\perp _\mu -g \tilde\cB_{S\perp}^{n\mu} S_n^T S_{\bar n} g\tilde\cB_{S\perp \mu}^{\bar n}     \right . \nn \\
&\left. -\frac{n^\mu \bar n^\nu}{2}   S_n^T ig\tilde G^{\mu \nu}_s S_{\bar n}    \right \}^{BC}    \,.
\end{align}
In the case that there are no soft emissions, we have $\cO_s^{BC} \to 8\pi \alpha_s \cP_\perp^2 \delta^{BC}$, and the operators ${\cal O}_{sn\bn}^{ij}$ reduce to a tree level Glauber potential, $1/\cP_\perp^2$ between collinear operators ${\cal O}_n^{iB}$ and ${\cal O}_\bn^{jC}$. With a single soft emission $\cO_s^{BC}$ reproduces the Lipatov vertex \cite{Kuraev:1976ge}. However, it is gauge invariant, and therefore also has Feynman rules with an arbitrary number of emissions. These are drawn as emissions from the central Glauber potential, such as
\begin{align}
\fd{4cm}{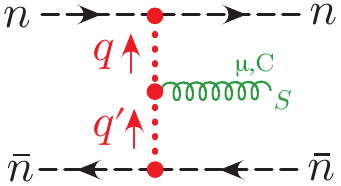}\qquad \text{and} \qquad \fd{4.5cm}{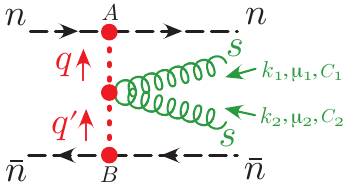}\,.
\end{align}
Explicit Feynman rules for the one and two emission vertices can be found in Ref.~\cite{Rothstein:2016bsq}.

The leading power Glauber Lagrangian also describes the operators that mediate collinear-soft forward scattering as shown in \Eq{eq:ns_scattering}.
The operators for $n$-$s$ scattering are
\begin{align}
\cO^{qq}_{ns}=\cO^{qB}_n \frac{1}{\cP_\perp^2} \cO_s^{q_n B}
 \,, \quad   
 \cO^{qg}_{ns}=\cO^{qB}_n \frac{1}{\cP_\perp^2} \cO_s^{g_n B}
 \,, \quad 
 \cO^{gq}_{ns}=\cO^{gB}_n \frac{1}{\cP_\perp^2} \cO_s^{q_n B}
 \,, \quad 
 \cO^{gg}_{ns}=\cO^{gB}_n \frac{1}{\cP_\perp^2} \cO_s^{g_n B}
 \,, 
\end{align}
with direct analogs for $\bar n$-$s$ scattering.
The collinear operators, $\cO^{iB}_n$ appearing here are the same as in \eq{Onops}. The soft operators are given by
\begin{align}
\cO_s^{q_n B}&=8\pi \alpha_s \left( \bar \psi_s^n T^B \frac{\Sl n}{2}  \psi^n_s  \right)\,, \nn\\
\cO_s^{g_n B}&= 8\pi \alpha_s \left(   \frac{\img}{2} f^{BCD} \cB^{nC}_{S\perp \mu} \frac{n}{2} \cdot (\cP +\cP^\dagger) \cB_{S\perp}^{nD\mu} \right)\,,
\end{align}
for quarks and gluons respectively. They are defined in terms of gauge invariant soft quark and gluon fields
\begin{align}\label{eq:gauge_soft}
 \psi_S^n = S_n^\dagger \psi_s 
 \,, \qquad
\cB_{S\perp}^{\bar n \mu}=\frac{1}{g}[S_\bn^\dagger \img D^\mu_{S\perp} S_{\bar n}]\,, \qquad \cB^{n\mu}_{S\perp}=\frac{1}{g} [S_n^\dagger \img D^\mu_{S\perp} S_n]\,.
\end{align}

Combining together these different potentials, we arrive at the leading power Glauber Lagrangian \cite{Rothstein:2016bsq}
\begin{align}\label{eq:Glauber_Lagrangian}
\cL_G^{\text{II}(0)} 
&= e^{-ix\cdot \cP} \sum\limits_{n,\bar n} \sum\limits_{i,j=q,g}   \cO_n^{iB} \frac{1}{\cP_\perp^2} \cO_s^{BC}   \frac{1}{\cP_\perp^2} \cO_{\bar n}^{jC}   + e^{-ix\cdot \cP} \sum\limits_{n} \sum\limits_{i,j=q,g} \cO_n^{iB}   \frac{1}{\cP_\perp^2} \cO_s^{j_n B}\,.
\end{align}
This Lagrangian is exact, i.e. it does not receive matching corrections, and it is not renormalized. 

\subsection{Reggeon $\neq$ Glauber Gluon}
\label{sec:neqglaub}

Having reviewed the structure of the EFT, in this section we discuss in considerable detail a number of distinctions between the treatment of the Regge limit in the EFT, as compared to conventional approaches that use Reggeon degrees of freedom. In particular, while Glaubers and Reggeons are naively similar, and the diagrams describing their interactions have similar topologies, they are in fact quite different. The similar topologies of the diagrams can lead to considerable confusion if interpreted incorrectly. This distinction is important for understanding the organization in the EFT. In particular, crossing symmetry has a different form in the EFT due to the fact that the Glauber graphs are non-analytic in the external momentum. The goal of this section is to highlight these differences, as well as how physics familiar from the standard Reggeon Field Theory picture arises in the EFT. Although some of the discussion of this section may seem overly pedantic, particularly for experts in Regge physics, we believe it is important so that statements made about diagrams in the EFT are not misunderstood in the language of Reggeons. It will also allow us to build up a physical intuition behind the organization of the EFT, and how this corresponds to the intuition of the Regge limit from other approaches.

\subsubsection*{Unitarity and Signature:}

An important consequence of the fact that diagrams in the EFT involving Glauber gluons describe a particular scaling limit of QCD, is that all diagrams in the EFT are Feynman diagrams of a true four dimensional QFT, and obey the associated properties, such as the Cutkosky rules and unitarity. This has a significant impact on the organization of diagrams in the EFT as compared with in the Reggeon field theory. 

In the Reggeon field theory, the Reggeons have a signature \cite{Gribov:1968fc}. The LL reggeization has a single Reggeon passed in the $t$-channel, which has negative signature. Since diagrams involving two Reggeons have a positive signature, they do not contribute to the negative signature channels, but rather to the positive signature Pomeron. The first contribution from multi-Reggeon exchange to the negative signature Reggeon therefore occurs with three Reggeon exchanges.

In the organization of the EFT, unitarity enforces that the two Glauber exchange box diagram must have an antisymmetric octet (negative signature) contribution
\begin{align}
 \left. \fd{3cm}{figsCO/EFT_loop1_qqqq}  \right |_{8_A} \propto  \img\pi \big|\cM^{(0)} \big|^2 \,,
\end{align} 
where $\cM^{(0)}$ is the tree-level single Glauber exchange,
since performing a $t$-channel cut, we obtain $\img\pi$ multiplying the square of the single exchange. This antisymmetric octet contribution therefore contributes to the Reggeization of the gluon, while the corresponding diagram involving two Reggeons does not. This makes clear that Reggeon $\neq$ Glauber. In the standard Reggeon field theory approach, these particular $\img\pi$ terms are obtained ``for free'' by exploiting crossing symmetry, while in the EFT, they are obtained from a distinct diagram.

\subsubsection*{Vanishing of Cross Diagrams:}

Another important difference between Reggeons and Glaubers is that graphs containing crossed Glaubers vanish. For example (see Ref.~\cite{Rothstein:2016bsq} for a detailed explanation)
\begin{align}
  \fd{3cm}{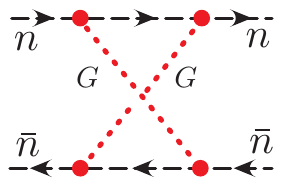}=0\,.
\end{align} 
The standard box with its associated color factor is given by
\begin{align}
\fd{3cm}{figsCO/EFT_loop1_qqqq}=
 2 g^4 
\left({\bf T}^2_{s-u}-\frac12 {\bf T}_t^2 \right) 
  {\bf T}^{\rm tree}  ( -\img\pi ) \Big(\frac{-\img }{4\pi}\Big) \int \!\!    \frac{ \ddslash\!^{d-2}k_\perp }{ 
    ({\vec k}_\perp^{\,2})({\vec k}_\perp\plus {\vec q}_\perp)^2 }
 \,.
\end{align}
This is consistent with the picture from unitarity, namely that the pure Glauber diagrams reproduce just a phase.  If the crossbox were not vanishing, then due to its propagator structure in $k^\pm$ it would need to be imaginary, which would imply that it has a non-vanishing cut, which it should not. Note that here, and throughout this section, we use the color space notation of Ref.~\cite{Catani:1996vz}, and define ${\bf T}_s={\bf T}_1+{\bf T}_2$, ${\bf T}_t={\bf T}_1+{\bf T}_4$, ${\bf T}_u={\bf T}_1+{\bf T}_3$, and ${\bf T}_{s-u}^2=({\bf T}_s^2-{\bf T}_u^2)/2$. 
We also use the notation ${\bf T}^{\rm tree}$ for the color (and spin) structure of the tree level amplitude.
See e.g. Refs.~\cite{DelDuca:2011ae,Bret:2011xm} for a detailed discussion.

On the other hand, for Reggeons, the box and cross box give identical kinematic results, and their color factors  are symmetrized over (see e.g. the textbook discussion in \cite{Forshaw:1997dc} or \cite{Caron-Huot:2013fea} for a more modern discussion)
\begin{align}
  \frac12\left(\hspace{-0.3cm} \fd{3cm}{figsCO/Reggeon_box}\!\! +\!\!  \fd{3cm}{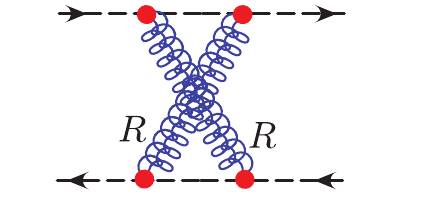}
  \hspace{-0.5cm}\right)
  =2 g^4 \left(  {\bf T}^2_{s-u}\right) 
  {\bf T}^{\rm tree} ( -\img\pi ) \Big(\frac{-\img}{4\pi}\Big) \int \!\!    \frac{ \ddslash\!^{d-2}k_\perp }{ 
    ({\vec k}_\perp^{\,2})({\vec k}_\perp\plus {\vec q}_\perp)^2 } \,.
\end{align}  
Here we see that the ${\bf T}^2_t$ color structure is explicitly absent, as compared to the result in the EFT. This symmetrization enforces the notion of signature. The color structure ${\bf T}^2_{s-u}$ is odd under the $s\leftrightarrow u$ exchange, and so only contains the $1$, $8_S$ and 27.
The same is true for higher numbers of Glauber exchanges \cite{Caron-Huot:2017fxr}.  Therefore, in general, crossed diagrams are drawn in the Reggeon EFT and symmetrized over, and the associated color factors differ  from the naively similar EFT diagrams. 

\subsubsection*{Glauber-Soft Couplings:}

At a more technical level, the separation between Glauber potential contributions and soft modes in the EFT also leads to some diagrammatic differences between Reggeon field theory diagrams and diagrams in the Glauber EFT. As an example, consider the Pomeron.  In the standard Reggeon picture (see e.g. Ref.~\cite{Kovchegov:2012mbw}) for a review), the LL pomeron is viewed as a ladder of Reggeized gluons, where each emission arises from a Lipatov vertex,
\begin{align}
\fd{4cm}{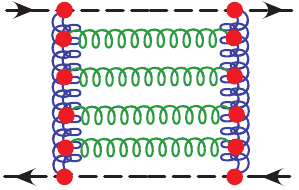}\,.
\end{align}
Here we have indicated the Reggeon by a blue double gluon and soft gluons in green.

In the EFT, the diagrams giving rise to the Pomeron are quite different. In particular, there do not exist diagrams in the EFT with multiple insertions of the Lipatov vertex on a Glauber line. Indeed, the Glauber line corresponds to an insertion of the Glauber potential operator, which due to their operator form can not be stacked up in the $t$-channel. The soft Glauber operator has Feynman rules for multiple soft emissions, giving rise to diagrams such as
\begin{align}
\fd{3cm}{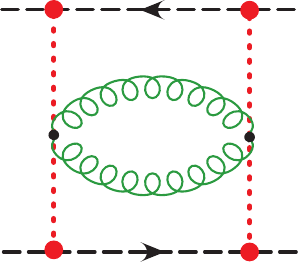}\,.
\end{align}
One can also have multiple emissions along the ladder, which are diagramatically more similar to those that appear in the Reggeon case. However, these arise from $T$-products of $n$-$s$ and $\bar n$-$s$ operators, such as
\begin{align}
\fd{2.5cm}{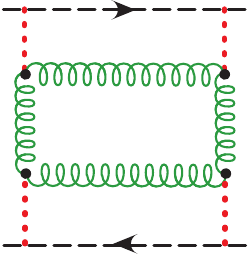}\,.
\end{align}
It is ultimately the combination of these two types of diagrams that reproduces the full Pomeron. For the LL Pomeron the tower of associated logarithms is reproduced in the EFT from the singlet renormalization group equation from graphs with two Glauber exchange with a single soft gluon loop (at one lower order in $\alpha_s$ than the graphs shown above). We will discuss this in much more detail later on. It is important to understand this difference in the topology of diagrams since it leads to somewhat different use of language.

Furthermore, in the Reggeon field theory approach, one gets complicated multi-Reggeon vertices, Pomeron loops, etc. Some of these Reggeon scattering vertices have been calculated, in particular by Bartels and collaborators \cite{Bartels:1978fc,Bartels:1980pe,Bartels:1991bh,Bartels:1994jj,Bartels:1999aw,Bartels:2002au,Bartels:2004hb,Bartels:2007dm,Bartels:2015gou,Bartels:2017nxb}. However, in the Glauber EFT, since one cannot have multiple insertions of the Lipatov vertex, these have no naive counterpart. Instead, we will argue in \Sec{sec:consistent} that these actually arise from the rapidity renormalization group, and it is the structure of the rapidity renormalization group that actually is more similar to the Reggeon Field Theory.

\subsubsection*{Soft-Collinear Consistency:}

Another feature of the EFT, is that since the rapidity regulator is introduced in the EFT only as a means of separating soft, collinear and Glauber contributions to an amplitude, the full amplitude must be independent of the rapidity cutoff. This implies that rapidity regulator dependence must exactly cancel between the $n$-collinear, $\bar n$-collinear and soft sectors. Therefore, all evolution equations, including the standard amplitude level Reggeization, or BFKL equations, can be derived either from purely collinear physics, or from purely soft physics, giving multiple perspectives on the same behavior. While this cutoff independence is quite common in the study of resummation for event shapes, it is less common in the forward scattering literature, where evolution equations are generally viewed in isolation. A large part of this paper will be devoted to understanding this collinear perspective on the Regge limit for amplitudes.

In summary, while the Reggeon Field Theory and the Glauber effective field theory aim to describe the same physics, they are quite distinct. In particular, the Glauber EFT is a genuine four dimensional field theory. The diagrams in the two theories are naively similar, but in fact quite distinct, and this should be taken into account when reading the paper. We hope that this serves as a warning to the reader. Much of the remainder of this paper will develop and better understand the differences between the two approaches.

\section{Multi-Glauber Operators and Soft-Collinear Consistency}
\label{sec:multi_glauber}

Our goal is to provide a factorized expression for the $2\to 2$ scattering amplitude in the forward limit,  to all orders in the coupling ($\alpha_s$), and at leading power in the $t/s$ expansion. It is well known from a variety of approaches that this requires a sum over an infinite number of Glauber (Regge) exchanges. We will derive a factorized expression for the amplitude as a sum over gauge invariant soft and collinear operators. By studying the renormalization group consistency in the effective field theory, we will then be able to derive the all loop mixing structure of these operators, which defines an infinite dimensional mixing matrix in the space of soft/collinear operators. We discuss connections between this picture, and other approaches to the Regge limit, including both the traditional Reggeon Field theory \cite{Gribov:1968fc,Abarbanel:1975me,Baker:1976cv}, and its more modern incarnations \cite{Caron-Huot:2013fea,Caron-Huot:2017fxr,Falcioni:2020lvv,Falcioni:2021buo,Falcioni:2021dgr,Caola:2021izf}.

\subsection{Factorization into Multi-Glauber Operators}
\label{sec:basis}

\begin{figure}
\begin{center}
\includegraphics[width=0.5\columnwidth]{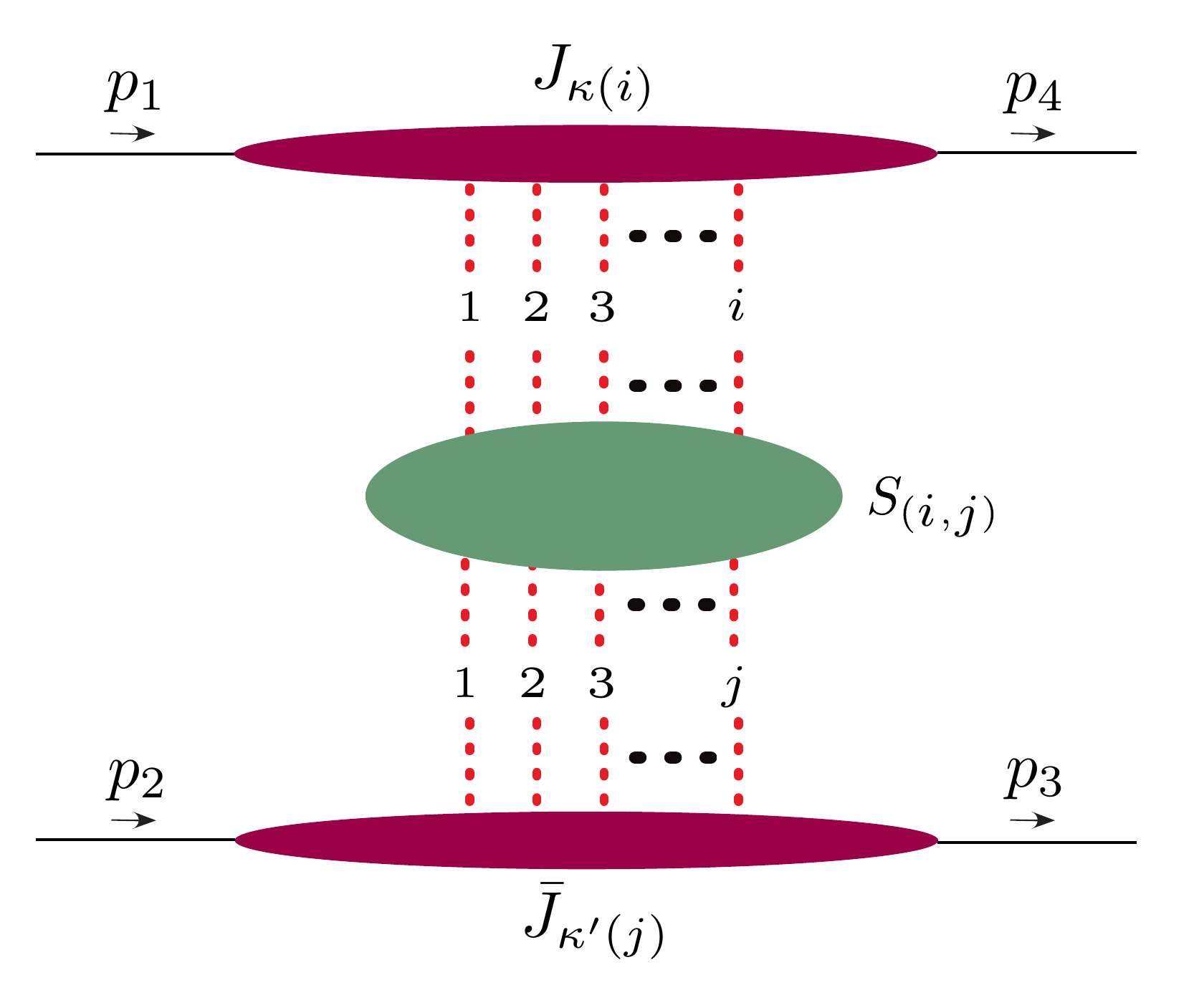}
\end{center}
        \vspace{-0.4cm}
        \caption{\setcaptionskip The factorized structure of the $2\to2$ forward scattering amplitude in the Glauber EFT. The amplitude is expressed as a sum over terms with different numbers of Glauber operators exchanges. For a fixed number of Glaubers,  the collinear factors describe the interaction of the projectiles with the Glaubers, while the soft factor describes radiative corrections to the Glauber potential. }
        \label{fig:fact_diagram}
        \setmainskip
\end{figure}

To factorize the $2\to2$ forward scattering amplitude into a convolution over a soft function and collinear functions in SCET, we begin by expanding the time evolution operator in powers of the Glauber Lagrangian. We organize this expansion  by the number of $n$-collinear and $\bn$-collinear operator insertions, $k$ and $k'$, respectively~\cite{Rothstein:2016bsq}. This results in the following formal expansion of the time evolution operator 
\begin{align} \label{eq:Uglabexp}
  T \exp \left[ \img\! \int\!\! d^4x\, {\cal L}_G^{\rm II(0)}(x)  \right]
   &=\bigg[ 1 + \img \int\!\! d^4y_1\: {\cal L}_G^{\rm II(0)}(y_1) + \frac{\img^2}{2!} T\!\! \int\!\! d^4y_1\, d^4y_2\: {\cal L}_G^{\rm II(0)}(y_1) {\cal L}_G^{\rm II(0)}(y_2) + \ldots \bigg]
  \nn\\
   &= 1 + 
   T \sum_{k=1}^\infty \sum_{k'=1}^\infty   
     \bigg[  \prod_{i=1}^{k} \int\!\! [dx_i^\pm]\!\! \int\!\! \frac{d^2q_{\perp i}}{q_{\perp i}^2}
      \big[{\cal O}_n^{q A_i}(q_{\perp i}) + {\cal O}_n^{g A_i}(q_{\perp i})\big](x_i) \bigg]
  \nn\\
  &\qquad\qquad\qquad \times 
  \bigg[ \prod_{i'=1}^{k'} \int\!\! [dx_{i'}^\pm]\!\!\int\!\! \frac{d^2q_{\perp i'}}{q_{\perp i'}^2} 
  \big[{\cal O}_\bn^{q B_{i'}}(q_{\perp i'}) +  {\cal O}_\bn^{g B_{i'}}(q_{\perp i'})\big](x_{i'}) \bigg]
  \nn\\
  &\qquad\qquad\qquad \times 
     O_{s(k,k')}^{A_1\cdot A_k,B_1\cdots B_{k'}}(q_{\perp 1},\ldots,q_{\perp k'})(x_1,\ldots,x_{k'})
  \nn\\
&\equiv 1 + \sum_{k=1}^\infty \sum_{k'=1}^\infty \ U_{(k,k')}
  \,, 
\end{align}
where $T$ is the time-ordering operation. 
At leading power, the full $2 \to 2$ connected amplitude for leading power forward scattering is given by 
\begin{align} \label{eq:Uglabexpn}
  \cM_{2\to 2}^{\kappa\kappa'} & = \img \left< p_3^{\kappa'}\, p_4^{\kappa} \right|  \sum_{i=1}^\infty \sum_{j=1}^\infty \ U_{(i,j)}
 \left| p_1^\kappa \, p_2^{\kappa'} \right>_{\rm conn} 
   \equiv  \sum_{i=1}^\infty \sum_{j=1}^\infty \cM_{(i,j)}^{\kappa\kappa'} 
  \,,
\end{align}
where $\kappa,\kappa'$ label the partons involved in scattering, i.e. $\kappa,\kappa'=q,\bar q,g$ for quarks, antiquarks, or gluons, respectively. Interactions involving the factorized ${\cal L}^{(0)}$ Lagrangian are left as part of the evolution of the states.
Since different collinear sectors are decoupled in the absence of Glauber exchanges (or hard interactions) the ``$1$'' in \eq{Uglabexp} does not contribute.  

Each term $\cM_{(i,j)}^{\kappa\kappa'}$ can be separately factorized into gauge invariant soft and collinear functions, leading ultimately to the expression for the $2\to2$ scattering amplitude shown in \fig{fact_diagram}.  We begin by illustrating this for $\cM_{(1,1)}^{\kappa\kappa'}$. At the bare level, the factorization of $\cM_{(1,1)}^{\kappa\kappa'}$ is given by 
\begin{align}
  \cM_{(1,1)}^{\kappa\kappa'}\! &= \!\img  \int\!\! \frac{\ddslash\!^{d'\!}\ell_{1\perp}}{\vec\ell_{1\perp}^2} \frac{\ddslash\!^{d'\!}\ell_{1\perp}'}{\vec\ell_{1\perp}^{\prime\,2}}\,
   J_{\kappa(1)}^A(\ell_{1\perp},\epsilon,\eta)\,
   S_{(1,1)}^{AA'}(\ell_{1\perp},\ell_{1\perp}',\epsilon,\eta)
    \bar J_{\kappa'(1)}^{A'}( \ell_{1\perp}',\epsilon,\eta)
   \deltaslash^{d'\!\!}(\ell_{1\perp}\!-\!q_\perp)
   \deltaslash^{d'\!\!}(\ell_{1\perp}'\!-\!q_\perp)
  \nn\\
   &= \frac{\img}{\left(\vec q_\perp^{\;2}\right)^{\!2}} J_{\kappa(1)}^A(q_{\perp},\epsilon,\eta)\ 
   S_{(1,1)}^{AA'}(q_{\perp},\epsilon,\eta)\ 
   \bar J_{\kappa'(1)}^{A'}(q_{\perp},\epsilon,\eta)
 \,.
\end{align}
Here $\eta$ is a rapidity regulator to regulate soft-collinear divergences\footnote{
  There are two types of rapidity regulators in this paper: a soft-collinear regulator $\eta$ and a Glauber regulator $\eta'$. The $\eta$ regulator in the form $|2k^z|^{-\eta}$, $|k^\pm|^{-\eta}$ (introduced in Refs.~\cite{Chiu:2011qc,Chiu:2012ir}) regulates soft-collinear divergences in the traditional sense; the $\eta'$ regulator in the form $|2 k^z|^{-\eta'}$ is put on Glauber momenta used to separate Glaubers in longitudinal position, and taking $\eta'\to0$ after calculation amounts to collapsing Glaubers to the same in longitudinal position (this is explained in more detail in Ref.~\cite{Rothstein:2016bsq}).
  In Ref.~\cite{Rothstein:2016bsq} these two types of regulators were not separated, and a common name $\eta$ was used. The explanation for the need to distinguish them beyond the lowest order loop graphs, and send $\eta'\to 0$ prior to $\eta\to 0$,  was given in Ref.~\cite{Moult:2022lfy}.
  
  While the $\eta'$ regulator is the only regulator that is known to work for Glauber exchanges,
  we note that the $\eta$ regulator can also be replaced by other soft-collinear regulators. In particular, we anticipate that other regulators will simplify calculations at higher collinear/soft loop orders. In this paper, we will use the $\eta$ regulator to carry out calculations when we are only interested in extracting the leading rapidity anomalous dimension equation from $1/\eta$ poles (calculations that can be done with any regulator). In contrast we will use a different regulator with improved analytic structure in \sec{vanish} to show that $1\to j>1$ soft transitions graphs vanish to all orders in soft loops.
},
$d'=d-2=2-2\epsilon$ is the dimension of the perpendicular integrals, and dimensional regularization is used to regulate infrared divergences. For notational convenience we define 
$\ddslash\!^{d'\!}\ell_\perp = \df^{d'\!}\ell_\perp/(2\pi)^{d'}$ and 
$\deltaslash^{d'\!\!}(\ell_\perp) = (2\pi)^{d'}\delta^{d'}(\ell_\perp)$. 
Each Glauber exchange generates a convolution integral over perpendicular momenta, a $1/\vec \ell_\perp^{\,2}$ factor, and the contraction of adjoint color indices $A, A'$ between a collinear and soft sector. 
There is also one overall momentum conserving $\delta$-function, and one momentum conserving $\delta$-function for the soft function, which can be rewritten in the form shown, with momentum transfer $q_\perp=p_{2\perp}-p_{3\perp}=p_{4\perp}-p_{1\perp}$ following the notation in \fig{fact_diagram}. 
Here $J_{\kappa(1)}^A$ is given by a purely $n$-collinear $\langle p_1| \mathcal{O}_n^{\kappa A} | p_4\rangle$ matrix element, $\bar J_{\kappa'(1)}^{A'}$ by an analogous purely $\bn$-collinear matrix element, and $S_{(1,1)}^{AA'}$ by a purely soft vacuum matrix element.
Note that we choose to define the collinear and soft functions such that the convolutions integrals are normalized in a manner that makes them dimensionless when $d'=2$, and that we can think of each of the $1/\vec\ell_{i\perp}^{\,2}$ factors as a propagator from a Glauber exchange between the $n-s$ or $s-\bn$ sectors. This slightly differs from the convention in Ref.~\cite{Rothstein:2016bsq} for the cross-section level $2$-to-$2$ scattering factorization, which was used to show how to derive the BFKL equation from within SCET.

This factorization can also be extended to the generic $(i,j)$ Glauber exchange term in the series pictured in \fig{fact_diagram}, which gives
\begin{align}\label{eq:fw_fact}
 \cM_{(i,j)}^{\kappa\kappa'}
 &=  
  \img \IInt_{\perp(i,\,j)}  
 J_{\kappa(i)}^{A_1\cdots A_i}( 
\ell_{1\perp},\, \dots, \, \ell_{i\perp},\epsilon,\eta)\, 
S_{(i,j)}^{A_1\cdots A_i\,A_1'\cdots A_j'}(\ell_{1\perp},\, \dots, \, \ell_{i\perp}; {\ell}^{\prime}_{1\perp} ,\, \dots , {\ell}^{\prime}_{j\perp},\epsilon,\eta)\, 
  \nn\\
 & \qquad\times {\bar J}_{\kappa' (j)}^{A_1'\cdots A_j'}({\ell}^{\prime}_{1\perp} ,\, \dots , {\ell}^{\prime}_{j\perp},\epsilon,\eta) 
  \, \nn\\
 &= \img \IInt_{\perp(i,\,j)}  
  J_{\kappa(i)}^{\{A_i\}}\bigl( \{\ell_{i\perp}\},\epsilon,\eta \bigr)\:
  S_{(i,j)}^{\{A_i\}\,\{A_j'\}}\bigl(\{\ell_{i\perp}\},\{\ell'_{j\perp}\},
   \epsilon,\eta\bigr)\:
  \bar J_{\kappa'(j)}^{\{A'_j\}}\bigl( \{\ell'_{j\perp}\},\epsilon,\eta \bigr)
  \,,
\end{align}
where in the second line we introduced a short hand for superscripts and momentum arguments.
The collinear $J$, $\bar J$ and soft $S$ functions are tied together through a convolution in transverse momentum, where we define the integral $\int\!\!\!\int_{\perp(i,\,j)}$ as
\begin{align} \label{eq:iint_perp}
  \IInt_{\perp(i,\,j)} 
  \equiv 
  \frac{(-\img)^{i+j}}{i!\, j!}
  \int \prod_{m=1}^{i}\! \frac{\ddslash\!^{d'}\!\ell_{m\perp}}{\vec\ell_{m\perp}^{\,2}} \, \deltaslash^{d'}\!\Bigl(\sum_m \ell_{m\perp}- q_\perp\Bigr) 
  \int\! \prod_{n=1}^{j} \frac{\ddslash\!^{d'}\!\ell'_{n\perp}}{{\vec\ell_{n\perp}^{\,\prime 2}}} \,  \deltaslash^{d'}\!\Bigl(\sum_n {\ell}^{\prime}_{n\perp}- q_\perp\Bigr) \,.
\end{align}
Note that the momentum conservation condition $\sum_m\ell_{m\perp}=\sum_n\ell'_{n\perp}=q_\perp$ imposed in the definition of $S_{(i,j)}$, 
implies that there are $i-1$ independent $\ell_{m\perp}$'s and $j-1$ independent $\ell'_{n\perp}$'s in the arguments of $S_{(i,j)}$.

Combining these terms together provides us with a factorized expression for the $2\to2$ scattering amplitude
\begin{align} \label{eq:factorization_summary}
  \cM_{2\to 2}^{\kappa\kappa'}
  & = \sum_{i=1}^\infty \sum_{j=1}^\infty \, \img \! \IInt_{\perp(i,\,j)}  
  J_{\kappa(i)}^{\{A_i\}}\bigl(\{\ell_{i\perp}\},\epsilon,\eta\bigr)\:
  S_{(i,j)}^{\{A_i\}\,\{A_j'\}}\bigl(\{\ell_{i\perp}\},\{\ell'_{j\perp}\},
   \epsilon,\eta\bigr)\:
  \bar J_{\kappa'(j)}^{\{A'_j\}}\bigl(\{\ell'_{j\perp}\},\epsilon,\eta\bigr)
  \,.
\end{align}
Note that in the Glauber picture the color space for decomposing objects is naturally in the t-channel, where the indices $\{A_i\}$ and $\{A_j'\}$ live. We will decompose these indices into irreducible representations later on.
At the level presented in \eq{factorization_summary}, this is a fairly formal looking expression. The main goal of this paper will be to make the meaning of these functions precise, and understand the renormalization, and renormalization group mixing structure of these functions.

Although \eq{factorization_summary} gives an expression for the amplitude as an infinite sum of soft and collinear factors, at any order in the expansion in the coupling, only a finite number of terms contribute. To see this, and to define our normalization for the soft and collinear operators, we can consider  $J$, $S$, $\bar J$ and $\cM_{(i,j)}$  order by order in the coupling constant, $\alpha_s=g^2/(4\pi)$. The lowest order $S_{(i,j)}$ is ${\cal O}(\gs^0)$ and is nonzero only for $i=j$,
\begin{align}\label{eq:S_LO}
  S_{(i,j)}^{[0]\{A_i\}\,\{A_j'\}}\bigl(\{\ell_{i\perp}\},\{\ell'_{j\perp}\}
  = 2\delta_{ij} \, j!\, \img^{\,j}  \prod_{m=1}^{j} \vec\ell_{m\perp}^{\;2}\, \delta^{A_m A_m'} \,  \prod_{n=1}^{j-1} \deltaslash^{d'}(\ell_{n\perp}-\ell'_{n\perp})\,. 
\end{align}
The lowest order $n$-collinear radiative jet functions for the quark, antiquark, and gluon channels start at ${\cal O}(g^i)$ and are
given by 
\begin{align}\label{eq:J_LO} 
J_{q(i)}^{[0]A_1\cdots A_i}
  &= \gs^i \Big[ \bar u_n  \frac{\bnslash}{2} T^{A_1}\cdots T^{A_i} u_n \Big] 
 \,, 
 \nn\\
 J_{\bar q(i)}^{[0]A_1'\cdots A_i'}
  &= \gs^i
    \Big[ \bar v_n  \frac{\bnslash}{2} \bar T^{A_1'}\cdots \bar T^{A_i'} v_n \Big] \,,
\nn\\
   J_{g(i)}^{[0]A_1\cdots A_i}
  &= \gs^i
    \Big[ \epsilon_{n\mu}^{*} \, b^{\mu\nu} \cT^{A_1}\cdots \cT^{A_i} \epsilon_{n\nu}\Big] 
  \,,
\end{align}
where $T^A_{\cdot\cdot}$, $\bar T^A_{\cdot\cdot}=-(T^A)^T_{\cdot\cdot}$, and $\cT^A_{\cdot\cdot} = \img f^{\cdot A \cdot}$ are color generators in the $3$, $\bar 3$ and $8$ respectively, where the dots indicate matrix indices.  For  forward scattering kinematics, $\bn\cdot p_1 = \bn\cdot p_4$ up to corrections suppressed by $\lambda$, and the gluon polarization vectors $\epsilon_n^\nu$ are contracted with
\begin{align} 
  b^{\mu\nu} &= \bn\cdot p_1 g_\perp^{\mu\nu} -\bn^\mu p_1^\nu - \bn^\nu p_4^\mu
    + \frac{p_{1\perp}\cdot p_{4\perp} \bn^\mu \bn^\nu}{\bn\cdot p_2 }
  \,.
\end{align}
Finally $\bar J_{\kappa'(j)} = \mbox{\raisebox{-0.1cm}{$\stackrel{\text{swap}}{\mbox{\scriptsize $n\leftrightarrow \bn$}}$}}\: J_{\kappa'(j)}$.
Notice that in \eq{J_LO} we do not have dependence on the internal transverse momentum arguments $\ell_{i\perp}$ at this order, but at higher orders they will appear.
Results in \eq{J_LO}  are expressed using the bare coupling $\gs$, which is related to the $\MSbar$ coupling $\gs(\mu)$ by 
$\gs = Z_g  \iota^{\epsilon/2} \mu^\epsilon  \gs(\mu) $  
where $\iota=e^{\gamma_E}/(4\pi)$ and $Z_g$ is the coupling renormalization factor.

In defining the normalization of our amplitude level soft and collinear functions, we have chosen to put all tree-level factors of $g$ from the Glauber exchanges into the collinear functions. This differs from the conventions of Ref.~\cite{Rothstein:2016bsq} where similar factorization formulas were considered for forward scattering cross sections, and tree-level factors of $g$ are in the soft function. 
Our convention for $g$'s is chosen because the lowest  order $S_{(i,j)}^{[0]}$ in \eq{S_LO} plays the role of a transition function. If we put $g$'s in this lowest order $S$, then we would need to have compensating $1/g$'s in the definition of the $\perp$-integrations. Therefore we have grouped the lowest order $\gs$ from the Glauber exchange with the collinear functions.  
Based on the results in Ref.~\cite{Rothstein:2016bsq}, we anticipate that the renormalized combination $J_{\kappa(i)}/g^i$ will be $\mu$ independent. 

Additionally, in defining the normalization for the transverse momentum integrals in \eq{iint_perp}, and correspondingly for the tree-level soft function in \eq{S_LO}, we have included specific factors of $\img$, $1/n!$, and $(2\pi)$. (Note that we exclusively make use of roman $\img$ for the imaginary number $\sqrt{-1}$, and use an italicized $i$ for indices.) To understand these factors it is useful to consider the scattering graph with $N$ Glauber exchanges, and jet and soft functions $J_{\kappa(N)}$ and $S_{(N,N)}$ at tree level, 
\begin{align}
  \cM_{2\to 2}\Big|_{N\text{-glauber}} 
 &= \img \! \IInt_{\perp(N,\,N)}  
  J_{\kappa(N)}^{[0]\{A_N\}} \:
  S_{(N,N)}^{[0]\{A_N\}\,\{A_N'\}}\bigl(\{\ell_{N\perp}\},\{\ell'_{N\perp}\}
   \bigr)\:
  \bar J_{\kappa'(N)}^{[0]\{A'_N\}}
 \nn\\
 &=
 J_{\kappa(N)}^{[0]A_1\cdots A_N} \bar J_{\kappa'(N)}^{[0]A_1\cdots A_N}\:
 \frac{2 (-\img)^{N-1}}{N!} 
 \!\int\!  \bigg[ \prod_{m=1}^{i}\! \frac{\ddslash\!^{d'}\!\ell_{m\perp}}{\vec\ell_{m\perp}^{\,2}} \bigg] \, \deltaslash^{d'}\!\Bigl(\sum_m \ell_{m\perp}- q_\perp\Bigr) \,.
\end{align} 
Here the last line agrees with Ref.~\cite{Rothstein:2016bsq}. For $(N-1)$ Glauber loops from $N$ Glauber exchanges, we see the expected factor of $(-\img)^{N-1}/N!$ associated with the fact that the sum of Glauber box graphs exponentiate into an eikonal phase. 
These same Glauber loop factors were built into our definition of 
$\int\!\!\!\int_{\perp(i,\,j)}$ in \eq{iint_perp}, via $(-\img)^{i+j}/(i! j!)$ for $i$ Glaubers on the upper projectile and $j$ Glaubers for the lower projectile, see \fig{fact_diagram}. When the soft function 
sets some number of the $i$ Glaubers on the upper projectile, equal to $j$ Glaubers on the lower projectile, then it also must cancel or modify these Glauber loop factors, since this identification yields less Glauber loops.  For example, this explains the presence of the $(j!\, \img^j)$ factor in the lowest order soft function in \eq{S_LO}  where the number of Glaubers attaching to the upper and lower projectiles, $i=j$.
In principal, if we only consider diagonal terms where $i=j$, then there is still some freedom in how one organizes factors of $\img$. However, we will see below that once off-diagonal $i\ne j$ terms come into play, the organization given in the definitions above is preferred. 

\subsection{Renormalization Group Evolution and Soft Collinear Consistency}\label{sec:consistent}

The key issue to providing an organizational principle for the  $2\to 2$ forward scattering amplitude is to understand the general structure of the mixing of the infinite set of collinear and soft operators in \eq{factorization_summary}. We will show that this can achieved by studying the renormalization group consistency of SCET for forward scattering amplitudes with Glauber operators.

The bare jet function $J_{\kappa(i)}^{A_1 \cdots A_i}(\ell_{1\perp}, \cdots, \ell_{i\perp},\epsilon,\eta)$ can be expressed in terms of the renormalized jet function $J_{\kappa(j)}^{B_1 \cdots B_j}(k_{1\perp},\cdots,k_{j\perp},\epsilon,\nu)$ by convolving over a counterterm factor  $Z_{J(j,i)}^{B_1\cdots B_j\, A_1\cdots A_i}$,
\begin{align}\label{eq:Jb_Z_Jr}
  & J_{\kappa(i)}^{A_1 \cdots A_i}(\ell_{1\perp}, \cdots, \ell_{i\perp},\epsilon,\eta)
  \\
   &\ =\, \sum_j \int_{\perp(j)}
  J_{\kappa(j)}^{B_1 \cdots B_j}(k_{1\perp},\cdots, k_{j\perp},\epsilon,\nu) \:
  Z_{J(j,i)}^{B_1\cdots B_j\, A_1\cdots A_i} ( k_{1\perp},\cdots, k_{j\perp},\ell_{1\perp},\cdots,\ell_{i\perp},\epsilon,\eta,\nu) 
 \nn\\
 &\ =\, \sum_j \int_{\perp(j)}
  J_{\kappa(j)}^{\{B_j\}}\bigl(\{k_{j\perp}\},\epsilon,\nu\bigr) \:
  Z_{J(j,i)}^{\{B_j\}\, \{A_i\}} \bigl( \{k_{j\perp}\},\{\ell_{i\perp}\},\epsilon,\eta,\nu\bigr) 
  \,, \nn
\end{align}
where we define a shorthand notation for arguments on the second line. Here, $\int_{\perp(j)}$ is given by
\begin{align}\label{eq:int_perp}
  \int_{\perp(j)} \equiv 
  \frac{(-\img)^{j}}{j!}  \int\bigg[  \prod_{m=1}^{j}\! \frac{\ddslash\!^{d'}\!k_{m\perp}}{\vec k_{m\perp}^{\,2}} \bigg]\, \deltaslash^{d'}  \!\Bigl(\sum_{m=1}^j k_{m\perp}- q_\perp\Bigr) 
 \,.
\end{align}
In \eq{Jb_Z_Jr} the counterterm $Z_J$ cancels $1/\eta$ divergences in $J_{\kappa (i)}(\eta)$ that arise from collinear momenta approaching the soft phase space region, leaving the rapidity finite $J_{\kappa (i)}(\nu)$. The $\epsilon$ arguments denote the potential presence of infrared divergences, and we will drop them below for simplicity.  Due to the symmetry $n\leftrightarrow \bn$ for the rapidity regulator used here, we have the same $Z_J$ renormalization factor for $\bar J$, which for later convenience we denote with left multiplication
\begin{align} \label{eq:Jbb_Z_Jbr}
  & \bar J_{\kappa'(j)}^{\{A_j'\}}\bigl(\{\ell'_{j\perp}\},\epsilon,\eta\bigr)
 =\, \sum_m \int_{\perp(m)}
  Z_{J(j,m)}^{\{A'_j\}\, \{B'_m\}} \bigl( \{\ell'_{j\perp}\},\{k'_{m\perp}\},\epsilon,\eta,\nu\bigr) \:
  \bar J_{\kappa'(m)}^{\{B'_m\}}\bigl(\{k'_{m\perp}\},\epsilon,\nu\bigr)  
 \,.
\end{align}

\begin{figure}
  \begin{center}
  \subfloat[]{\label{fig:Jn_Z}
  \includegraphics[width=0.33\columnwidth,valign=t]{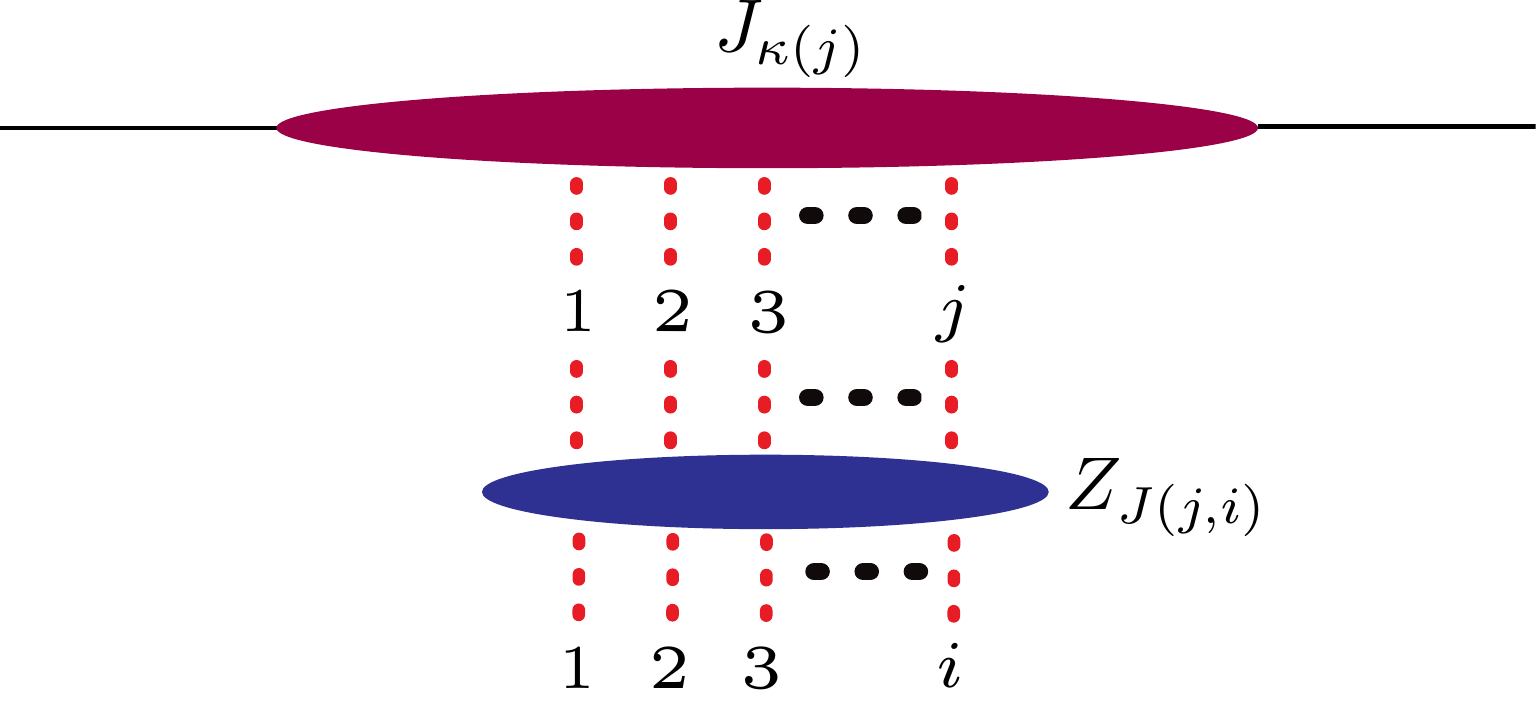}
  }
  \subfloat[]{\label{fig:Jn_Z1j}
  \includegraphics[width=0.33\columnwidth,valign=t]{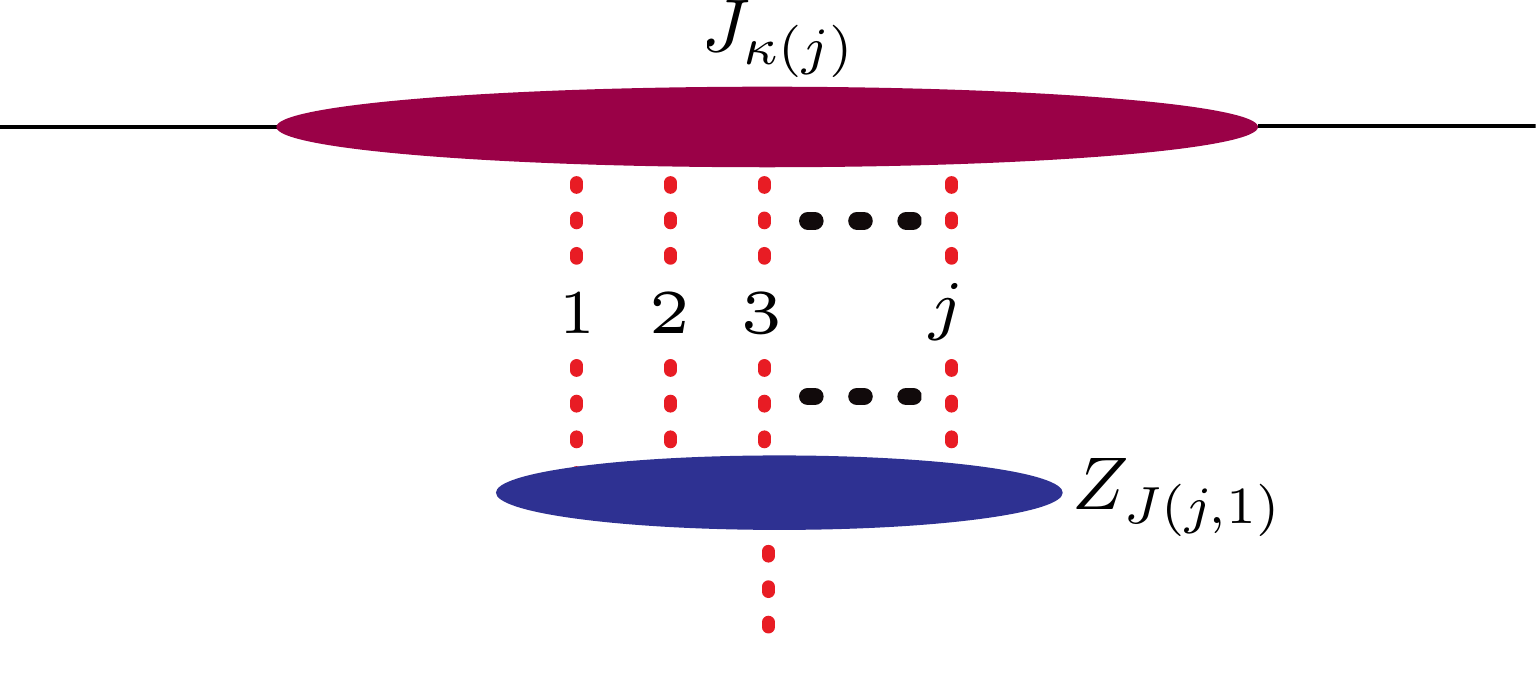}
  }
  \subfloat[]{\label{fig:Jn_Zi1}
  \includegraphics[width=0.33\columnwidth,valign=t]{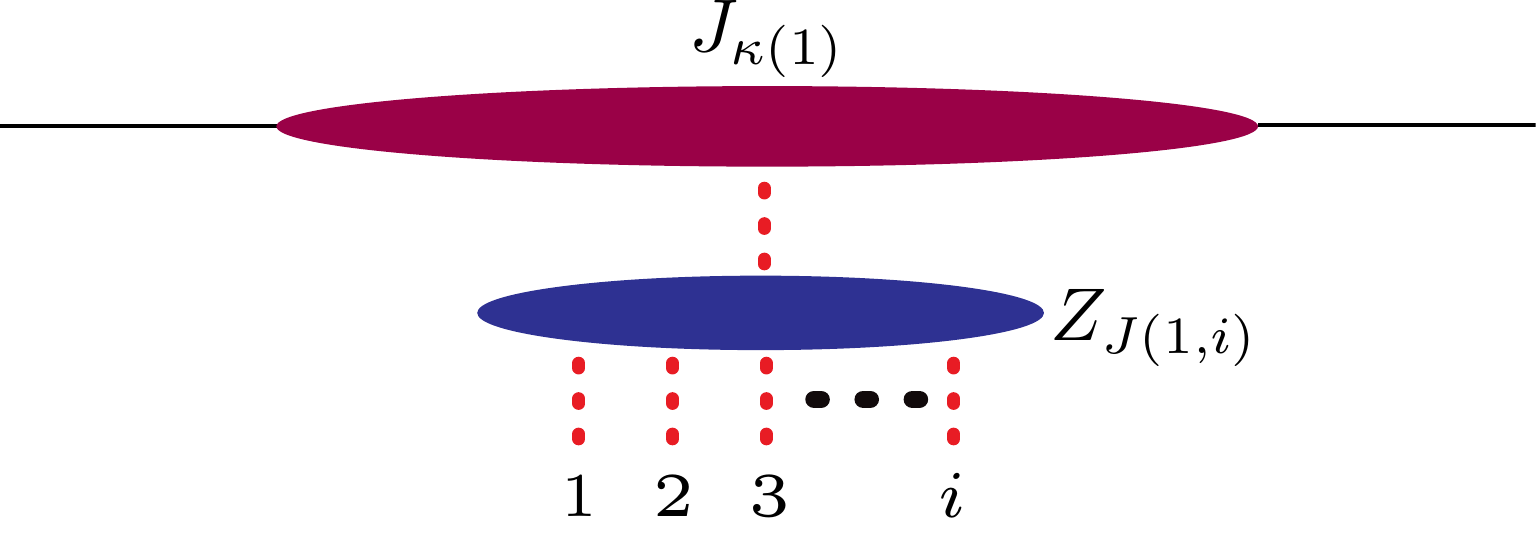}
  }
  \end{center}
  \caption{(a) The renormalization factors $Z_{J(j,i)}$ can be thought of as interaction among Glaubers. 
  (b) $Z_{J(j,1)}$ with $j>1$ vanishes as shown in \Sec{sec:vanish}.
  (c) $Z_{J(1,i)}$ with $i>1$ vanishes as shown in \Sec{sec:vanish}.
  }
  \label{fig:Z_factor}
\end{figure}

In both \eqs{Jb_Z_Jr}{Jbb_Z_Jbr} the $Z_{J(j,i)}^{\{B_j\}\{A_i\}}$ can be thought as an interaction connecting $i$ Glaubers to $j$ Glaubers, as shown in \fig{Jn_Z}.
Diagrammatically, Glaubers are illustrated by the ``vertical'' lines, while the action of the counterterms are illustrated as ``horizontal'' blobs.
In the transverse renormalization integral in \eq{int_perp} we include the same Glauber loop factor, $(-\img)^j/j!$, that appeared in the transverse integral for the factorization theorem in \eq{iint_perp}.  Later in this section we will demonstrate that \emph{this factor must be included} in \eq{int_perp}, as it is constrained by RGE consistency of the EFT. 
While the diagrams in \fig{Jn_Z} do not technically arise from Feynman diagrams in our EFT, the presence of the same Glauber loop factors indicates that there is no harm in interpreting them in this manner, and hence we use the same dashed red line notation for the transverse momentum and color index exchanges.

\begin{figure}
  \begin{center}
  \includegraphics[width=0.3\columnwidth]{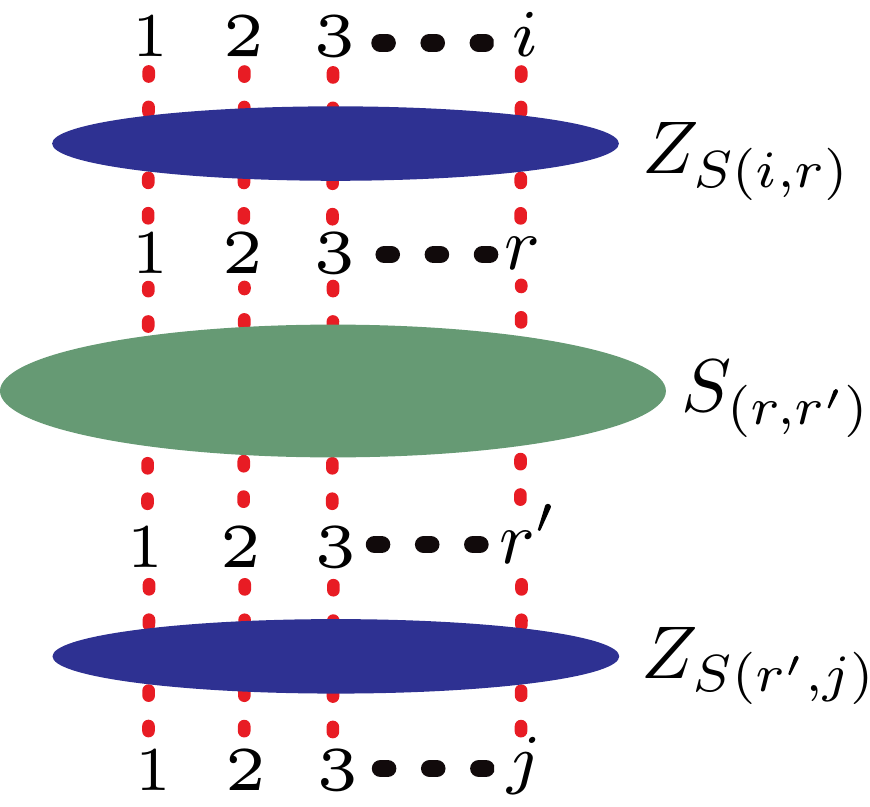}
  \end{center}
          \vspace{-0.4cm}
          \caption{\setcaptionskip An illustration of the renormalization group structure for multi-Glauber soft functions. The $Z$ factors from rapidity renormalization can change the number of Glaubers on either the top or bottom of the soft function.}
          \label{fig:Z_S}
          \setmainskip
  \end{figure}

Similar to the collinear case, the bare soft function can be renormalized by $Z_S$ factors. Unlike the collinear case the $Z_S$ factors appear on both sides, since they absorb $1/\eta$ poles from soft momenta approaching the $n$-collinear or $\bn$-collinear limits respectively,
\begin{align}\label{eq:ZS}
S_{(i,j)}^{\{A_i\}\{A_j'\}}\bigl(\{\ell_{i\perp}\}, \{{\ell}^{\prime}_{j\perp}\},\eta\bigr)\, 
  &\!=\!
  \sum_{r,r'}\! \IInt_{\perp(r,r')} \!
  Z_{S(i,r)}^{\{A_i\}\{B_r\}}\bigl(\{\ell_{i\perp}\}, \{k_{r\perp}\},\eta,\nu\bigr)
  \, S_{(r,r')}^{\{B_r\}\{B'_{r'}\}}\bigl(\{k_{r\perp}\},\{k'_{r'\perp}\},\nu\bigr)
  \nn\\
  &\qquad\times
   Z_{S(r',j)}^{\{B'_{r'}\}\{A_j'\}}\bigl(\{k'_{r'\perp}\}, \{{\ell}^{\prime}_{j\perp}\},\eta,\nu\bigr) 
  \,.
\end{align}
Here we have omitted $\epsilon$ arguments for simplicity. By symmetry, the same $Z_S$ appears on the left and right hand side of $S_{(r,r')}$. This renormalization group structure is illustrated in \fig{Z_S}.

We can write \Eqs{eq:Jb_Z_Jr}{eq:ZS}  as abstract matrix equations in the infinite space of soft and collinear functions
\begin{align}
  \mathbf{J}_\kappa(\eta)&=\mathbf{J}_\kappa(\nu)\cdot \mathbf{Z_J}(\eta,\nu)\,,
  \\
  \mathbf{S}(\eta)&=\mathbf{Z_S}(\eta,\nu)\cdot \mathbf{S}(\nu)\cdot \mathbf{Z_S}(\eta,\nu)\,,
  \nn \\
  \mathbf{\bar J}_{\kappa'}(\eta)&= \mathbf{Z_J}(\eta,\nu)\cdot \mathbf{\bar J}_{\kappa'}(\nu)
   \,, \nn
\end{align}
where the vector $\mathbf{J}_\kappa$ is defined as $\mathbf{J}_\kappa\equiv (J_{\kappa(1)}, J_{\kappa(2)},J_{\kappa(3)}\,,\dots)$, and we take $S_{(i,j)}$ as an element of the matrix $\mathbf{S}$ and similarly for $\mathbf{Z_J}$ and $\mathbf{Z_S}$. Multiplication in the equations above includes summing over the number of Glaubers, carrying out the transverse convolutions $\int_\perp$, and the color contractions. 

The $2\to 2$ forward scattering amplitude is independent of the rapidity cutoff $\eta$ and the rapidity factorization scale $\nu$, and has the same factorized form at the bare and renormalized level. 
\begin{align}
 -\img\,\cM_{2\to2}^{\kappa\kappa'}=\mathbf{J}_\kappa(\eta)\cdot \mathbf{S}(\eta)\cdot \mathbf{\bar J}_{\kappa'}(\eta) =\mathbf{J}_\kappa(\nu)\cdot \mathbf{S}(\nu)\cdot \mathbf{\bar J}_{\kappa'}(\nu)\,.
\end{align}
This relies on a property of SCET, namely renormalization group consistency between the soft and collinear sectors. In particular it implies that
\begin{align} \label{eq:ZSisZJ}
  \mathbf{Z_S}=\mathbf{Z_J}^{\!\!\!-1}\,.
\end{align}
This equation is important in that it places strong constraints on the structure of the renormalization, and implies that rapidity renormalization group equations in the EFT can be derived \emph{either} from the soft or collinear sectors.  Since there are advantages to each of the soft and collinear sector calculations, this correspondence will be useful.

The rapidity anomalous dimensions for $\mathbf{J}_\kappa$ are defined as 
\begin{align}
  \mathbf{\Gamma} \equiv -(\nu \partial_\nu \mathbf{Z_J})\cdot \mathbf{Z_J}^{\!\!\!-1}\,,
\end{align}
where $\partial_\nu = \partial/\partial\nu$. 
When needed we will denote the components of $\mathbf{\Gamma}$ as $\gamma_{(i,j)}$.
Using the $\nu$ independence of $\mathbf{J}_\kappa(\eta)$, $\mathbf{\bar J}_{\kappa'}(\eta)$ and  $\mathbf{S}(\eta)$, plus \eq{ZSisZJ}, it is then easy to derive that $\mathbf{J}_\kappa(\nu)$, $\mathbf{S}(\nu)$, and $\mathbf{\bar J}_{\kappa'}(\nu)$ satisfy the following RRGEs
\begin{align}
  \nu\partial_\nu\mathbf{J}_\kappa(\nu)=\mathbf{J}_\kappa(\nu)\cdot\mathbf{\Gamma}\,,
  \qquad
  \nu\partial_\nu\mathbf{S}(\nu)=-\mathbf{\Gamma}\cdot\mathbf{S}(\nu)-\mathbf{S}(\nu)\cdot\mathbf{\Gamma}\,,
  \qquad
  \nu\partial_\nu \mathbf{\bar J}_{\kappa'} = \mathbf{\Gamma} \cdot \mathbf{\bar J}_{\kappa'}(\nu)
\,.
\end{align}
In general these equations have a matrix structure with mixing in both the space of the number of Glauber exchanges and color space, and a convolution structure in momentum space.

Consider the matrix structure in the number of Glauber exchanges $i=1,2,3,\ldots$ and $j=1,2,3,\ldots$. Since this space is infinite dimensional, it is interesting to understand the order in $\gs$ for the entries. The functions $\mathbf{J}$, $\mathbf{S}$, $\mathbf{Z_J}$, $\mathbf{\Gamma}$ have the following counting 
\begin{align}\label{eq:J_gs_order}
  \mathbf{J}_\kappa &= \bigl[\cO(\gs^1),\cO(\gs^2), \cO(\gs^3), \cO(\gs^4),\dots \bigr]\,,\\
  \mathbf{S} &= \begin{bmatrix}
    1+\cO(\gs^2) & 0 & 0 & 0 & 0 & \cdots \\
    0 & 1+\cO(\gs^2) & \cO(\gs^3) & \cO(\gs^4) &\cO(\gs^5) &\cdots \\
    0 & \cO(\gs^3) & 1+\cO(\gs^2) & \cO(\gs^3) &\cO(\gs^4) &\cdots \\
    0 & \cO(\gs^4) & \cO(\gs^3) & 1+\cO(\gs^2) &\cO(\gs^3) &\cdots \\
    0 & \cO(\gs^5) & \cO(\gs^4) & \cO(\gs^3) &1+\cO(\gs^2) &\cdots  \\
    \vdots &  \vdots &\vdots&\vdots&\vdots&\ddots 
    \end{bmatrix}
  \,,  \\
  \label{eq:Z_gs_order}
  \mathbf{Z_J} &= \begin{bmatrix}
    1+\cO(\gs^2) & 0 & 0 & 0 & 0 & \cdots \\
    0 & 1+\cO(\gs^2) & \cO(\gs^3) & \cO(\gs^4) &\cO(\gs^5) &\cdots \\
    0 & \cO(\gs^3) & 1+\cO(\gs^2) & \cO(\gs^3) &\cO(\gs^4) &\cdots \\
    0 & \cO(\gs^4) & \cO(\gs^3) & 1+\cO(\gs^2) &\cO(\gs^3) &\cdots \\
    0 & \cO(\gs^5) & \cO(\gs^4) & \cO(\gs^3) &1+\cO(\gs^2) &\cdots  \\
   \vdots &  \vdots &\vdots&\vdots&\vdots&\ddots 
    \end{bmatrix}
    \,, \\
    \label{eq:Gamma_gs_order}
    \mathbf{\Gamma} &= \begin{bmatrix}
      \cO(\gs^2) & 0 & 0 & 0 & 0 & \cdots \\
      0 & \cO(\gs^2) & \cO(\gs^3) & \cO(\gs^4) &\cO(\gs^5) &\cdots \\
      0 & \cO(\gs^3) & \cO(\gs^2) & \cO(\gs^3) &\cO(\gs^4) &\cdots \\
      0 & \cO(\gs^4) & \cO(\gs^3) & \cO(\gs^2) &\cO(\gs^3) &\cdots \\
      0 & \cO(\gs^5) & \cO(\gs^4) & \cO(\gs^3) &\cO(\gs^2) &\cdots  \\
   \vdots &  \vdots &\vdots&\vdots&\vdots&\ddots 
      \end{bmatrix}
      \,. 
\end{align}
The counting for the collinear functions tracks the lowest order result in \eq{J_LO}. 
Here, 1's appearing on the diagonal terms of $\mathbf{S}$ and $\mathbf{Z_J}$ are defined to be the leading order expression of $S_{(i,i)}$ as in \eq{S_LO}. The order counting in $\gs$ is easiest to see in the soft sector. For example, the first non-trivial transition is the $2\to 3$ transition arising from the so called ``tennis court'' graph
\begin{align} \label{eq:tenniscourt}
\fd{5cm}{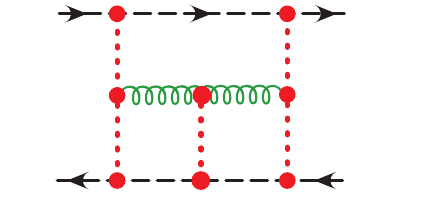}\,,
\end{align}
which is $\cO(\gs^3)$ in the soft sector from the three Glauber-soft couplings.\footnote{This diagram again illustrates the distinction between Glaubers and Reggeons. In the Reggeon Field such a diagram would violate signature.} The generalization to higher $n\to m$ transitions is straightforward.

These matrices are similar to what appears for the Regge $W$ operator's evolution in Ref.~\cite{Caron-Huot:2013fea}. The closest analogy is to the evolution of our $\mathbf{S}$, since both equations have a similar matrix form. There are however some important differences, related to the fact that the Glauber exchanges considered here {\em are not} equivalent to Regge exchanges. In particular in the Reggeon picture the exchanges are constrained from the start by signature symmetry, which forbids odd transitions, like $n\to n+1$. However, for the Glauber exchanges many such entries are non-zero and necessary. Indeed they have a fairly simple form that is constrained by unitarity and crossing symmetry.  In contrast, in the Glauber matrix all the $1\to n$ transitions vanish, as we will show in \Sec{sec:vanish}. However, $1\to n$ transitions with $n$ odd do not vanish for the Reggeon. As we will explain later on, this means that in the Glauber picture the single exchange is a true eigenstate of the Hamiltonian. 

A key reason for this difference is that Glauber exchange diagrams have cut rules that are just like regular Feynman diagrams.  Therefore the two Glauber exchange diagram in the octet has to encode the square of one Glauber exchange in the octet channel through its imaginary part. In fact  multiple Glauber exchange diagrams are entirely determined by the terms needed to maintain  unitarity in the forward scattering, and exponentiate into an eikonal scattering phase~\cite{Rothstein:2016bsq}. 

The implication of there being no $n\to 1$ or $1\to n$ interactions with $n>1$ means that $Z_{J(j,1)}=Z_{J(1,i)}=0$, which are the graphs shown in  \fig{Jn_Z1j} and \fig{Jn_Zi1}.
From the collinear perspective, this result is natural, since, as illustrated in \fig{Jn_Zi1}, with only a single Glauber attaching to the collinear sector there can be no non-trivial cuts of the graph as the two sides of any such cut do not correspond to a physical scattering process in forward scattering kinematics.  Given the vanishing of these terms, renormalization group consistency them implies a corresponding result in the soft sector. In \Sec{sec:vanish} we show that $S_{(1,n)}$ and $S_{(n,1)}$ vanish to all orders.

Since these equations are rather abstract, we now make several comments on the structure of the RRRG, and its simplification at fixed logarithmic orders. First, we note that for the case of single Glauber exchange, the RRGE of $J_{\kappa(1)}$ and $S_{(1,1)}$ is unentangled with the rest of the components of the infinite dimensional matrix. Also, since it has no transverse momentum convolution, and only the single $8_A$ color channel, it is purely multiplicative. Therefore the all-orders RGE equations are
\begin{align}
  \nu\partial_\nu J_{\kappa(1)}&=J_{\kappa(1)} \gamma_{(1,1)}\,,
  \\
  \nu\partial_\nu S_{(1,1)}&=-\gamma_{(1,1)} S_{(1,1)} - S_{(1,1)}  \gamma_{(1,1)}=-2\gamma_{(1,1)} S_{(1,1)} \,,
 \nn \\
  \nu\partial_\nu \bar J_{\kappa'(1)}&=\gamma_{(1,1)} \bar J_{\kappa'(1)}
 \nn \,,
\end{align}
with simple multiplication of factors.
These equations immediately gives rise to an exponential solution (a simple pole in Laplace space), which is the so-called gluon Regge pole. It is therefore natural in the Glauber based approach to define $\gamma_{(1,1)}$ as the gluon Regge trajectory. We note that this is an all orders, gauge invariant definition. This definition of the anomalous dimension agrees with the value computed in Refs.~\cite{Falcioni:2020lvv,Falcioni:2021buo,Falcioni:2021dgr,Caola:2021izf}, however, the effective field theory provides an operator that this is the anomalous dimension of. We will comment on its relation to other definitions in the literature below. One interesting feature of $\gamma_{(1,1)}$ that can be deduced from the renormalization group consistency equations is that it is independent of the partonic label, $\kappa$, to all orders in perturbation theory. This is as physically expected, since it is supposed to be describing the dressing of the $t$-channel gluon, but here we are also able to see this property emerge from the collinear perspective.  We anticipate that the ability to calculate the Regge trajectory purely from collinear physics will provide new ways of computing it at higher perturbative orders.

Let us now discuss the generic all order form of the RGEs in more detail.
For $i,~j>1$, we can write the RRGEs of $J_{\kappa(i)}$, $\bar J_{\kappa'(j)}$ and $S_{(i,j)}$ as 
\begin{subequations}
\begin{align}
 \label{eq:J_RGE}
  \nu\partial\nu\, J_{\kappa(i)}=& \sum_{i'=2}^\infty J_{\kappa(i')} \otimes \gamma_{(i',i)}\,,\\
  \label{eq:S_RGE}
  \nu\partial\nu\, S_{(i,j)}=& -\sum_{i'=2}^\infty \gamma_{(i,i')}\otimes S_{(i',j)} - \sum_{j'=2}^{\infty}S_{(i,j')}\otimes \gamma_{(j',j)}\,,\\
  \nu\partial\nu\, \bar J_{\kappa'(j)}=& \sum_{j'=2}^\infty \gamma_{(j,j')} \otimes \bar J_{\kappa'(j')} \,,
\end{align}
\end{subequations}
where the color indices are contracted implicitly, and $\otimes$ denotes the  convolution for a certain number of transverse integrals (here $i'$ or $j'$ of them) as in \eq{int_perp}.
We note that although there are an infinite number of terms on the right hand side, to a given order of $\alpha_s$, the number of terms appearing in these RRGEs will be truncated.
To be specific, from the $\gs$ order counting shown in Eqs.~\eqref{eq:J_gs_order} and \eqref{eq:Gamma_gs_order}, we see that all the $i'\le i$ terms in the $J_{\kappa(i)}$ RRGE are equally important (leading logarithmic), while the $i'> i$ terms start with an $\cO(\alpha_s^{i'-i})$ suppression, i.e., first appear at N$^{i'-i}$LL order.
We can express the truncated versions of the RRGEs to N$^r$LL accuracy for $J$, $\bar J$ and $S$ are
\begin{subequations} \label{eq:JSJ_RGE_truncated}
\begin{align}\label{eq:J_RGE_truncated}
  \nu\partial\nu J_{\kappa(i)}
  &= \sum_{i'=2}^{i+r} J_{\kappa(i')} \otimes \gamma_{(i',i)} +\cO\bigl(\gs^{i+2r+4}\bigr) \,,\\[.1in]
  \label{eq:S_RGE_truncated}
  \nu\partial\nu S_{(i,j)}
  &= -\hspace{-0.5cm}\sum_{j'=\max(2,\,j-r)}^{j+r} \hspace{-0.5cm} \gamma_{(i,j')}\otimes S_{(j',j)} - \hspace{-0.5cm}\sum_{i'=\max(2,\,i-r)}^{i+r}\hspace{-0.5cm}  S_{(i,i')}\otimes \gamma_{(i',j)}+\cO(\gs^{\abs{i-j}+2r+4})\,,\\
  \nu\partial\nu \bar J_{\kappa'(j)}&= \sum_{j'=2}^{j+r} \gamma_{(j,j')} \otimes \bar J_{\kappa'(j')} +\cO\bigl(\gs^{j+2r+4}\bigr) \,.
\end{align}
\end{subequations}
It is important to note that this logarithmic counting corresponds to the logarithmic counting for the specific $N$-Glauber operators, not the logarithmic counting of the overall amplitude, which is often higher order. For example, the leading logarithmic evolution of the two-Glauber exchange operator corresponds to NLL counting for the amplitude, since its single Glauber loop with the factors of $g$ from the tree-level $J$ and $\bar J$ does not contribute any logarithm.

It is useful to consider explicit examples beyond $i=1$.  For two Glauber exchange $i=2$, taken with $r=0$, we have LL equations which give us access to the evolution of the singlet Pomeron, the $8_S$, and other color channels.  
Using \eq{JSJ_RGE_truncated} they are
\begin{align}\label{eq:consistency_LL2}
  \nu\partial\nu J_{\kappa(2)} &=  J_{\kappa(2)} \otimes \gamma_{(2,2)}
  \,, \\
  \nu\partial\nu S_{(2,2)} &= -\gamma_{(2,2)}\otimes S_{(2,2)} - S_{(2,2)}\otimes \gamma_{(2,2)}
  \,, \nn \\
  \nu\partial\nu \bar J_{\kappa'(2)} &=  \gamma_{(2,2)} \otimes \bar J_{\kappa'(2)} 
  \,. \nn
\end{align}
Here there is no mixing between different numbers of Glauber exchanges (such mixings do occur for larger $r$), but there are multiple color channels. The corresponding (suppressed) color indices correspond to the number of Glauber $t$-channel exchanges. For the $i=2$ case distinct color channels appear at most once, so it is easy to carry out the diagonalization into individual equations for each channel. This decomposition is left to the next section. 
In addition \eq{consistency_LL2} now has a non-trivial convolution in transverse momenta.  For the $8_A$ channel the $i=2$ convolutions in these equations still have a power law (pole) solution, since this solution is related  by unitarity and crossing symmetry to the $i=1$ solution for $8_A$. The same is true for the $8_S$. For the other color channels the convolutions remain non-trivial.

Next consider results where there are three Glauber exchanges for either $i$ of $j$ in \eq{J_RGE_truncated}, again at lowest order, $r=0$. Here the factorization theorem first begins to have off-diagonal terms, taking the form
\begin{align}
- \img\,\cM_{2\to 2}^{\kappa\kappa'} \Big|_{\text{3-Glauber}} 
  &=  J_{\kappa(2)} \otimes S_{(2,3)} \otimes \bar J_{\kappa'(3)}
   +  J_{\kappa(3)} \otimes S_{(3,2)} \otimes \bar J_{\kappa'(2)}
   +  J_{\kappa(3)} \otimes S_{(3,3)} \otimes \bar J_{\kappa'(3)}
  \,.
\end{align}
At this $r=0$ level we already have the RG equation for $J_{\kappa(2)}$ and $\bar J_{\kappa'(2)}$ in \eq{consistency_LL2}. For the remaining objects,
\eq{JSJ_RGE_truncated} gives
\begin{align}\label{eq:consistency_LL3}
  \nu\partial\nu J_{\kappa(3)}=
   &  J_{\kappa(2)} \otimes \gamma_{(2,3)}
   +  J_{\kappa(3)} \otimes \gamma_{(3,3)} 
  \,, \\
  \nu\partial\nu S_{(2,3)}=
   & - \gamma_{(2,3)}\otimes S_{(3,3)} 
     - S_{(2,2)}\otimes \gamma_{(2,3)}
  \,, \nn \\
\nu\partial\nu S_{(3,3)}=& -\gamma_{(3,3)}\otimes S_{(3,3)} - S_{(3,3)}\otimes \gamma_{(3,3)}
  \,, \nn \\
  \nu\partial\nu \bar J_{\kappa'(3)}=&  \gamma_{(3,2)} \otimes \bar J_{\kappa'(2)} + \gamma_{(3,3)} \otimes \bar J_{\kappa'(3)}
  \,. \nn
\end{align}
For these 3-Glauber objects we see that off-diagonal terms with $\gamma_{(2,3)}$ are required already at lowest order in the coupling expansion.  Furthermore, in the equation for $\nu\partial_\nu S_{(2,3)}$ we see that two types of $\perp$ convolution integral appear, the first is an $\otimes_3$ with three $\perp$ integrals, and the second is an $\otimes_2$ with two $\perp$ integrals.  Since the two terms are generated by the same Feynman diagram in \Eq{eq:tenniscourt}, they each must have the same number of $\img$ factors. From \eq{S_LO} we see that  the $S_{(2,2)}$ and $S_{(3,3)}$ contribute a relative factor of $\img$, already at tree level, which must be compensated by the factors of $\img$ in $\otimes_2$ versus $\otimes_3$.  This explains the convention in \eq{int_perp}, which needs its factor of $(-\img)^j$.  
Furthermore, when calculating the tennis court diagram in \Eq{eq:tenniscourt}, the divergence which could contribute to these two terms in the anomalous dimension equation are from soft loop momentum $k^-\to 0$ and $k^-\to \infty$. This provides an example which enables us to understand our convention for the choice of $j!$ in \eq{int_perp}, since these factors also differ in $S_{(2,2)}$ and $S_{(3,3)}$, and must again be compensated for in $\otimes_2$ and $\otimes_3$.

\subsection{Vanishing of  $1\to j$ Glauber Transitions in the EFT}
\label{sec:vanish}

In \Sec{sec:consistent} we presented the general structure of the rapidity RG equations under the assumption that the $1\to n$ Glauber transitions vanish in the EFT. This fact does not follow from the consistency of the RG alone, but we believe that it is a very nice feature of the RG. We therefore devote this section to proving this fact. As we will describe in \Sec{sec:relate}, we believe that the vanishing of the $1\to n$ transitions can be interpreted as
identifying a simple set of degrees of freedom for the Reggeon Field Theory.  In addition to contributing to the general understanding of the structure of the EFT of forward scattering, to prove the vanishing of the $1\to n$ transitions at higher loops, we introduce a new approach to regulating rapidity divergences in the EFT. In Ref.~\cite{Moult:2022lfy} it was shown that rapidity regularization of the forward limit should be done with a $\eta'$ regulator for Glauber divergences, that is distinct from the $\eta$ regulator used for soft and collinear rapidity divergences. Building on this, here we note that it is advantageous to use a regulator for the soft/collinear rapidity divergences that allows contour deformation,\footnote{One might also consider an exponential regulator~\cite{Li:2016axz}. But obtaining convergence for the triangle graph makes it more complicated.}
such as $(k^- \pm \img 0)^{-\eta}$~\cite{Becher:2011dz} or $(k^z \pm \img 0)^{-\eta}$.
Non-analyticity of the rapidity regulator, such as through $|2k^z|^{-\eta'}$, is only needed for the Glauber loops. We will see that this greatly clarifies the structure of proofs in the EFT, since non-analyticity is only introduced by the Glauber regulator. Additionally, analytic regulators are much simpler for higher order calculations \cite{Li:2016ctv,Moult:2022xzt,Duhr:2022yyp}, so we anticipate that this will simply calculations in the EFT of forward scattering. 

\begin{figure}[t!]
  \begin{center}
    \raisebox{2cm}{
    \hspace{-1.9cm}
    a)\hspace{9cm} 
    b)\hspace{3cm} 
     } \\[-55pt]
  \hspace{-0.55cm}
  $-\img\cM_{\rm{\Delta}}\equiv  \fd{3cm}{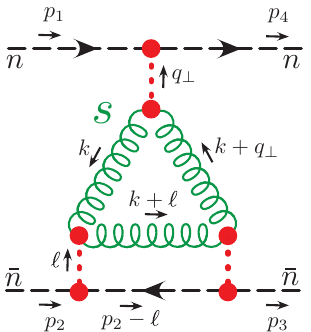}\quad=\quad\fd{3cm}{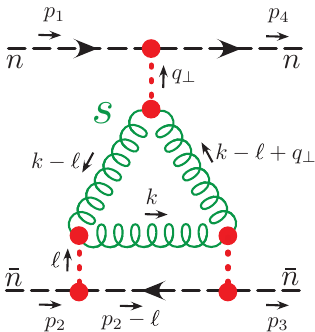}
  $
  \hspace{0.6cm}
$-\img\cM_{\rm{h}}\equiv \fd{3cm}{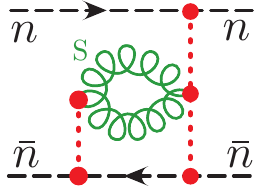}$
  \end{center}
    \vspace{-0.4cm}
    \caption{\setcaptionskip The one loop soft $1\to2$ transition graphs include the ``triangle'' $-\img\cM_\Delta$, and ``horse'' graphs $-\img\cM_{\rm h}$, and their mirror images (not shown). The triangle graph is shown with two distinct momentum routings.  
    These graphs vanish as explained in the text. 
    }
    \label{fig:diagrams_1to2}
    \setmainskip
  \end{figure}

We now show that all the $1\to j$ soft transition graphs vanish for $j>1$.
At the lowest order, the $1\to2$ transition via a soft loop arises from the graphs shown in \fig{diagrams_1to2}, namely the ``triangle graph'' and the ``horse graph'' (as well as its mirror image).
We will begin by explicitly showing that the soft triangle graph is zero, and that the horse graph is zero.
Then we will argue that the same logic can be applied to any $1\to j>1$ transition to any loop order.  The intuition behind why the triangle graph is zero is that in the presence of the Glauber $\eta'$ regulator   (before taking $\eta'\to 0$), the two bottom $\bn$-$s$ Glaubers are separated in lightcone time. The $\bn$-$s$ Glauber vertices then emit (absorb) two soft gluons which both have positive (negative) $\bar n$-momenta of $\cO(\lambda)$, violating momentum conservation.

Let's first look at the triangle graph with the first momentum routing in \fig{diagrams_1to2}a.
Using the SCET Glauber Feynman rules, we have $-\img\cM_\Delta =  -\frac38 C_A (8\pi\alpha_s)^3 [T^A \frac{\bnslash}{2} \otimes \bar T^A \frac{\nslash}{2}] \cM_\Delta^L$, where the loop integral for the triangle graph is
\begin{align}\label{eq:M_Delta}
  &\cM_\Delta^L
  =\int\! \frac{\dbar^d\ell\, \dbar^d k 
  \Bigl[
  k^+(k^-)^2d'-2k^{\!-}\bigl(\vec k_\perp^2+(\vec k_\perp+\vec q_\perp)^2\bigr)
  +\frac{4[\vec k_\perp\cdot (\vec k_\perp+\vec q_\perp)]^2}{k^+\pm\img0}
  \Bigr]
  R_{\eta'}(\ell^-,\ell^+)\, R(k^-,k^+,\ell^+)
  }
  {[k^+k^-\!-\!\Delta_1\!+\!\img0]
  [k^+k^-\!-\! \Delta_2+\img0]
  \bigl[k^-(k^+ \!+\! \ell^+)-\Delta_3+\img0\bigr]
  \bigl[p_2^- \!-\! \ell^- \!-\! \Delta_4/p_2^+ \!+\! \img0\bigr]
  D_G
  }
  \,,
\end{align}
with $d'=d-2$, 
$D_G=\vec q_\perp^2\,\vec \ell_\perp^2\, (\vec\ell_\perp-\vec q_\perp)^2$, and  $\Delta_1=\vec k_\perp^2$,
$\Delta_2=(\vec k_\perp +\vec q_\perp)^2$,
$\Delta_3=(\vec k_{\perp}+\vec\ell_\perp)^2$,
and $\Delta_4=(\vec p_{2\perp}-\vec\ell_\perp)^2$.
Recall that the $\eta'$ regulator $R_{\eta'}$ is to regulate the Glauber loop,
\begin{align}
  R_{\eta'}=\abs{\ell^--\beta_{\bar ns}\,\ell^+}^{-2\eta'}\,,
\end{align}
where $\beta_{\bar ns}\sim\cO(\lambda)$ is an arbitrary positive boost factor that makes the regulator homogeneous in power counting.
The regulator $R$ regulates the soft loop, which depends on the $\pm$ components of the soft momenta $k$ and $k+\ell$; in other words, it depends on $k^-$, $k^+$ and $\ell^+$ (due to the power counting $\ell\sim  (\lambda,\lambda^2,\lambda)$ no $\ell^-$ dependence appears in $R$).
Here we choose $R$ to be an analytic regulator, for example 
a power like analytic regulator given by\footnote{
  This example of regulator comes from a product of terms appearing at different vertices of the following form: both the left and the right vertices give $(k^z+\ell^z+\img0)^{-\eta/2} (-k^z + \img0)^{-\eta/2}$;
  the middle vertex gives $(-k^z + \img0)^{-\eta}$.
}
\begin{align} \label{eq:Rv2}
  R= (-\ell^+ -k^+ +k^- +\img 0)^{-\eta}  (k^+  - k^-+ \img 0)^{-2\eta} 
  \,.
\end{align}
In the following calculation, we will show that the Glauber regulator $R_{\eta'}$ plays an essential role in making the integral vanish.

Using Eq.~(B.4) of Ref.~\cite{Rothstein:2016bsq}, we first do the $\ell^-$ integral, and define the result to be $\tilde R_{\eta'}(\ell^+)$,
\begin{align}  
  &\int_{-\infty}^{+\infty} \!\!\!  \ddslash\! \ell^- \:
   \frac{   |\ell^--\beta_{\bn s}\ell^+|^{-2\eta'} }{
    p_2^- \!-\! \ell^- \!-\! \Delta_4/p_2^+ \!+\! \img0 }
    \nn\\
    \label{eq:tilde_R_eta'}
  =\,& \frac12 \csc(2\pi \eta') 
  \Big[ (-\beta_{\bn s} \ell^++a+\img0)^{-2\eta'} - (\beta_{\bn s} \ell^+-a-\img0)^{-2\eta'} \Big]
  \equiv \tilde R_{\eta'}(\ell^+)
    \,,
\end{align}
where $a=p_2^--\Delta_4/p_2^+$.

We now evaluate the $\ell^+$ integral.  The branch cut from \eq{tilde_R_eta'} is above the real axis. The regulator $R$ in \eq{Rv2} also gives a branch cut for $\ell^+$ that starts above the real axis. 
Thus we can choose to close the $\ell^+$ contour below.\footnote{Notice that the regulator~\eqref{eq:tilde_R_eta'} ensures that the $\ell^+$ contour integral converges for the large semi-circular contour that needs to be added to obtain a closed contour, and that this piece vanishes as $R^{-2\eta'}$ when the radius $R\to \infty$. Therefore, although there is only one pole for $\ell^+$, we can close the contour.}
If $k^-<0$, the pole for $\ell^+$ is above the real axis, so the integral vanishes. 
If $k^->0$, the pole for $\ell^+$ is below the real axis and the integral is proportional to the residue of this pole. Therefore, the $\ell^+$ integral gives a $\theta(k^-)$, and we have
\begin{align}\label{eq:M_Delta_tmp}
  \cM_\Delta^L
  =&-\!\frac\img2\!
  \int\! \frac{\dbar^{d'}\ell_\perp \dbar^d k 
  \Bigl[
  -2\bigl(\vec k_\perp^2+(\vec k_\perp\!+\!\vec q_\perp)^2\bigr) + k^+k^-d'
  +\frac{4[\vec k_\perp\cdot (\vec k_\perp+\vec q_\perp)]^2}{k^-(k^+\pm\img0)}
  \Bigr]\theta(k^-)
  \dbtilde R_{\eta'}(k^+)\, \tilde R(k^-,k^+)
  }
  {(k^+k^-\!-\!\Delta_1\!+\!\img0)
  (k^+k^-\!-\! \Delta_2+\img0)
  D_G
  }
  \,.
\end{align}
The $\eta'$ regulator now becomes
\begin{align}
  \dbtilde R_{\eta'}(k^+)
  &\equiv \tilde R_{\eta'}(\ell^+=\Delta_3/k^--k^+-\img0)
  \nn\\
  \label{eq:tilde_R_eta'_k}
  &= \frac12 \csc(2\pi \eta') 
   \Big[ (\beta_{\bn s} k^++b+\img0)^{-2\eta'} - (-\beta_{\bn s}k^+-b-\img0)^{-2\eta'} \Big]\,,
\end{align}
where $b=p_2^--\Delta_4/p_2^+ -\beta_{\bn s}\Delta_3/k^-$. Similarly, $\tilde R(k^-,k^+)$ is 
\begin{align}
  \tilde R(k^-,k^+)\equiv R(k^-,k^+, \ell^+=\Delta_3/k^--k^+-\img0).
\end{align}

Next we consider the $k^+$ integral in \eq{M_Delta_tmp}.
We will argue that the $\theta(k^-)$ puts all the singularities on the same side of the real axis, while sufficient regulator factors are present to eliminate singular contributions.   
First we note that the $\dbtilde R_{\eta'}(k^+)$ regulator in \eq{tilde_R_eta'_k} introduces branch cuts in $k^+$ below the real axis, and ensures that there is convergence on a contour that extends out to complex infinity even in the presence of a single pole. Furthermore, when we start with \eq{Rv2}, the regulator $\tilde R(k^-,k^+)$ introduces branch cuts below the real axis.
Therefore we will consider the $k^+$ integral on a contour closed above the real axis. We consider the three terms in the square bracket of the numerator separately.
For the first term, notice that due to the $\theta(k^-)$ the two poles for $k^+$ are below the real axis, so there is no singular structure in the upper half plane and the integral gives $0$.
For the second term, we partial fraction to give
\begin{align}
  \frac{k^+k^-}{(k^+k^-\!-\!\Delta_1\!+\!\img0)
  (k^+k^-\!-\! \Delta_2+\img0)}=
  \frac{1}{\Delta_1-\Delta_2}
  \left(\frac{\Delta_1}{k^+k^-\!-\!\Delta_1\!+\!\img0}-\frac{\Delta_2}{k^+k^-\!-\!\Delta_2\!+\!\img0}\right)
  \,.
\end{align}
Both of these have a single pole below the real axis and hence also have no singular structure in the upper half plane.
The third term requires some further discussion, since a nontrivial zero-bin needs to be subtracted. We will show that after zero-bin subtraction, the result is zero independent of the sign of $\img0$ in the eikonal propagator $\frac{1}{k^+\pm\img0}$:
\begin{align}\label{eq:I_pm}
  I=\tilde I^{(\pm)}-I_{\rm 0-bin}^{(\pm)}=0\,.
\end{align}
Here, $\tilde I^{(\pm)}$ is the naive integral without zero bin subtraction, and $I_{\rm 0-bin}^{(\pm)}$ is the Glauber zero bin,
\begin{align}\label{eq:I_tilde}
  \tilde I^{(\pm)} &= \int\! \frac{\dbar^{d'}\ell_\perp\, \dbar^d k 
  \,\bigl[\vec k_\perp\cdot (\vec k_\perp+\vec q_\perp)\bigr]^2 \theta(k^-)
  \,   \dbtilde R_{\eta'}(k^+)
  \tilde R(k^-,k^+)
  }
  {(k^+k^-\!-\!\Delta_1\!+\!\img0)
  (k^+k^-\!-\! \Delta_2+\img0)
  (k^+\pm\img0)k^-
  }
  \,,
  \\[.1in]
  \label{eq:I_0bin}
  I_{\rm 0-bin}^{(\pm)} &=\int\! \frac{\dbar^{d'}\ell_\perp\, \dbar^d k 
  \,\bigl[\vec k_\perp\cdot (\vec k_\perp+\vec q_\perp)\bigr]^2 \theta(k^-)
  \dbtilde R_{\eta'}(k^+)
  \tilde R(k^-,k^+)
  }
  {\Delta_1\Delta_2
  (k^+\pm\img0)k^-
  }
  \,.
\end{align}
If the sign for $\img0$ is plus, then for both \eq{I_tilde} and \eq{I_0bin}, the singularities are below the real axis. Thus we have
\begin{align}
  \tilde I^{(+)}= I_{\rm 0-bin}^{(+)}=0\,.
\end{align}
If the sign for $\img0$ is minus, then for both \eq{I_tilde} and \eq{I_0bin}, we evaluate the integral by picking the single pole in the upper half plane $k^+=\img0$, and find
\begin{align}
  \tilde I^{(-)}&= \frac\img2\int\! \frac{\dbar^{d'}\!\ell_\perp\, \dbar^{d'}\! k_\perp\, \dbar k^- 
  \,\bigl[\vec k_\perp\cdot (\vec k_\perp+\vec q_\perp)\bigr]^2 \theta(k^-)
  \,   \dbtilde R_{\eta'}(\img 0)
  \tilde R(k^-,\img 0)
  }
  {\Delta_1 \Delta_2 k^-
  }
  \nn\\
  &= I_{\rm 0-bin}^{(-)}
  \,.
\end{align}
In conclusion, we see that the vanishing result in \eq{I_pm} is indeed correct.

Notice that if one had used the second momentum routing in \fig{diagrams_1to2}, then the loop momentum $\ell$ passes through two soft propagators. 
The loop integral for this routing can be obtained from the first routing by taking $k\to k-\ell$.
In this case after doing the $\ell^-$ integration, all poles and cuts for $\ell^+$ are on the same side unless one has $\theta(-k^-)$, which is opposite to the result to the $\theta(k^-)$ which appeared from the first momentum routing. Furthermore, with $k^- <0$ the contributions from the two $\ell^+$ poles give contributions with equal value and opposite sign, up to $R_{\eta'}$ regulator factors.  In fact one gets a difference of $R_{\eta'}$ regulators that vanish as ${\cal O}(\eta')$ as $\eta'\to 0$. This corresponds to the fact that the graph should vanish due to the  Glauber collapse rule, whereby graphs vanish when a vertex interrupts the collapse of Glauber vertices to equal longitudinal position~\cite{Rothstein:2016bsq}. Here the presence of the upper Glauber vertex interrupts the collapse of the lower two Glauber rungs, which is made manifest by this choice of momentum 
routing.\footnote{In general one must be careful with this collapse argument in the presence of multiple Glauber loops, since performing each loop causes a collapse of propagators, and collapsed propagators with their associated vertices do not count for the purpose of interrupting another Glauber loop. This is not an issue for the triangle graph where there is only a single Glauber loop.}
If one does not expand in $\eta'$, then one finds that with this momentum routing after performing the $\ell^\pm$ integrals, all poles and cuts in $k^+$ are now in the upper half plane, and closing in the lower half plane gives zero. 
This gives an alternate argument which again shows that the triangle diagram is zero.

Next consider the horse graph in \fig{diagrams_1to2}b which involves an $n$-$\bn$ Glauber exchange from top to bottom, with two soft gluons attached. Here the Glauber collapse argument cannot be used because there is never an intermediate vertex. However, the same argument that made the triangle graph vanish with the first routing in \fig{diagrams_1to2} applies equally well here. With a similar routing as the first graph in \fig{diagrams_1to2}, the horse graph has the same relativistic propagators as the triangle, except the $k+q_\perp$ propagator is absent. The regulator factors $R_{\eta'}$ and $R$ also have the same form. The remaining factors for the horse include  purely transverse momentum dependent numerators together with eikonal propagators of four types: $(n\cdot k+\img 0)$, $(n\cdot(\ell+k)-\img 0)$, $(n\cdot\ell-\img 0)(\bn\cdot k+\img 0)(n\cdot k+\img 0)$, or $(n\cdot\ell-\img 0)(\bn\cdot k+\img 0)(n\cdot (\ell+k)-\img 0)$.  Without loss of generality, we can assume that the $\pm \img 0$ in these eikonal denominators enter with the signs shown, since other sign choices will give the same answer after including zero-bin subtractions.  Performing integrals in the same order as we did for the triangle, the $\ell^-$ integral is identical to \eq{tilde_R_eta'}, and the $\ell^+$ integral is again only non-zero with a $\theta(k^-)$ from closing below around the pole $\ell^+ = \Delta_3/k^- -k^+ -\img 0/k^-$. Depending on the eikonal denominators this may induce additional poles in $k^+$ in the lower half plane. However, once again there are no poles in $k^+$ in the upper half plane, and there are suitable regulator factors to ensure convergence of contours at infinity and remove singular contributions, so performing the $k^+$ integral gives zero for the horse graph.

\begin{figure}[t!]
  \begin{center}
    \raisebox{2cm}{
    \hspace{1.9cm}
    a)\hspace{6.5cm} 
    b)\hspace{7cm} 
     } \\[-55pt]
  \hspace{-0.55cm}
  $ \fd{3cm}{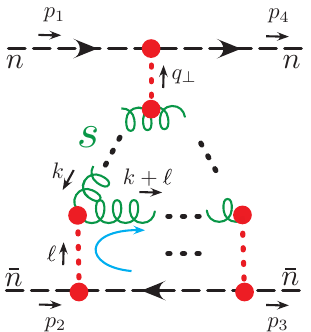}$
  \hspace{3.6cm}
  $\fd{3cm}{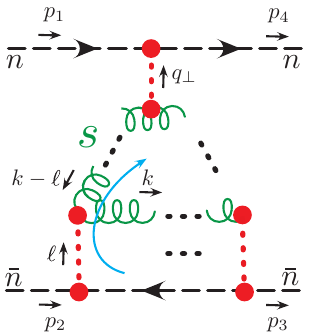}$
  \end{center}
    \vspace{-0.4cm}
    \caption{\setcaptionskip A general $1\to j$ soft graph with two different momentum routings, where $j>1$. The blue arrow indicates the direction of the loop momentum $\ell$.  
    These general $1\to j>1$ graphs vanish for the reasons explained in the text.
 }
    \label{fig:diagrams_1toj}
    \setmainskip
  \end{figure}

With a similar argument, we now show that any $1\to j >1$ soft graph to any loop order is zero.
Since there is only one Glauber connecting to the $n$-collinear sector, 
at least either the leftmost or the rightmost Glauber at the bottom must connect the soft sector through a $\bn$-$s$ Glauber interaction (i.e., not through soft gluons attaching to an $n$-$\bn$ Glauber like the rightmost Glauber in the horse graph in \fig{diagrams_1to2}b). Taking the leftmost 
Glauber to be $\bn$-$s$, then \fig{diagrams_1toj} is a general graph to consider (the soft gluon in the figure may be changed to a soft quark without changing the argument).  In \fig{diagrams_1toj}a we let the Glauber loop momentum $\ell$ go through various soft propagators from the left most $\bn$-$s$ Glauber rung to the rightmost $\bn$-$s$ Glauber, without passing through the upper $n$-$s$ Glauber rung vertex. In contrast, in \fig{diagrams_1toj}b the loop momentum $\ell$ also passes through this $n$-$s$ vertex.  In both graphs the loop momentum $k$ is defined such that it passes through all three of these Glauber vertices. 
Using the momentum routing in \fig{diagrams_1toj}a, we get a $\theta(k^-)$ after doing the $\ell^\pm$ integrals. Then, for the $k^+$ integral, all the singularities are in the lower half plane, and suitable regulators are present to ensure convergence for the contour in the lower half plane and to eliminate singular contributions, and thus the integral vanishes.   The alternate routing in \fig{diagrams_1toj}b can be obtained by taking $k\to k-\ell$ in the momentum routing of \fig{diagrams_1toj}a.
Here doing the $\ell^\pm$ integrals leads to a $\theta(-k^-)$.  Likewise all singularities in the $k^+$ variable have switched sides, and are now in the upper half plane (while convergence is ensured for closing in the lower half plane). Thus closing the $k^+$ contour below gives zero for these graphs with this alternate routing.

Note that this same argument does not apply to $i \to j$ transitions with $i,j >1$ (or with $i=j=1$), which do not generically vanish in the EFT. For $1\to j>1$ soft graphs there is no additional minus momentum flowing into the soft loop, while for $i \to j$ transitions with $i,j >1$, there is an additional minus momentum in the soft loop.  This changes the above steps since there is a region with $k^+$ poles on both sides of the real axis.
The tennis court graph in \Eq{eq:tenniscourt}, which mediates a $2\to 3$ transition, provides an explicit example of a non-vanishing Glauber transition graph.  For $i=j=1$ the argument does not apply because there is not an $R_{\eta'}$ regulating factor, nor is there a Glauber loop. This case corresponds to the soft-eye graph evaluated in Ref.~\cite{Rothstein:2016bsq} which is non-zero. Performing the soft $k$ loop integration by contours using $k^\pm$ variables is known to be tricky for this special case~\cite{Collins:2018aqt}, and leads to a non-zero result.

\subsection{Relationship with Other Approaches}\label{sec:relate}

The RG consistency equations provide considerable insight into the way in which the EFT organizes the all orders structure of the Regge limit. In particular, it significantly clarifies the relation to approaches based on ``Reggeons'', such as the Reggeon Field Theory \cite{Gribov:1968fc,Abarbanel:1975me,Baker:1976cv}, and its more modern incarnations, such as the approach of Ref.~\cite{Caron-Huot:2013fea}, and its subsequent iterations \cite{Caron-Huot:2017fxr,Falcioni:2020lvv,Falcioni:2021buo,Falcioni:2021dgr,Caola:2021izf}.

\begin{figure}
  \begin{center}
  \includegraphics[width=0.3\columnwidth]{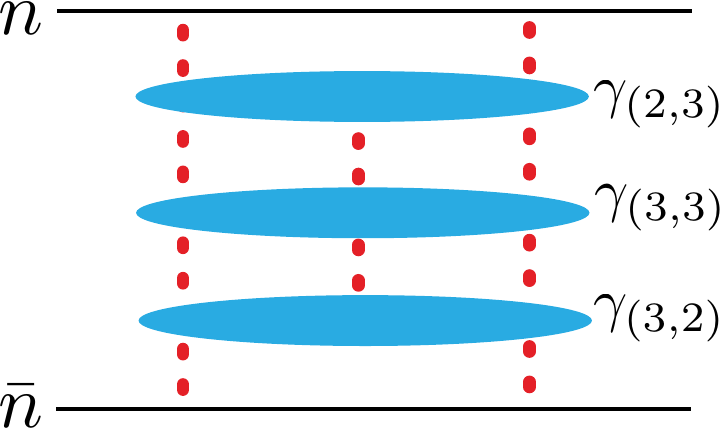}
  \end{center}
          \vspace{-0.4cm}
          \caption{\setcaptionskip An illustration of the iteration of $\gamma$ factors (shown in blue) which arise from the rapidity renormalization group evolution in a forward scattering amplitude.
          Each $\gamma$ factor comes along with a $\ln\frac{s}{-t}$.
          This evolution can be illustrated by diagrams representing interactions in the two-dimensional transverse space, analogous to the Reggeon Field Theory. The blobs for the leading order collinear and soft functions, which appear in the convolution for the full amplitude, have been suppressed.}
          \label{fig:iteration}
          \setmainskip
  \end{figure}

We first note a general distinction between the EFT of forward scattering and Reggeon Field Theory. Reggeon Field Theory is supposed to be a $2+1$ dimensional field theory, with the two-dimensions being the transverse dimensions of the scattering, and the additional dimension being the rapidity. This field theory is supposed to describe the dynamics of Reggeons. On the other hand, the EFT for forward scattering is a genuine four dimensional field theory with soft and collinear degrees of freedom, whose interactions are mediated by Glauber potentials.  Reggeon Field Theory is not associated with the EFT for forward scattering, but rather with the rapidity renormalization group equations derived from this theory. Due to the power counting of the EFT, these involve only convolutions in the two transverse dimensions, and describe evolution in rapidity. As we have seen above, these RG equations involve an infinite set of $Z$ factors, or anomalous dimensions $\gamma$, describing interactions and transitions between Glauber operators with different numbers of operators. This evolution can be illustrated by diagrams representing the interactions in the two-dimensional transverse plane, an example of which is shown in \Fig{fig:iteration}. We believe that these diagrams representing iterative solutions of the rapidity renormalization group evolution are related to the vertices of a Reggeon effective theory.  We therefore see that what the EFT for forward scattering provides is an intermediate step between QCD and a Reggeon Field Theory, which enables the efficient calculation of these vertices, and most importantly, guarantees their consistency, since they are derived from a rapidity renormalization of a genuine field theory.

With this understanding, we can now compare how the rapidity renormalization group of the EFT organizes the Glauber interactions, as compared to the Reggeon Field Theory. Here an important difference is that the rapidity renormalization group is diagonal for the single Glauber operator, and does not involve $1\to n$ Glauber transitions. 
A key difference between the organizations is that standard approaches are organized in terms of ``Reggeons'', which have a definite signature under crossing, while the ``Glaubers'', do not have a definite signature. Operators of a definite signature necessarily involve an infinite number of Glauber operators. This can clearly be seen, since even at lowest order crossing for the single Glauber links it to the antisymmetric octet of the two Glauber exchange. Alternatively, the approach of Ref.~\cite{Caron-Huot:2017fxr}, which shows how the $\img\pi$ can be absorbed, is loosely equivalent in the EFT to absorbing some part of the two-Glauber operator (or its exponential) into the single Glauber operator. However, once this is done, then since the two-Glauber operator can mix with the three-Glauber operator, it enables $1\to 3$ mixing. It will be interesting to explicitly see how this works out by performing higher order calculations in the Glauber EFT.

We can also compare our mixing matrices with those in Ref.~\cite{Caron-Huot:2013fea}. The general structure of the counting in our matrices is similar to the mixing structure appearing in Ref.~\cite{Caron-Huot:2013fea}, with the main exception mentioned above that we do not have $1\to n$ mixing. However, as mentioned above, if we absorb part of the two-Glauber operator into the single Glauber operator, then we will induce such $1\to n$ transitions, with exactly the counting in the coupling appearing in the mixing matrices of Ref.~\cite{Caron-Huot:2013fea}. It will be particularly interesting to compare the calculation of the three Glauber structure in the EFT to see how it reproduces the recent calculations in this framework \cite{Falcioni:2020lvv,Falcioni:2021buo,Falcioni:2021dgr,Caola:2021izf}.

More speculatively, we believe that the organization of the rapidity renormalization equations could be a good starting point for a Reggeon Field Theory \cite{Gribov:1968fc,Abarbanel:1975me,Baker:1976cv}. Indeed, it seems to us to be quite natural that the rapidity renormalization group equation for the single Glauber operator is diagonal, since this implies that it is an eigenstate in the Reggeon Field Theory. All the mixing process of the rapidity renormalization group can then be interpreted as genuine scattering processes, $n\to m$, with $n,m\geq 2$. It is well known that  in the case that the number of Reggeons does not change, high energy scattering is famously integrable \cite{Lipatov:1993yb,Faddeev:1994zg} (see also Refs.~\cite{Korchemsky:1994um,Korchemsky:2001nx,Korchemsky:2003rc}). It would be nice to derive this directly from the rapidity evolution equations of the EFT. Mixings in the number of Reggeons/Glaubers are supposed to lift this to a QFT in $2+1$ dimensions, since Reggeons can be produced/annihilated. In our case, there will be no decays, only scatterings, which could simplify the analysis. Some of these Reggeon scattering vertices have been calculated, in particular by Bartels and collaborators \cite{Bartels:1978fc,Bartels:1980pe,Bartels:1991bh,Bartels:1994jj,Bartels:1999aw,Bartels:2002au,Bartels:2004hb,Bartels:2007dm,Bartels:2015gou,Bartels:2017nxb}. This physics should be described by the transitions in the mixing equations of our rapidity evolution equation. It will be interesting to push calculations in the Glauber EFT to higher orders to make further connections with the Reggeon Field Theory.

\section{Off-Forward BFKL Equations from a Collinear Perspective}
\label{sec:BFKL}

In this section we derive the LL rapidity RGE (RRGE) for $J_{\kappa(2)}(\ell_{1\perp}, \ell_{2\perp}) $
\begin{align}
  \fd{5cm}{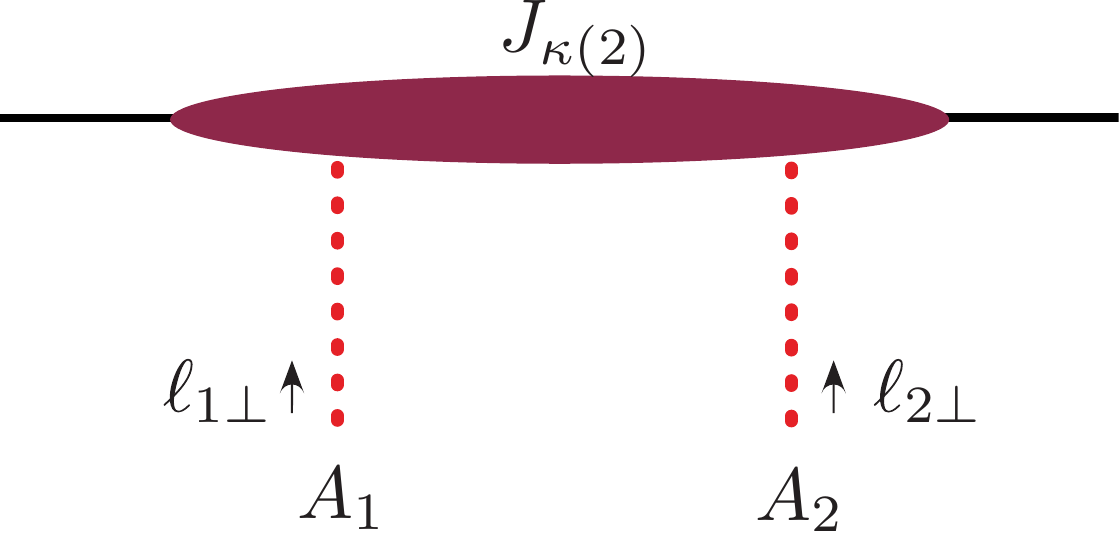} \nn\,.
\end{align}
In \sec{BFKL_consistency} we review the renormalization group consistency equation for two-Glauber exchange truncated to LL accuracy.
In \sec{BFKL_graphs}, we describe the general procedure and list all the contributing graphs,
and calculate graphs with one collinear loop.
In \sec{unraveling}, we decompose the color structures for these collinear graphs to the ``base'' structures.
With all the preparation above, in \sec{BFKL_RGE}, we read off the RRGEs.
In \sec{BFKL_color}, we decompose RRGEs into different irreducible color channels.

Note that throughout this section we use the $\eta$ regulator for collinear rapidity divergences. While we have advocated in \Sec{sec:vanish} that different analytic regulators are in fact more convenient for understanding certain properties of the EFT, for the simple one-loop collinear graphs considered in this section, there is little difference. We have therefore chosen to use the $\eta$ regulator for a more direct comparison with \cite{Rothstein:2016bsq}. We have checked that both regulators give the same result at the level of $1/\eta$ poles needed to determine the rapidity RGE. 

\subsection{Renormalization Group Consistency for Two-Glauber Exchange}
\label{sec:BFKL_consistency}

Using the counting in $\gs$ presented in \sec{consistent}, we can restrict the general consistency equations to leading logarithmic accuracy, and obtain the following equations for two-Glauber exchange
\begin{align}\label{eq:consistency_LL}
  \nu\partial\nu J_{\kappa(2)}=&  J_{\kappa(2)} \otimes \gamma_{(2,2)}\,,\\
  \nu\partial\nu S_{(2,2)}=& -\gamma_{(2,2)}\otimes S_{(2,2)} - S_{(2,2)}\otimes \gamma_{(2,2)}\,,\\
  \nu\partial\nu \bar J_{\kappa'(2)}=&  \gamma_{(2,2)} \otimes \bar J_{\kappa'(2)} \,.
\end{align}
As is well known, there does not exist any mixing at LL order. The novelty of this result is that it exactly relates the anomalous dimensions appearing in the soft sector with the anomalous dimensions appearing in the collinear sector, allowing us to calculate the BFKL equation from a collinear operator. 
 In \app{soft}, we calculate the same anomalous dimension $\gamma_{(2,2)}$ from the soft perspective so that the two can be directly compared.

\subsection{Two-Glauber Collinear Diagrams}
\label{sec:BFKL_graphs}

Since the RRGE does not depend on state $\kappa$, in the explicit calculations in this section, we consider only quark-antiquark collisions, i.e. we take $\kappa=q$, $\kappa'=\bar q$, and we will omit indices $\kappa$, $\kappa'$ for simplicity.\footnote{The use of different external partons excites different color projections for the Glauber states,  e.g. there is no $t$-channel 27 exchange with $\kappa=q$ since $3\otimes\bar 3=1\oplus 8$. But, since we derive the general operator $Z_{J(2,2)}^{B_1B_2\,A_1A_2}$ or $\gamma_{(2,2)}^{B_1B_2\,A_1A_2}$ before color projection, we are still able to determine results for these channels.}
In addition, for simplicity, we set $\ell_{1\perp}=\ell_\perp$, $\ell_{2\perp}=q_\perp-\ell_\perp$, and write $J_{(2)}$ with one argument, $J_{(2)}(\ell_\perp)=J_{(2)}(\ell_{1\perp}=\ell_{\perp},\ell_{2\perp}=q_\perp-\ell_{\perp})$, and similarly for $\bar J_{(2)}$ and $S_{(2,2)}$. 

\begin{figure}
  \begin{center}
    $-\img\cM_{(2,2)}^{[0]}\equiv\fd{4cm}{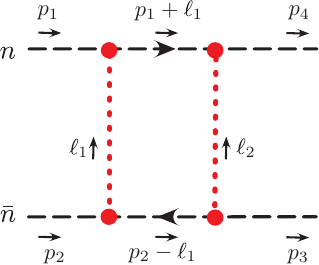}$ 
  \end{center}
          \vspace{-0.4cm}
          \caption{\setcaptionskip 
          The two-Glauber exchange box diagram.
          }
          \label{fig:box_graph}
          \setmainskip
  \end{figure}

We need to calculate all the two-Glauber exchange diagrams with one $n$-collinear loop, 
matched with the leading order Glauber box diagram as shown in \fig{box_graph}.
$-\img\cM_{(2,2)}^{[0]}$ can be thought of convoluting the LO $n$-collinear function with the LO soft function and the LO $\bn$-collinear function. Similarly, the sum of the graphs with one $n$-collinear loop is a convolution of the NLO $n$-collinear function with the LO soft function and the LO $\bn$-collinear function.
Thus we have
\begin{align}
  -\img\cM_{(2,2)}^{[0]} = J_{(2)}^{[0]}\otimes S_{(2,2)}^{[0]}\otimes \bar J_{(2)}^{[0]}\,, \qquad
  \sum -\img\cM_{(2,2)}^{\text{one $n$ loop}} = J_{(2)}^{[1]}\otimes S_{(2,2)}^{[0]}\otimes \bar J_{(2)}^{[0]}\,.
\end{align}
Comparing $-\img\cM_{(2,2)}^{[0]}$ with the $\eta$ pole of sum of one-$n$-collinear-loop graphs, we can then read off the rapidity RGE for $J_{(2)}$ at the lowest order, i.e., LL.

The calculation of the box graph can be found in Ref.~\cite{Rothstein:2016bsq}. Here we present the result,
\begin{align}  \label{eq:Gboxes0}
-\img\cM_{(2,2)}^{[0]} &= - g^4 \bigl(T^{A_1} T^{A_2}\bigr) \otimes \bigl(\bT^{A_1} \bT^{A_2}\bigr)\, \cS^{n\bn}  \int\! \frac{\ddslash\!^{d'}\!\ell_\perp}{\vec\ell_\perp^{\,2} \prpsqm{q}{\ell}}
\\
&= \frac{1}{4} \int\!\! \frac{\dbar^{d'}\!\ell_\perp\, \dbar^{d'}\!\ell'_{\perp}}{\vec\ell_\perp^{\,2} \prpsqm{q}{\ell}\, \vec\ell^{\prime2}_\perp \prpsqm{q}{\ell'}} \,
J_{(2)}^{[0]A_1 A_2}(\ell_\perp)  S_{(2,2)}^{[0]A_1 A_2\, A_1' A_2'}(\ell_{\perp}, \ell^{\prime}_{\perp} ) \bar J_{(2)}^{[0] A_1'A_2'}(\ell_\perp^{\prime})  \,, \nn
\end{align}
where 
the spinor factor is
\begin{align} \label{eq:Snbn}
   \cS^{n\bn} = \Big[ \bar u_n  \frac{\bnslash}{2} u_n \Big]\Big[ \bar v_\bn  \frac{\nslash}{2} v_\bn \Big] \,.
\end{align} 
In the second line, we write $-\img\cM_{(2,2)}^{[0]}$ as convolution of the leading order collinear and soft functions, which, according to \eqs{S_LO}{J_LO}, are 
\begin{align}
  & J_{(2)}^{[0]A_1 A_2}(\ell_\perp) = g^2 (T^{A_1} T^{A_2}) \Bigl[\bar u_n  \frac{\bnslash}{2} u_n \Bigr]
  \,,\quad
  \bar J_{(2)}^{[0]A_1'A_2'}(\ell_\perp^{\prime}) = g^2 (\bT^{A_1'} \bT^{A_2'}) \Bigl[ \bar v_\bn  \frac{\nslash}{2} v_\bn \Bigr]\,,\\[.2cm]
  & S_{(2,2)}^{[0]A_1 A_2 A_1'A_2'}(\ell_\perp, \ell_\perp^{\prime}) = -4\, \deltaslash^{d'}\!(\ell_\perp-\ell_\perp^{\prime}) \delta^{A_1 A_1'}\delta^{A_2A_2'}\,\vec\ell_\perp^{\;2} \prpsqm{q}{\ell}\,.
\end{align}

\begin{figure}[t!]
  \begin{center}
    \raisebox{2cm}{
    \hspace{1.4cm}
    a)\hspace{2.7cm} 
    b)\hspace{2.8cm} 
    c)\hspace{2.7cm} 
    d)\hspace{6.2cm} 
     } \\[-55pt]
  \hspace{-0.55cm}
  \includegraphics[width=0.16\columnwidth]{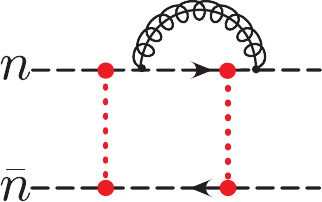}
  \hspace{0.2cm}
  \includegraphics[width=0.16\columnwidth]{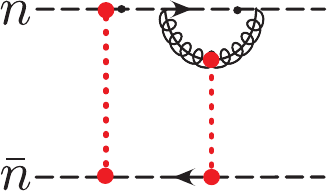}
  \hspace{0.2cm}
  \includegraphics[width=0.16\columnwidth]{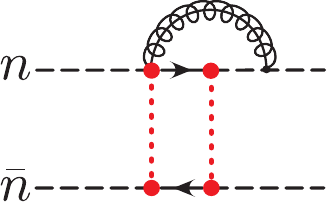}
  \hspace{0.2cm}
  \includegraphics[width=0.17\columnwidth]{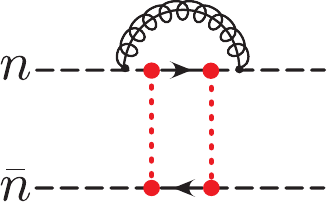}
  \\[5pt]
  \end{center}
    \vspace{-0.4cm}
    \caption{\setcaptionskip Graphs with one collinear loop and two Glaubers that do not contribute to the rapidity renormalization of the two Glauber collinear operator. The first two graphs vanish by the Glauber collapse rule, the third graph is scaleless, and the fourth graph is not rapidity divergent.}
    \label{fig:vanishing_diagrams}
    \setmainskip
  \end{figure}

\begin{figure}[t!]
  \begin{center}
    \raisebox{2cm}{
    \hspace{-1.9cm}
    a)\hspace{5.4cm} 
    b)\hspace{4.8cm} 
    c)\hspace{1.4cm} 
     } \\[-55pt]
  \hspace{-0.55cm}
  $-\img\cM_{\rm{P}}\equiv  \fd{3.5cm}{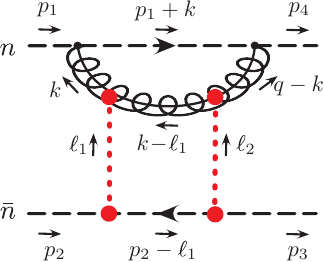}\:
  $
$-\img\cM_{\rm{NP}}\equiv \fd{3.5cm}{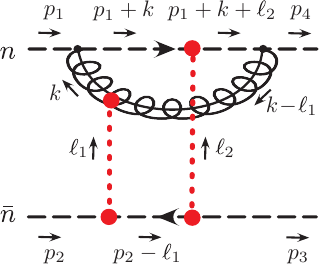}\:$  
$-\img\cM_{\rm{NP}}'\equiv\fd{3.5cm}{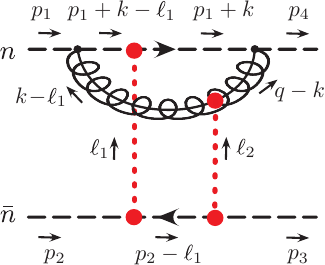}$  \\[5pt]
  \end{center}
    \vspace{-0.4cm}
    \caption{\setcaptionskip The three one-loop collinear diagrams that contribute to the rapidity renormalization group evolution of the two-Glauber state. The notation ``P'' and ``NP'' denotes planar and non-planar, respectively. The two non-planar graphs are mirror images of each other.}
    \label{fig:diagrams_1col_loop}
    \setmainskip
  \end{figure}

We now consider the two-Glauber graphs with one $n$-collinear loop. A crucial simplification arises from the \textit{Glauber collapse rule} \cite{Rothstein:2016bsq}, where any obstruction to the ``collapse'' of two adjacent Glauber emissions results in zero amplitude. In Ref.~\cite{Moult:2022lfy}, this was shown to continue to hold at higher loop level for a consistently constructed rapidity regulator. For example, (a) and (b) in \fig{vanishing_diagrams} vanish due to the collapse rule.
Notice that the collapse rule does not force similar graphs with two collinear loops vanish, such as
 \begin{align}
  \fd{3.5cm}{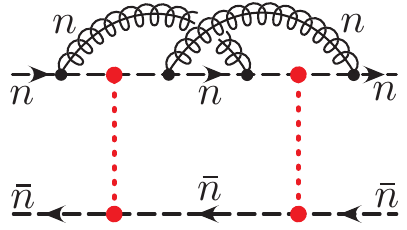} \ =\ 
  \fd{3.5cm}{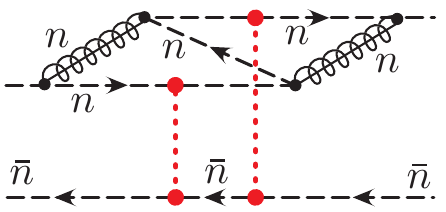}\,.
 \end{align}
They should be considered when deriving the NLL RGE for $J_{(2)}$, which is beyond the scope of this work. Additionally,  Graph (c) in \fig{vanishing_diagrams} (and similar collinear Wilson line emission loops) results in vanishing scaleless integrals. Finally, graph (d) is not rapidity divergent, i.e., does not have a $1/\eta$ pole (the calculation is similar to the one for $\cM_{\rm P}$~\eqref{eq:diagram_planar}, except that there is no divergence in the last step doing $k^-$ integral). The only three non-vanishing diagrams with rapidity divergences are illustrated in \fig{diagrams_1col_loop}. We now calculate these three graphs in details.
Again, for simplicity, we will take $\ell_1=\ell$, $\ell_2=q_\perp-\ell$ in calculations.

We begin with the planar graph $\cM_{\rm{P}}$. We define its numerator $N_\text{P}$, prefactor $P_{\text{P}}$ including color and external spinor factors, and Glauber denominator $D_{\text{P}}^\perp$ as\footnote{
  The $\eta$ regulator in $N_\text{P}$ here (and similarly in $N_\text{NP}$ below) can be replaced by other regulators. For example, with an analytic regulator like $(k^-\pm\img0)^{-\eta}$ used in \sec{vanish}, our result does not change at the level of $1/\eta$ poles needed to determine the rapidity RGE.
  One may also consider using the exponential regulator $\exp{\bigl(-\tau e^{-\gamma_E}k^0\bigr)}$~\cite{Li:2016axz}: this amounts to changing the $k^-$ integral in \eq{diagram_planar}, giving a $\ln\tau$ when expanding $\tau\to0$ instead of a $1/\eta$ pole.
},
\begin{align}
	N_\text{P}(k) &= -4 \img (64 \pi^3 \as^3) \left( 4 \prp{k} \cdot (\prp{q} - \prp{k})\right) w^2 \left|\frac{k^-}{\nu}\right|^{-\eta}+\cdots\,,
	\nn \\
	P_\text{P} &= \prnth{-\img f^{A_1B_1C}} \prnth{-\img f^{A_2B_2C}} \prnth{T^{B_1} T^{B_2}} \otimes \prnth{\bT^{A_1} \bT^{A_2}} \cS^{n\bn}\,,
	\nn \\
	D_\text{P}^\perp(q_\perp,\ell_\perp) &= \vec\ell_\perp^{\;2} \prpsqm{q}{\ell}\, .
\end{align}
Here, $\cdots$ in $N_\text{P}(k)$ includes terms with higher power in $k^-$, which are not rapidity divergent in the later $k^-$ integral;
the only rapidity divergent term as shown in $N_\text{P}(k)$ comes from both the two Glauber-collinear-gluon interactions to be the second term in the regulated Feynman rule~\eqref{eq:collinear_regulator}. The two $\abs{k^-}^{-\eta/2}$ combine to give $\abs{k^-}^{-\eta}$. The regulated Feynman rule in \eq{collinear_regulator} is fixed by the requirement that 1-$n$-collinear-loop $i$-Glauber-exchange graphs exponentiate, which is explained in \app{regulator}.
Alternatively, this diagram could be regulated using the exponential regulator \cite{Li:2016axz}, which we believe would greatly simplify higher order calculations.

Evaluating this graph, we find
\begin{align}\label{eq:diagram_planar}
-\img\cM_{\rm{P}}
 &= P_\text{P} \!\!
	\int\!\!\frac{\dbar^d\ell \; \dbar^dk \; w'^4 \left| \frac{2 \ell^z}{\nu'} \right|^{-2 \eta'} \, N_\text{P}(k)}{
	\prnth{k^2 + \img0} \prnth{(k-\ell)^2 + \img0} \prnth{(k-q)^2 + \img0} \left[\frac{\prnth{p_1 +k}^2+\img0}{p_1^- + k^-}\right] 
	\left[\frac{\prnth{p_2 - \ell}^2+\img0}{p_2^+}\right] D_\text{P}^\perp(q_\perp,\ell_\perp)
}
\nn \\
&= \frac{P_\text{P}}{4}
\!\!\int\!\!
\frac{\dbar^{d'}\ell_\perp \; \dbar^{d'}\!k_\perp \, \dbar k^- \dbar k^+ \;   \theta(-k^-) N_\text{P}(k)}{
k^-\!\! \prnth{k^+k^- \!-\!\vec k_\perp^2 \!+\! \img0} \! \prnth{k^+k^- \!-\! (\vec k_\perp \!-\! \vec q_\perp)^2 \!+\! \img0} \! \left[p_1^+\! +\! k^+ \!-\!\frac{(\vec p_{1\perp} \!+\! \vec k_\perp)^2-\img0}{p_1^-+k^-}\right] \!
D_\text{P}^\perp(q_\perp,\ell_\perp) 
}+\cO(\eta')
\nn \\[.1in]
&= -\img \frac{P_\text{P}}{16}
\int\!\!\frac{\dbar^{d'}\!\ell_\perp \; \dbar^{d'}\!k_\perp \; dk^- \; \theta(-k^-) \, \theta(p_1^-+k^-)\, N_\text{P}(k)}{
k^-\!\! \prnth{k^+k^- \! -\! \vec k_\perp^2 + \img0}\! \prnth{k^+k^- \! -\! (\vec k_\perp\!-\!\vec q_\perp)^2 + \img0}
\! D_\text{P}^\perp(q_\perp,\ell_\perp) 
}\Biggr|_{k^+=\frac{(\vec p_{1\perp} +\vec k_\perp)^2-\img0}{p_1^-+k^-}-p_1^+} +\cO(\eta')
\nn\\
&= -\frac{(32 \pi^2 \as^3) w^2 }{2} P_\text{P}
	\int\frac{\dbar^{d'}\!\ell_\perp \; \dbar^{d'}\! k_\perp \left( 4 \prp{k} \cdot (\prp{q} - \prp{k})\right)}{
	\vec\ell_\perp^{\,2} \prpsqm{q}{\ell} \prpsq{k} \prpsqm{k}{q}
}   \int_{-p_1^-}^{0} \!\frac{d k^- \,\nu^\eta}{\abs{k^-}^{1+\eta}} + \cO(\eta^0)+\cO(\eta')
\nn\\
 &=  \frac{(32 \pi^2 \as^3) w^2}{\eta} \left(\frac{\sqrt{s}}{\nu} \right)^{-\eta}   P_\text{P}
	\int\!\! \frac{\dbar^{d'}\!\ell_\perp}{\vec\ell_\perp^{\,2} \prpsqm{q}{\ell}}
  \int\!\! \frac{\dbar^{d'}\!k_\perp \; \vec q_\perp^{\,2}}{
	\prpsq{k} \prpsqm{k}{q}
} + \cO(\eta^0) +\cO(\eta')
\, .
\end{align}
Here to get the second equality, we performed the integrals for $\ell^z$ and $\ell^0$ making use of Eq.~(B.5) and Eq.~(B.6) in Ref.~\cite{Rothstein:2016bsq}. To get the third line, we carry out the $k^+$ integral by contours. To get the fourth line, we keep only the leading term in the $k^-\to0$ limit, given that all the next-to-leading terms are less singular than $\cO(\eta^{-1})$. To get the last line, we notice that $2 \prp{k} \cdot (\prp{q} - \prp{k})=\vec q_\perp^{\,2}-\prpsq{k}-\prpsqm{q}{k}$, and that $\prpsq{k}$ and $\prpsqm{q}{k}$ would make the integral for $ k_\perp$ scaleless and thus vanish.
Notice that we didn't do the final transverse integrals.
The $\ell_\perp$ integral can be understood as a convolution with the LO soft and $\bn$-collinear function, while the $ k_\perp$ integral is the convolution that will appear in the rapidity RGE. This agrees with the general expectation that the dynamics for the rapidity divergent terms should be described by a field theory living in the two-dimensional transverse plane \cite{Verlinde:1991iu}.

Moving on to the non-planar graph, we define its numerator, color factor, and denominator as
\begin{align}
	N_\text{NP}(k, \ell) &= 4\img (64 \pi^3 \as^3) \left( 4 \frac{\prp{k} \cdot (\prp{\ell} - \prp{k})}{k^-}\right)  w^2 \left|\frac{k^-}{\nu}\right|^{-\eta}+\cdots
	\,, \nn \\
	P_\text{NP} &= \prnth{-\img f^{B_1A_1C}} \prnth{T^{B_1} T^{A_2} T^{C}} \otimes \prnth{\bT^{A_1} \bT^{A_2 }} \cS^{n\bn}
  \,,\nn \\
D_\text{NP}^\perp(q_\perp,\ell_\perp) &= \vec\ell_\perp^{\;2} \prpsqm{q}{\ell}\, .
\end{align}
Here, unlike the planar graph, the eikonal denominator $k^-$ in $N_\text{NP}$, as well as the rapidity divergence, comes from a Wilson line Feynman rule, i.e., the third term in the regulated Feynman rule~\eqref{eq:collinear_regulator};
again, $\cdots$ in $N_\text{NP}$ are terms with high power in $k^-$ which are not rapidity divergent.
We note also that we may re-express $P_{\text{NP}}$ as follows, 
\begin{align} \label{eq:cnp1}
	P_\text{NP} 
  &= \prnth{-\img f^{B_1A_1C}} \Bigl[ \prnth{T^{B_1} T^{C} T^{A_2}} 
  +
  \prnth{\img f^{B_2A_2C}}
  \prnth{T^{B_1} T^{B_2}} \biggr]
  \otimes \prnth{\bT^{A_1} \bT^{A_2 }} \cS^{n\bn}
  \nn\\
	&= -\left(\frac{N_c}{2} C_{\delta}^{B_1B_2\,A_1A_2} + C_{\text{H}}^{B_1B_2\,A_1A_2} \right) \prnth{T^{B_1} T^{B_2}} \otimes \prnth{\bT^{A_1} \bT^{A_2}}  \cS^{n\bn}  \,,	
\end{align}
where we define color factors $C_\delta$ and $C_{\rm H}$ as
\begin{align}
  C_\delta^{B_1B_2\,A_1A_2}=\delta^{B_1 A_1} \delta^{B_2 A_2}\,,\qquad C_{\rm H}^{B_1B_2\,A_1A_2} = \prnth{-\img f^{B_1 A_1 C}} \prnth{-\img f^{B_2A_2C}}  \,.
\end{align}
Then $-\img\cM_{\rm NP}$ can be calculated as
\begin{align} 
 -\img\cM_{\rm{NP}}
  =\,& P_\text{NP} 
	\int\!\frac{\dbar^dk \; \dbar^d\ell \; w'^4 \left| \frac{2 \ell^z}{\nu'} \right|^{-2 \eta'} \, N_\text{NP}(k,\ell)}{
	\prnth{k^2 + \img0} \prnth{(k-\ell)^2 + \img0} \left[\frac{\prnth{p_1 +k}^2+\img0}{p_1^- + k^-}\right] 
	\left[\frac{\prnth{p_1 + k + q-\ell}^2+\img0}{p_1^- + k^-}\right]
  \left[\frac{\prnth{p_2 - \ell}^2+\img0}{p_2^+}\right]
  D_\text{NP}^\perp(q_\perp,\ell_\perp)
}
\nn \\
=\,& \frac{(32 \pi^2 \as^3) w^2}{\eta} \left(\frac{\sqrt{s}}{\nu} \right)^{-\eta} \, P_\text{NP}
	\int\!\! \frac{\dbar^{d'}\!\ell_\perp}{\vec\ell_\perp^{\,2} \prpsqm{q}{\ell}}
  \int\!\! \frac{\dbar^{d'}\!k_\perp \; \vec\ell_\perp^{\,2}}{
	\prpsq{k} \prpsqm{\ell}{k}
} + \cO(\eta^0)+ \cO(\eta')
 \,.
\end{align}
The steps to carry out these integrals are similar to those for \eq{diagram_planar} except that we used Eq.~(B.8) in Ref.~\cite{Rothstein:2016bsq} to do the $\ell^\pm$ integrals. Note that in this graph, the eikonal denominator comes from a Wilson line Feynman rule, unlike for the planar graph, where it arises from the collapse of the propagator between the two Glaubers.

We now compute the mirror-image $\cM_{\rm{NP}}'$ of the non-planar graph $\cM_{\rm{NP}}$. We define the color and external spinor prefactor to be $P_{\text{NP}}'$,
\begin{align}
  P_\text{NP}' &= -\prnth{-\img f^{B_2 A_2 C}}\prnth{T^{C} T^{A_1} T^{B_2}} \otimes \prnth{\bT^{A_1} \bT^{A_2 }} \cS^{n\bn} \,.
\end{align}
By doing the similar decomposition as \eq{cnp1}, we note that $P_{\text{NP}}' = P_{\text{NP}}$; but we choose to explicitly retain the factor of $P_{\text{NP}}'$ to distinguish them for clarity of which graphs contribute to which terms. The mirror image graph gives
\begin{align}
 -\img\cM_{\rm{NP}}'
&=\frac{(32 \pi^2 \as^3) w^2}{\eta} \left(\frac{\sqrt{s}}{\nu} \right)^{-\eta} \,  P'_\text{NP}
 \int\!\! \frac{\dbar^{d'}\!\ell_\perp}{\vec\ell_\perp^{\,2} \prpsqm{q}{\ell}}
 \int\!\! \frac{\dbar^{d'}\!k_\perp \; \prpsqm{q}{\ell}}{
 \prpsqm{k}{\ell} \bigl(\prp{q} - \prp{k}  \bigr)^2
} + \cO(\eta^0)
\,,
\end{align}
where we have expanded with $\eta'\to0$ and omitted the $\cO(\eta')$ terms.

Adding all contributions together, we find
\begin{align}
&-\img\cM_{\rm{P}}-\img\cM_{\rm{NP}}-\img\cM_{\rm{NP}}'\nn\\[.2cm]
=\,& \frac{(32 \pi^2 \as^3) w^2}{\eta} \left(\frac{\sqrt{s}}{\nu} \right)^{-\eta} 
 \int\!\! \frac{\dbar^{d'}\!\ell_\perp}{\vec\ell_\perp^{\,2} \prpsqm{q}{\ell}}
 \int\!\!\dbar^{d'}\!k_\perp \nn\\
 &\qquad\qquad\times
 \left(\frac{P_\text{P}\; \vec q_\perp^{\,2}}{\prpsq{k} \prpsqm{q}{k}}
 + \frac{P_\text{NP} \; \vec\ell_\perp^{\,2}}{\prpsq{k} \prpsqm{\ell}{k}}
 + \frac{P_\text{NP}' \; \prpsqm{q}{\ell}}{\prpsqm{k}{\ell} \prpsqm{q}{k} }\right)+\cO(\eta^0)
 \,.
\end{align}
The final transverse integral over $\ell_\perp$ is frozen, since it is understood to be convolved with the leading order $S_{(2,2)}^{[0]}$ and $\bar J_{(2)}^{[0]}$
\begin{align}
  & -\img\cM_{\rm{P}}-\img\cM_{\rm{NP}}-\img\cM_{\rm{NP}}'\\ 
  =\,& -\frac14 \int\!\! \frac{\dbar^{d'}\!\ell_\perp\, \dbar^{d'}\!\ell'_{\perp} }{\vec\ell_\perp^{\,2} \prpsqm{q}{\ell}\, \vec\ell^{\prime2}_\perp \prpsqm{q}{\ell'}} \,
J_{(2)}^{[1]A_1 A_2}(\ell_\perp)  S_{(2,2)}^{[0]A_1 A_2\, A_1' A_2'}(\ell_{\perp}; \ell^{\prime}_{\perp} ) \bar J_{(2)}^{[0] A_1' A_2'}(\ell_\perp^{\prime}) +\cO(\eta^0)\,, \nn
\end{align}
determining the structure of the renormalization.

Using the definitions of $P_{\text{P}}$ and $P_{\text{NP}}$, we can write down $J_{(2)}^{[1]}$ as
\begin{align}\label{eq:J1bare}
  &J_{(2)}^{[1]A_1 A_2}(\ell_\perp)= \frac{2 \alpha_s w^2}{\eta} \left(\frac{\sqrt{s}}{\nu} \right)^{-\eta}
 \Big[ \bar u_n  \frac{\bnslash}{2} u_n \Big] T^{B_1} T^{B_2}
  \int\!\!\dbar^{d'}\!k_\perp \nn \\
  & 
  \qquad\times
  \Bigg(  - \frac{ \vec q_\perp^{\,2}C_{\rm H}^{B_1B_2\,A_1A_2}}{\prpsq{k} \prpsqm{q}{k}}
  + \frac{ \vec\ell_\perp^{\,2} (C_{\rm H}+N_c C_\delta/2)^{B_1B_2\,A_1A_2} }{\prpsq{k} \prpsqm{\ell}{k}}
  + \frac{  \prpsqm{q}{\ell} (C_{\rm H}+N_c C_\delta/2)^{B_1B_2\,A_1A_2} }{\prpsqm{k}{\ell} \prpsqm{q}{k} }\Bigg)
  \nn \\
  &
  \qquad+\cO(\eta^0)\,.
\end{align}
This provides our final result for the one-loop bare collinear corrections to two-Glauber exchange.

\subsection{Unraveling Color Mixing}
\label{sec:unraveling}

In this section, we explain how to interpret the color structures present in the one-loop collinear function $J_{(2)}^{[1]A_1A_2}$ in terms of renormalization group mixing. To interpret the renormalization group mixing, it will be necessary to decompose the collinear functions in a specific basis of color structures, corresponding to those of the tree level Glauber operators. These color structures will be equivalent to those derived from the anomalous dimensions in the soft sector, as expected from soft collinear consistency. In Sec. \ref{sec:BFKL_RGE} we will then use these to derive the evolution equations for specific color channels.

Naively, one may think that the two non-planar graphs $\cM_{\rm NP}$ and $\cM_{\rm NP}'$ would contribute to $Z_{J(3,2)}$, in violation with \eq{consistency_LL}.  This is not true, because they would give rise to interactions that look like
\begin{align*}
  \includegraphics[width=.3\textwidth]{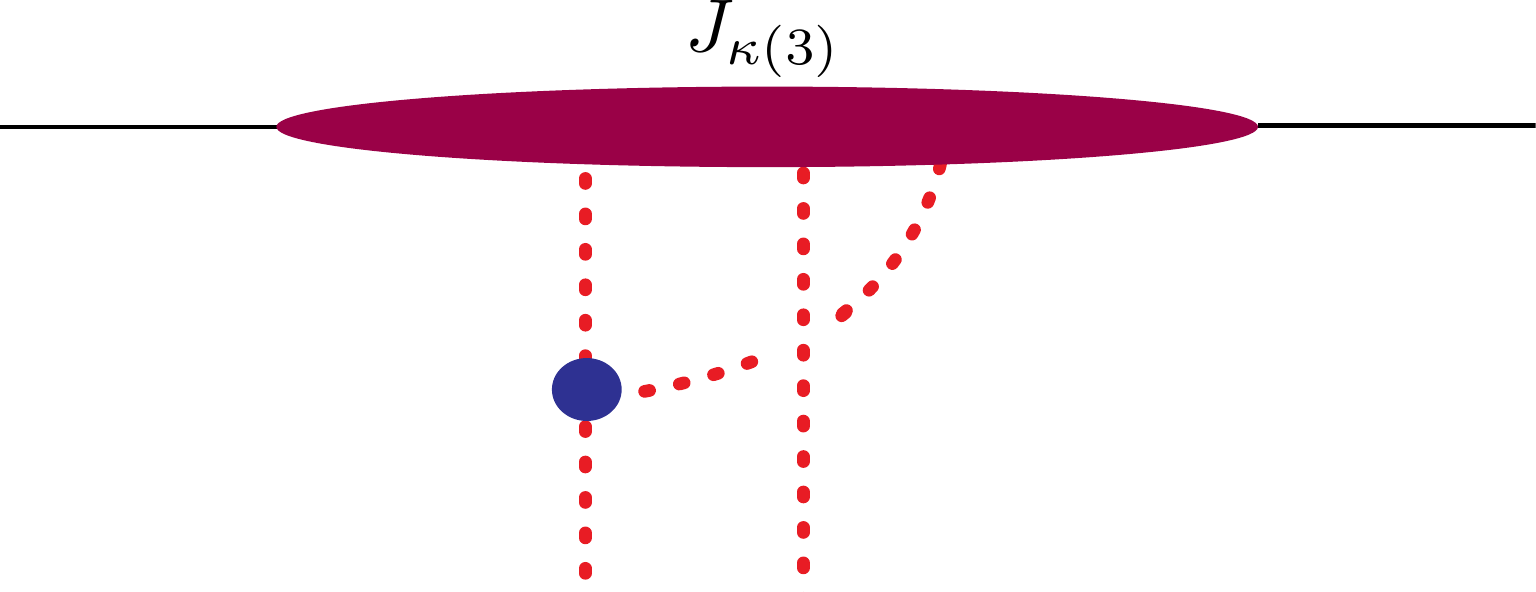}
\end{align*}
which does not make sense: Glaubers should be vertical and cannot ``bend'' since they are virtual and do not propagate. In other words, we need to decompose the color structure of these two graphs into graphs that only have vertical (denoting Glauber) and horizontal (interaction between/among Glaubers) lines.
For $\cM_{\rm NP}$ and $\cM_{\rm NP}'$, this corresponds to decomposing the color structure in $P_{\rm NP}$ as in \eq{cnp1}, and similarly for $P_{\rm NP}'$. 
Graphically, we have 
\begin{align}
  \cC:\; \fd{1.5cm}{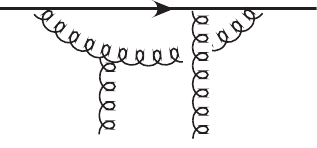} &=  \fd{1.5cm}{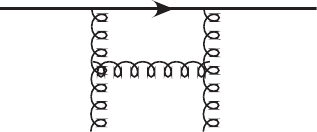} + \fd{1.5cm}{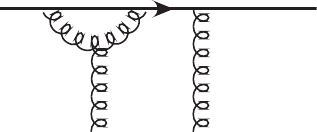} =  \fd{1.5cm}{figsCO/color_planarH.pdf} - \frac{N_c}{2}\, \fd{1.5cm}{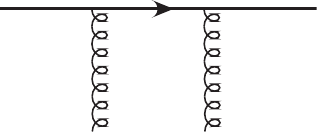}\,,\\[.1in]
  \cC:\; \fd{1.5cm}{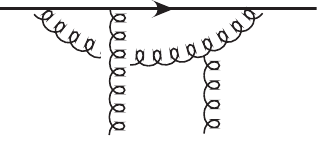} &=  \fd{1.5cm}{figsCO/color_planarH.pdf} + \fd{1.5cm}{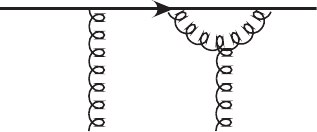} =  \fd{1.5cm}{figsCO/color_planarH.pdf} - \frac{N_c}{2}\, \fd{1.5cm}{figsCO/color_LO.pdf}\,,
\end{align}
where $\cC$ indicates that these diagrams should be interpreted in terms of color flow, not momentum flow. In this way, we decompose colors of collinear graphs into ``fundamental'' base objects, by which we mean that there are only ``horizontal'' and ``vertical'' lines. This decomposition also makes the comparison between the collinear and soft pictures manifest, as shown in \app{soft}.

\begin{figure}[t!]
  \begin{center}
    \raisebox{2cm}{
    \hspace{-1.9cm}
    a)\hspace{5.4cm} 
    b)\hspace{4.8cm} 
    c)\hspace{1.4cm} 
     } \\[-55pt]
  \hspace{-0.55cm}
  $-\img\cM_{\rm{H}}\equiv \! \fd{3.5cm}{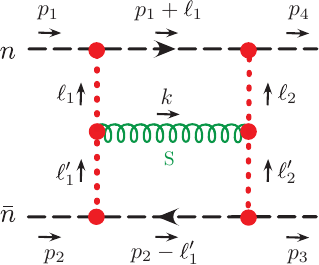}\:
  $
$-\img\cM_{\rm{SE}}\equiv\! \fd{3.5cm}{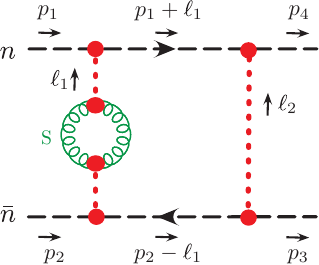}\:$ 
$-\img\cM_{\rm{SE}}'\equiv\!\fd{3.5cm}{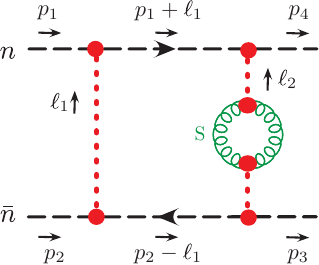}$  \\[5pt]
  \end{center}
    \vspace{-0.4cm}
    \caption{\setcaptionskip The three one-loop soft diagrams that have rapidity divergences and therefore contribute to the rapidity renormalization group evolution of the two-Glauber state. The notation ``SE'' denotes soft eye. We will present the calculations of these graphs in \app{soft}.}
    \label{fig:diagrams_1soft_loop}
    \setmainskip
  \end{figure}

From such color unraveling, we see that the LL RRGE for $J_{(2)}$ receives contributions from $Z_{J(2,2)}^{B_1B_2\,A_1A_2}$ (or $\gamma_{(2,2)}^{B_1B_2\,A_1A_2}$), which have two color structures
\begin{align}
  \fd{1.2cm}{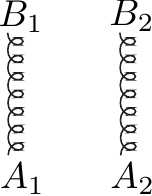} = C_{\delta}  \equiv  \delta^{A_1 B_1} \delta^{A_2 B_2} \,,\qquad
  \fd{1.2cm}{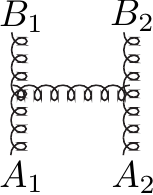} &= C_{\rm H}  \equiv   (- \img f^{A_1 B_1 C} )(-\img f^{A_2 B_2 C}) \,.
  \end{align}
This is of course familiar from the standard calculation of the BKFL pomeron from the soft sector, where the three graphs in \fig{diagrams_1soft_loop}
manifestly have this form. However, here we have seen how it is reproduced from the collinear sector, as is required from soft-collinear consistency. The complete calculation of the anomalous dimension from the soft sector of the effective theory is provided in \app{soft}.

\subsection{Rapidity RGE for Two Glauber Exchange}
\label{sec:BFKL_RGE}

In this section we combine the one-loop collinear functions calculated in \sec{BFKL_graphs} and the color decomposition procedure outlined in \sec{unraveling} to extract the off-forward BFKL equation at LL. We begin by writing \eq{J1bare} as a convolution over the lower order $J_{(2)}^{[0]A_1A_2}$,
\begin{align}
J_{(2)}^{[1]A_1A_2}(\ell_\perp) &=  \frac1{\eta} w^2 \left(\frac{\sqrt{s}}{\nu} \right)^{-\eta} \bigg[
  \frac2{N_c}\prnth{-\img f^{A_1B_1C}} \prnth{-\img f^{A_2B_2C}} \int\! \dbar^{d'}\!k_\perp \, J_{(2)}^{[0]B_1 B_2 }(k_\perp)\,  K_\text{NF}(\ell_\perp, k_\perp)
    \nn \\
    & \qquad - J_{(2)}^{[0]B_1B_2}(\ell_\perp)\, \delta^{A_1B_1}\delta^{A_2B_2}  \bigl(\omega_G(\ell_\perp) + \omega_G(\ell_\perp-q_\perp)\bigr) 
   \bigg] + \mathcal{O}(\eta^0)\,,
\end{align}
with
\begin{align}
    & \omega_G(\ell_\perp) = -\as N_c \int \frac{\dbar^{d'}\!k_\perp \; \vec\ell_\perp^{\,2}}{\prpsq{k} \prpsqm{\ell}{k}} \,,
    \nn \\
    & K_\text{NF}(\ell_\perp, k_\perp) = 
    \as N_c
    \prnth{-\frac{\vec q_\perp^{\,2}}{\prpsq{k} \prpsqm{q}{k}} + \frac{\vec\ell_\perp^{\,2}}{\prpsq{k}\prpsqm{\ell}{k}} + \frac{\prpsqm{q}{\ell}}{ \prpsqm{q}{k} \prpsqm{k}{\ell}} } \, .
\end{align}
Here, the notation $\w_G$ and $K_{\rm NF}$ follows from Ref.~\cite{Kovchegov:2012mbw}, and NF stands for non-forward (we have removed a factor of $4\pi^2$ compared with Eq.~(3.103) in Ref.~\cite{Kovchegov:2012mbw}).
Recall that $J_{(2)}^{[0]}$ does not depend on its argument. We put arguments $k_\perp$ and $\ell_\perp$ for the two terms in the equation above so that they match the graphic below.

The $\eta$-divergence in $J_{(2)}^{[1]A_1A_2}$ must be canceled by an appropriate counterterm $Z_{J(2, 2)}^{A_1 A_2\, B_1 B_2}$. The LL RRGE can thus be found by differentiating with respect to $\log \nu$,
\begin{align} \label{eq:collRRGE}
 \nu\frac{\partial}{\partial \nu}J_{(2)}^{A_1A_2}(\ell_\perp,\nu) 
  &= \int\! \dbar^{d'}\!k_\perp\, 
      J_{(2)}^{B_1B_2}(k_\perp,\nu) \biggl[\frac2{N_c}\prnth{-\img f^{A_1B_1C}} \prnth{-\img f^{A_2B_2C}} K_\text{NF}(\ell_\perp, k_\perp)
      \nn \\
  & \qquad - \delta^{A_1B_1}\delta^{A_2B_2}  \deltaslash^{d'}\!(\ell_\perp- k_\perp) \prnth{\omega_G(\ell_\perp) + \omega_G(\ell_\perp-q_\perp)} \biggr]
      \\
  &= -\frac12\int\! \frac{\dbar^{d'}\!k_\perp}{\vec k_\perp^2 \prpsqm{q}{k}}\, 
      J_{(2)}^{B_1B_2}(k_\perp,\nu) \,
      \gamma_{(2,2)}^{B_1B_2\,A_1A_2}(k_\perp,\ell_\perp)\,.
\end{align}
Here in the second line, we have pulled out some factors to be consistent with the definition of $\int_{\perp(j)}$ as in \eq{int_perp}. We then see that $\gamma_{(2,2)}$ is
\begin{align}\label{eq:gamma22}
  \gamma_{(2,2)}^{B_1B_2\,A_1A_2}(k_\perp,\ell_\perp)
  &=-4\alpha_s\prnth{-\img f^{A_1B_1C}} \prnth{-\img f^{A_2B_2C}} \prnth{-\vec q_\perp^{\,2}+ \frac{\vec\ell_\perp^{\,2} \prpsqm{q}{k} 
  + \prpsq{k}\prpsqm{q}{\ell}
  }{\prpsqm{\ell}{k}} } 
      \nn \\
  &  + 2\delta^{A_1B_1}\delta^{A_2B_2}\, \vec\ell_\perp^{\;2}\prpsqm{q}{\ell}\, \deltaslash^{d'}\!(\ell_\perp- k_\perp) \Bigl(\omega_G(\ell_\perp) + \omega_G(\ell_\perp-q_\perp)\Bigr)\,,
\end{align}
which is symmetric under $k_\perp\leftrightarrow\ell_\perp$. 

Graphically, we may express these equations as
\begin{align}
  &\frac{\partial}{\partial\nu} \fd{3.5cm}{figsCO/Jn.pdf} = \fd{3.5cm}{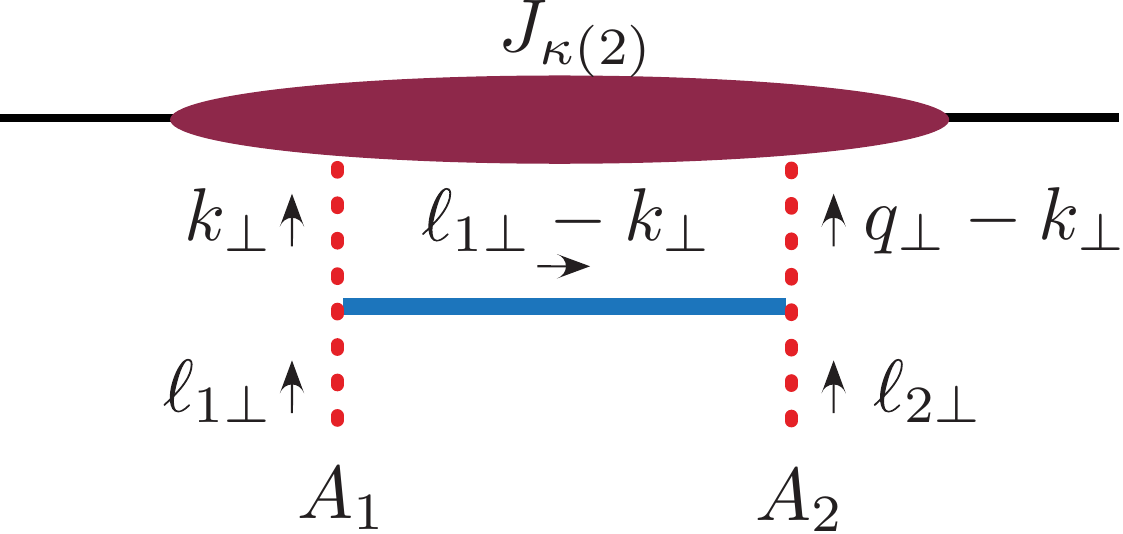}+\fd{3.5cm}{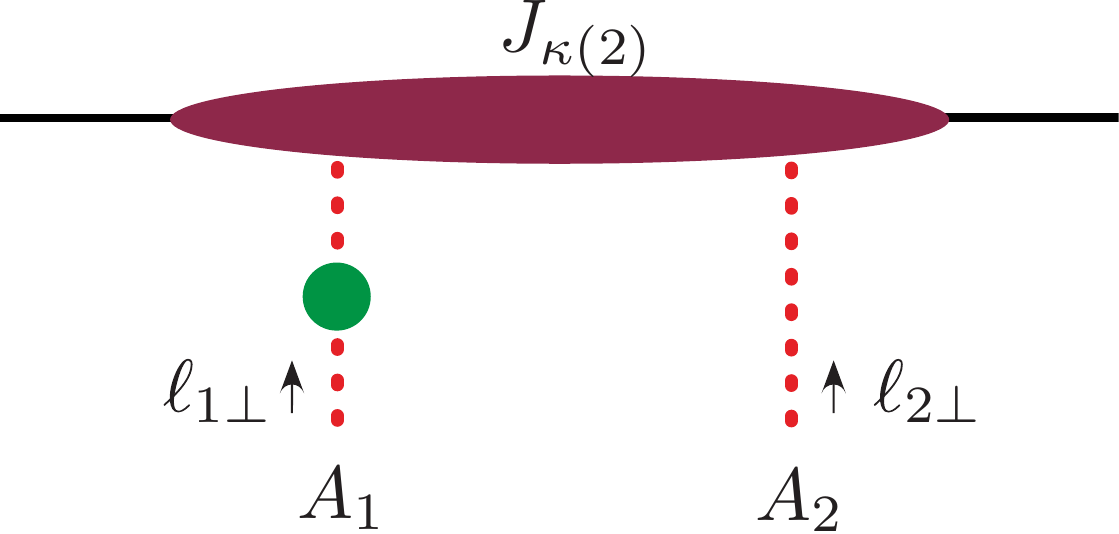}+\fd{3.5cm}{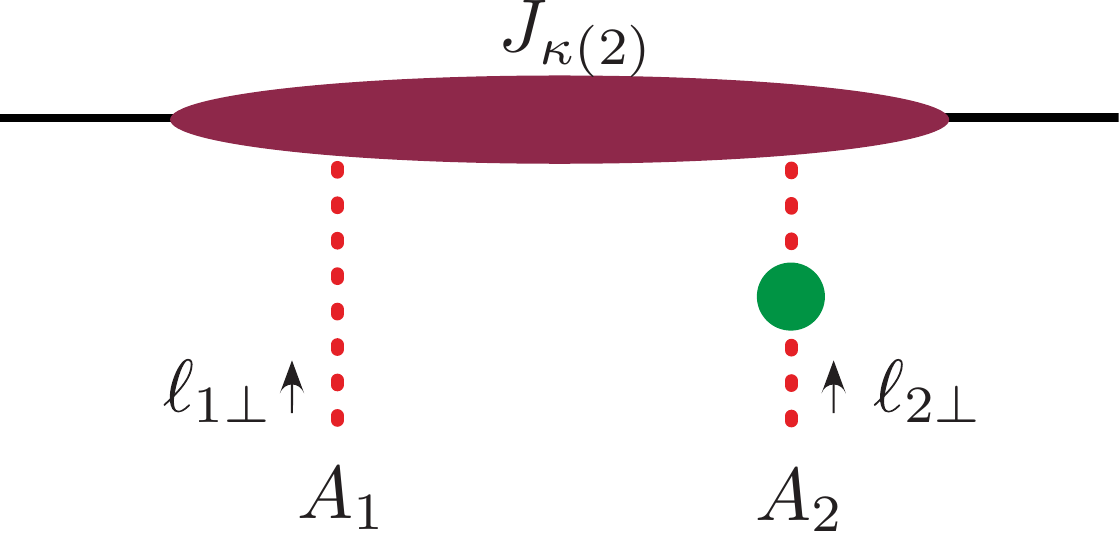}\,,\nn
  \\[.2in]
  &\fd{3.3cm}{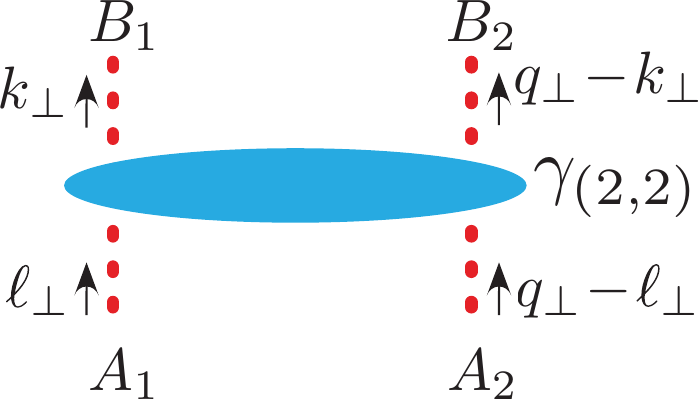} \quad = \quad \fd{3.3cm}{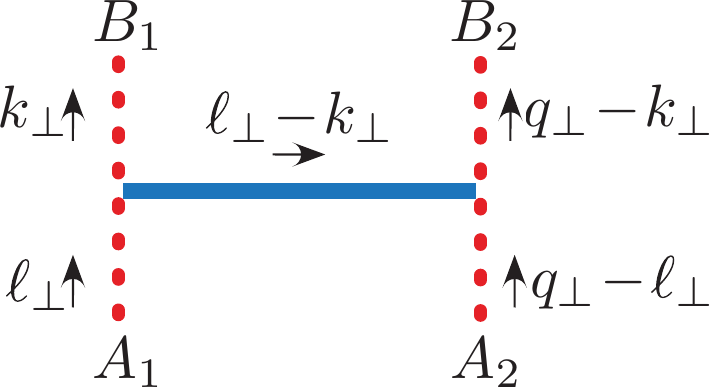} ~+~ \fd{3.3cm}{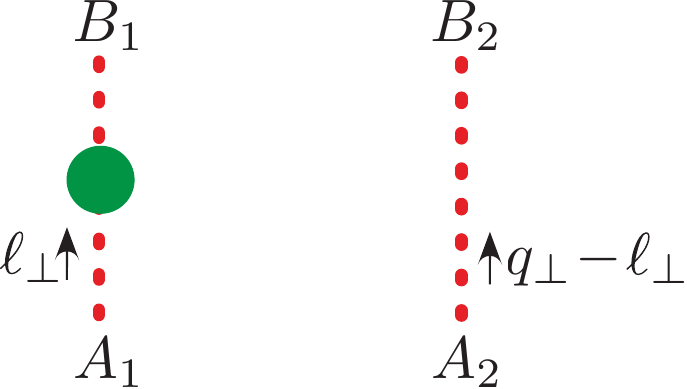} ~+~ \fd{3.3cm}{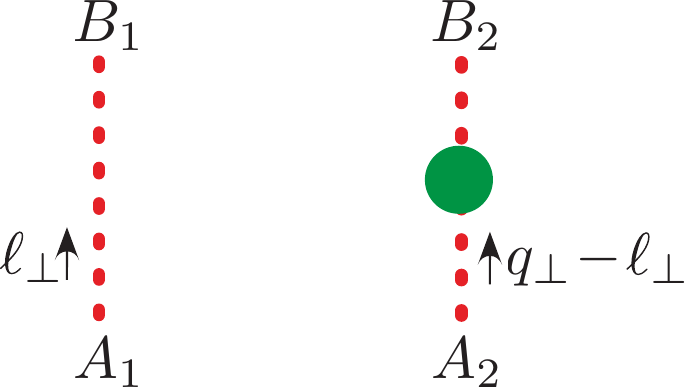}\,.
\end{align}
Here, the horizontal blue line indicates convolution with $K_\text{NF}$ and the green blob indicates a multiplicative factor $\w_G$ (Regge trajectory). This is of course the standard BFKL equation \cite{Fadin:1975cb,Lipatov:1976zz,Kuraev:1976ge,Kuraev:1977fs,Balitsky:1978ic,Lipatov:1985uk}. However, the novelty of our approach is that we have been able to derive this entirely from collinear calculations. As mentioned, we anticipate that this will allow a better understanding of the relation between BFKL and DGLAP, as well as facilitate calculations at higher perturbative orders.

In \app{soft}, we derive the same result from the standard soft perspective, showing that the results for the rapidity anomalous dimensions agree up to a factor of $-1$, exactly as required by soft-collinear consistency. 

\subsection{Decomposition into Irreducible Color Representations}
\label{sec:BFKL_color}

We can now decompose the LL RRGE for $J_{(2)}$ into different irreducible representations (irreps) $R$ of $SU(N_c)$. Due to color conservation, the RRGE does not mix different irreps.
We begin with a general discussion of color decomposition, and then we show that by projecting onto various color channels, we obtain the standard results for gluon Reggeization ($8_A$), the BFKL Pomeron $(1)$, and the other color channels.  

Suppose that we wish to decompose the color flow of $m$ gluons in the $t$-channel into different irreps.~$R$. This decomposes the tensor product as a direct sum,
\begin{align}
\bigotimes_{i = 1}^m 8 = \bigoplus_{R \in \rm{irrep}(SU(N_c))} R.
\end{align}
Let $\alpha = A_1 \cdots A_m$ be a multi-index for these $m$ gluons. We can concretely express the decomposition into irreps. in terms of projectors $P_{m\, R}^{\alpha \alpha'}$, whose explicit expressions can be found in Ref.~\cite{Keppeler:2012ih}. These projectors satisfy a completeness relation
\begin{align}\label{eq:proj_compl}
  \delta_{m}^{\alpha\alpha'} = \sum_R P_{m\, R}^{\alpha\alpha'}\,,
\end{align}
as well as an orthogonality relation
\begin{align}
  P_{m\, R}^{\alpha_1 \alpha_2}\, P_{m\, R'}^{\alpha_2\alpha_3} = \delta_{RR'} P_{m\, R}^{\alpha_1 \alpha_3}.
\end{align}
Intuitively, $P_{m\, R}^{\alpha \alpha'}$ projects a $m$-gluon color state onto a subspace $R$.

By inserting projectors, we can decompose $J_{(i)}^{\alpha} \otimes S_{(i,j)}^{\alpha\beta} \otimes \bar J_{(j)}^{\beta}$ into different irreps.
\begin{align}
  J_{(i)}^{\alpha} \otimes S_{(i,j)}^{\alpha\beta} \otimes \bar J_{(j)}^{\beta} = \sum_{R,R'} J_{(i)}^{R\,\alpha} \otimes S_{(i,j)}^{RR'\,\alpha\beta} \otimes \bar J_{(j)}^{R'\,\beta}\,,
\end{align}
where $\alpha$, $\beta$ are color indices for $i$, $j$ gluons respectively, and $J_{(i)}^{R\,\alpha}$, $S_{(i,j)}^{RR'\,\alpha\beta}$ are defined as 
\begin{align}\label{eq:J_R}
  J_{(i)}^{R\,\alpha} &\equiv J_{(i)}^{\alpha'} P_{i\, R}^{\alpha'\alpha}\,, \\
  \label{eq:S_R}
  S_{(i,j)}^{RR'\,\alpha\beta} &\equiv P_{i\,R}^{\alpha\alpha'} \, S_{(i,j)}^{\alpha'\beta'} \, P_{j\,R'}^{\beta'\beta}\,.
\end{align}
Note that irreps.~of different dimensions (eg.~8 versus 10) are always orthogonal, even for different number of gluons, so
\begin{align}
  P_{i\,R}^{\alpha\alpha'} \, S_{(i,j)}^{\alpha'\beta'} \, P_{j\,R'}^{\beta'\beta} = 0\,,\qquad 
   \text{for } {\rm dim}\: R\neq {\rm dim}\: R'\,.
\end{align}
In other words, the soft function does not take a color state in one subspace (eg.~$R=8$) to another (eg.~$R'=10$). Neither does $Z_{J(i,j)}^{\alpha\beta}$ mix  irreps.~of different dimensions, 
\begin{align}
  P_{i\,R}^{\alpha\alpha'} \, Z_{J(i,j)}^{\alpha'\beta'} \, P_{j\,R'}^{\beta'\beta} = 0\,,\qquad 
  \text{for } {\rm dim}\: R\neq {\rm dim}\: R' \,.
\end{align}

In what follows, we will restrict our attention to the 2-gluon case (i.e. $i = j = 2$) at LL order.  
We wish to project the RRGE in~\eqref{eq:collRRGE} onto different color channels to develop an RRGE for each $J_{(2)}^{R\,A_1A_2}$. To do this, we decompose the two 2-gluon indices as
\begin{align}\label{eq:2gluon_decomp}
  8\otimes8=1\oplus 8_A\oplus 8_S\oplus 10\oplus \overline{10}\oplus 27\oplus 0\,.
\end{align}
For this case none of the irreps.~can cause mixing, so we always have a single element for each color channel with $R=R'$. 
The pattern of representations for the three Glauber exchange is more interesting. It can be decomposed as $8\otimes8\otimes8=1^2 \oplus 8^8 \oplus 10^4 \oplus \overline{10}^4 \oplus 27^6 \oplus 35^2 \oplus \overline{35}^2 \oplus 64$, from which we see that many subspaces are multi-dimensional and the pattern of possible mixing may be more interesting.

Since the two color structures $C_\delta$ and $C_{\rm H}$ in \eq{collRRGE} (or in $Z_{J(2,2)}$) do not mix different irreps., they can be written as linear combination of projectors (identity operator within $R$'s). For $C_\delta$, this is obvious from \eq{proj_compl}
\begin{align}\label{eq:C_d_proj}
  C_\delta^{A_1 A_2\, B_1 B_2} = \delta^{A_1B_1}\delta^{A_2B_2} = \sum_R P_{2 \, R}^{A_1 A_2\, B_1 B_2}\,.
\end{align}
For $C_{\rm H}$, writing
\begin{align}\label{eq:C_H_proj}
  \frac{2}{N_c}C_{\rm H}^{A_1 A_2\, B_1 B_2}=\frac{2}{N_c} (-\img f^{A_1A_2C})(-\img f^{B_1B_2C}) = \sum_R c_R P_{2\, R}^{A_1 A_2\, B_1 B_2}\,,
\end{align}
and simplifying to the case of $N_c=3$, we explicitly calculate $c_R$ to be
\begin{align}
 c_1 = -2, \qquad c_{8_A} = c_{8_S} = -1\,, \qquad
 c_{10} = c_{\overline{10}} = 0\,, \qquad
 c_{27} = 2/3\,.
\end{align}

Plugging \eqs{C_d_proj}{C_H_proj} into \eq{collRRGE}, we obtain the RRGE for $J_{(2)}^{R\,A_1A_2}$,
\begin{align} \label{eq:collRRGE2}
 \nu\frac{\partial}{\partial \nu}J_{(2)}^{R\, A_1 A_2}(\ell_\perp,\nu) 
      &= \int\! \dbar^{d'}\!k_\perp \biggl[ -c_R  K_\text{NF}(\ell_\perp, k_\perp)
+ \deltaslash^{d'}\!(\ell_\perp- k_\perp) \prnth{\omega_G(\ell_\perp) + \omega_G(\ell_\perp-q_\perp)} \biggr]
 \nn\\
& \qquad \times J_{(2)}^{R\, A_1 A_2}(k_\perp,\nu).
\end{align}
We see that there is no operation on the color indices $A_1 A_2$ or on the irrep.~$R$, which is due to the fact that there is no mixing between the irreps.~in \eq{2gluon_decomp}.
A parallel equation for $S_{(2,2)}$ is given in \app{soft}.

For the antisymmetric octet, substituting $c_{8_A} = -1$ we have
\begin{align} \label{eq:collRRGE3}
 \nu\frac{\partial}{\partial \nu}J_{(2)}^{8_A\, A_1 A_2}(\ell_\perp,\nu) 
      &= \int\! \dbar^{d'}\!k_\perp \biggl[ K_\text{NF}(\ell_\perp, k_\perp)
+ \deltaslash^{d'}\!(\ell_\perp- k_\perp) \prnth{\omega_G(\ell_\perp) + \omega_G(\ell_\perp-q_\perp)} \biggr] 
  \nn\\
&\qquad \times J_{(2)}^{8_A\, A_1 A_2}(k_\perp,\nu) \,.
\end{align}
It is easy to see that a single pole solution $e^{\omega_G(q_\perp)}$ satisfies \eq{collRRGE3}, exactly as we expect for gluon Reggeization at NLL \cite{Fadin:1993wh}. (Note that this result is LL from the perspective of the evolution of the two-Glauber operator, but NLL for the amplitude).  For the BFKL pomeron, we substitute $c_R = c_1 = 2$, and our result agrees with Ref.~\cite{Fadin:1993wh}.
Results for the $8_S$ and $27$ also agree with Ref.~\cite{Ioffe:2010zz}.
The vanishing of the RGE for the $10$ and $\overline{10}$ at this order is also known\footnote{For $\kappa=q,\bar q$, there is no projection on to the $10$ and $\overline{10}$ color channels, since $3\otimes\bar 3=1\oplus 8$. For $\kappa=g$, the leading order gluon collinear function~\eqref{eq:J_LO} contains a color structure $C_{\rm H}$, which also has no projection onto the $10$ or $\overline{10}$.}.
Therefore, we have seen that we have been able to reproduce all the standard BFKL evolution equations for two-Glauber exchange purely from a collinear perspective.

\section{Simplifications in the Planar Limit from Glauber Instantaneity}
\label{sec:planar}

In this paper we have understood the generic structure of the renormalization group evolution of multi-glauber soft and collinear operators. These exhibit a complicated infinite dimensional mixing matrix exciting operators with increasing number of Glauber exchanges at higher orders in the perturbative expansion. One aspect of these equations, which is not at all transparent is how they simplify in the planar limit. In the planar limit, it is expected that amplitudes in the Regge limit reduce to Regge poles.  The lack of cuts in planar graphs was proven long ago by Mandelstam \cite{Mandelstam:1963cw} (see also Ref.~\cite{Eden:1966dnq}). It also is known through explicit calculation of the amplitude that in planar $\cN=4$ super Yang-Mills \cite{Drummond:2007aua}, and its conformal fishnet theory deformation \cite{Korchemsky:2018hnb} the amplitude reggeizes into a single pole to all loop order. General arguments have also been given in Ref.~\cite{Caron-Huot:2013fea}. Here we would like to show how this emerges transparently from the collinear perspective on the Regge limit, combined with the Glauber collapse rules of Ref.~\cite{Rothstein:2016bsq} (which was shown to hold at higher loop orders with a correctly chosen regulator in Ref.~\cite{Moult:2022lfy}). This argument will hold in generic asymptotically free gauge theories, irrespective of their matter content.

Consider an $N$-loop correction to the N-Glauber state.  It was proven in Ref.~\cite{Rothstein:2016bsq} that the only non-vanishing diagrams are those for which the Glaubers can collapse to an instantaneous exchange occurring at a single spacetime point without begin interrupted by an intermediate vertex. It is then straightforward to see that there are only two possibilities: Case 1: The graph is planar, and all the Glaubers attach to the same parton and are uninterrupted, as in 
\begin{align}
  \fd{3cm}{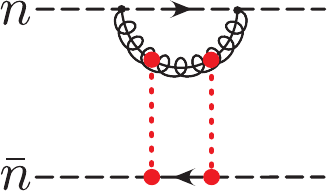}\,.
\end{align}
This implies that the loop correction is independent of the number of Glaubers, and has no convolutions. It is therefore a pole, with anomalous dimension equal to that of the single Glauber exchange. Case 2: The graph is non-planar as in
\begin{align}
 \fd{3cm}{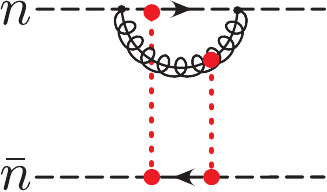}\,.
\end{align}
In this case the Glaubers resolve the multi-collinear structure, leading in general to convolutions. Therefore we immediately see the link between planarity, Glauber instantaneity, and Reggeization. This is physically similar to the picture discussed in  Ref.~\cite{Caron-Huot:2013fea}, which used a combination of planarity and light-cone ordered perturbation theory.

Therefore, we see that in the planar limit, the Regge limit organizes into a simple pole solution. While this has long been known, and understood diagramatically, what is difficult diagramatically is to systematically organize the structure of the Regge limit beyond the planar case.  The organization presented in this paper in terms of the renormalization group structure of multi-Glauber operators provides a field theoretic framework for understanding the more general non-planar case. It would be interesting to study it in a $1/N_c$ expansion to see if it provides any simplification to the infinite dimensional mixing structure.

\section{Conclusions and Outlook}
\label{sec:conc}

In this paper we have studied the renormalization group structure of multi-Glauber operators in the effective field theory for forward scattering. We derived the general structure of the renormalization group consistency equations, which relate anomalous dimensions for the soft (Glauber) operators to those for the collinear operators. In addition to providing a general understanding of the operator mixing structure of multi-Glauber operators, this also allows evolution equations in the forward limit, e.g. the BFKL equation, to be derived from a purely collinear perspective. This contrasts with the standard perspective where they are derived from the behavior of soft/Glauber/Reggeon degrees of freedom that interact between the two high energy partons.

We used this collinear perspective to rederive the standard LL BFKL evolution equations for the $1$, $8_S$, and $27$ color channels from a purely collinear perspective. A particularly convenient aspect of calculations in the collinear sector is that they make manifest the connection between non-planarity and Regge cuts. Using the instantaneity of Glauber exchanges, we used this to give a simple proof of Reggeization in a planar gauge theory. While this is a known result, our proof is constructive, and enables non-planar contributions to be systematically incorporated.

Using renormalization group consistency of the effective field theory, we have been able to constrain the all orders structure of the rapidity renormalization group equations governing the forward limit. In addition to organizing the structure for higher order calculations, this provides significant insight into how the effective field theory organizes the physics of the Regge limit, and makes clear the relation between the effective field theory approach, and previous Reggeon Effective Field Theory approaches. We have found that it is the rapidity renormalization group evolution equations of the EFT, which act only in the plane perpendicular to the scattering, are a form of Reggeon Field Theory, while the Glauber EFT is a genuine four dimensional field theory, from which these rapidity evolution equations can be derived. However, the form of the Reggeon Field Theory arising from the rapidity renormalization group equations of the EFT is slightly different than the standard Reggeon Field Theory,  since we have found that the single Glauber is an eigenstate of the rapidity evolution, namely there are no $1\to n$ Glauber transitions. All mixing in the rapidity RG can be interpreted as $n\to m$ scatterings, with $n,m\geq 2$ of the degrees of freedom in the Reggeon Field Theory. It would be interesting to develop this further, to see if one could develop a field theory underlying the rapidity evolution equations of the EFT. We believe that this considerably sharpens the understanding of the organization of the physics of forward scattering in the EFT, and how it relates to the organization of the Reggeon Field Theory. To achieve this understanding, we also introduced several technical simplifications to the EFT of forward scattering. In particular, building on \cite{Moult:2022lfy}, where we introduced a modified rapidity regularization of the forward limit with an  $\eta'$ regulator for Glauber divergences, and an $\eta$ regulator for soft and collinear rapidity divergences, in this paper we showed that it is advantageous in certain situations to also use other regulators for rapidity divergences in soft/collinear loop graphs.
Apart from simplifying proofs in the effective theory due to the analytic structure of the regulator, we anticipate that this will also be beneficial in simplifying higher order calculations.

We believe that the collinear perspective, as well as our ability to give gauge invariant operator definitions of the Regge/pomeron trajectories purely in terms of collinear operators will enable a number of new calculations in the Regge limit. In a future publication we will use it to derive the evolution equation for the Regge octet cut contribution that arises with the exchange of three Glaubers. For the study of the non-planar limit, calculations in the collinear sector cleanly separate Regge pole and Regge cut contributions. We also believe that our definition of the Regge trajectory as the rapidity anomalous dimension of a collinear operator will facilitate perturbative calculations to higher orders since collinear calculations involving a single sector are typically much simpler than soft calculations involving multiple sources. We intend to further pursue these directions in future work.

\begin{acknowledgments}
We thank Simon Caron-Huot, Einan Gardi and Aniruddha Venkata for useful discussions. 
This work was supported by the U.S.~Department of Energy, Office of Science, Office of Nuclear Physics, from DE-SC0011090. I.M. was supported by start-up funds from Yale University. I.S.~was also supported in part by the Simons Foundation through the Investigator grant 327942. We also thank the Erwin-Schr\"odinger Institute and {\it QFT at the Frontiers of the Strong Interaction} workshop for support and hospitality while parts of this work were completed.
\end{acknowledgments}

\appendix

\section{Rapidity Regulators for Collinear Loop Graphs}
\label{app:regulator}

To properly compute graphs with both collinear loops and Glauber exchanges, we need the following regulated Feynman rule,
\begin{align}\label{eq:collinear_regulator}
\fd{3cm}{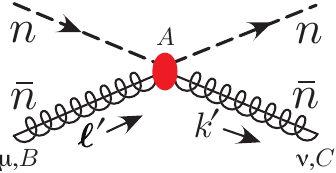} & = \frac{-8\pi \alpha_s f^{ABC} }{(\vec \ell^\prime_\perp-\vec k^\prime_\perp)^2 }
     \Big[ \bar u_n \frac{\bnslash}{2} T^A u_n  \Big]  
     \nn \\
     & \qquad \times \Biggl[ n\mcdot k^\prime\, g_\perp^{\mu\nu} 
     	- \left(n^\mu \ell^{\prime\nu}_\perp + n^\nu k^{\prime\mu}_\perp \right) w \left| \frac{n \cdot k'}{\nu}\right|^{-\eta/2}
      + \frac{\ell^\prime_\perp\cdot k^\prime_\perp n^\mu n^\nu}{n\cdot k^\prime} w^2 \left| \frac{n \cdot k'}{\nu}\right|^{-\eta} \Biggr]    \, ,
\end{align}
which comes from the regulated operators
\begin{align}
{\cal O}_n^{g B} & = \frac{\img}{2} f^{BCD}  {\cal B}_{n\perp\mu}^C \,
	\frac{\bn}{2}\cdot (\cP\!+\!\cP^\dagger)
	w^2 \left|\frac{\bn}{2} \cdot \frac{(\cP\!+\!\cP^\dagger)}{\nu}\right|^{\eta}  
	{\cal B}_{n\perp}^{D\mu} 
	\nn\\
W_n &= \sum_{\rm perms} \exp\bigg\{ \frac{-g}{n\cdot \cP} \bigg[
	w^2 \left|\frac{\bn \cdot \cP}{\nu}\right|^{-\eta} \bar n \cdot A_n\bigg] \bigg\} 
	\,, \nn\\
{\cal B}_{n\perp}^\mu &=  \left(A_{n\perp}^\mu w \left|\frac{\bn\cdot \cP}{\nu}\right|^{-\eta/2} - \frac{k_\perp^\mu}{\bn\cdot k} \bn\cdot A_{n,k} 
	w^2 \left|\frac{\bn\cdot \cP}{\nu}\right|^{-\eta}\right) + \ldots  \,.
\end{align}
Here $w$ is a formal bookkeeping parameter which satisfies
\begin{align}
    \nu\frac{\partial}{\partial \nu} w^2(\nu) =-\eta~ w^2(\nu) \,, \qquad \lim_{\eta \to 0}\, \omega(\nu)=1\,.
\end{align}
For convenience we set $w=1$ throughout our calculations since it can be trivially restored. Note that the $\eta$ regulator here can be replaced by other regulators, such as the exponential regulator~\cite{Li:2016axz}.

The powers of $\eta$ appearing in the regulator factors in \eq{collinear_regulator} are determined by requiring that the following 1-collinear-loop $i$-Glauber-exchange planar graphs,
\begin{align}\label{eq:1niG}
  -\img\cM_{1ni\text{G}} \equiv
  \fd{3cm}{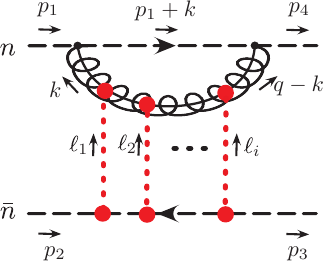}\,,
\end{align}
exponentiate in a similar way to the exponentiation of the box diagrams discussed in section~9.1 of Ref.~\cite{Rothstein:2016bsq}. (Note that $\eta$ there corresponds to $\eta'$ here.)

Let's start by looking at the single Glauber exchange case,
\begin{align}\label{eq:Tproduct_picture}
  &-\img\cM_{1n1\text{G}}=\fd{3cm}{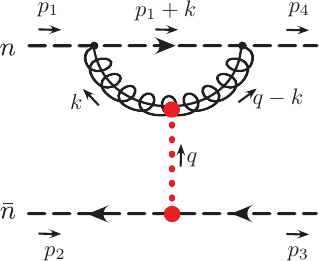}\\[.2cm] 
 &\sim \img \int\!\! \frac{\dbar^d k \,
 \left[\abs{k^-}^{-\eta} \, \vec k_\perp\cdot (\vec q_\perp-\vec k_\perp)/k^-\right]
 }{\bigl(k^- k^+ -\vec k_\perp^2 +\img0\bigr)
 \bigl[k^-k^+ -(\vec k_\perp-\vec q_\perp)^2 + \img0\bigr]
 \Bigl[p_1^++k^+ - \frac{(\vec p_{1\perp}+\vec k_\perp)^2 - \img0}{p_1^-+k^-}\Bigr] } 
 +\cO(\eta^0)
 \nn \\
 &\sim -\img \int\! \frac{dk^- \dbar^{d'}\!k_\perp \,\theta(-k^-)\theta(p_1^-+k^-)
 \left[\abs{k^-}^{-\eta} \, \vec k_\perp\cdot (\vec q_\perp-\vec k_\perp)/k^-\right]
 }{\bigl(k^- k^+ -\vec k_\perp^2 +\img0\bigr)
 \bigl[k^-k^+ -(\vec k_\perp-\vec q_\perp)^2 + \img0\bigr]}
 \Biggr|_{k^+ =-p_1^+ + \frac{(\vec p_{1\perp}+\vec k_\perp)^2 - \img0}{p_1^-+k^-}}+\cO(\eta^0)\nn\\
 &\sim -\img \frac{1}{\eta} 
  \int\! \frac{\dbar^{d'}\!k_\perp\, \vec q_\perp^{\,2}
 }{\vec k_\perp^2 (\vec k_\perp-\vec q_\perp)^2}+\cO(\eta^0)
 \nn\\
 & \underset{{\rm F.T.}_\perp}{\implies} \frac{1}{\eta}\img\phi(b_\perp)
 \,.\nn
\end{align}
Here in the second line, we keep only the rapidity divergent term in the numerator, which comes from the third term in \eq{collinear_regulator};
the remaining terms have higher powers of $k^-$, which are not rapidity divergent in the $k^-$ integral.
To get the third line, we integrate $k^+$ by contours. The third line is log-divergent in the limit that $k^-$ goes to zero, corresponding to a soft scaling of the collinear loop momentum. Such log divergences can be regulated by $\abs{k^-}^{-\eta}$; further integrating $k^-$ gives an $\eta$ pole as expected. The Fourier transform operation, according to Ref.~\cite{Rothstein:2016bsq}, is defined to be
\begin{align}
  \underset{{\rm F.T.}_\perp}{\implies} \ 
   = \int\!\! \ddslash\!^{d-2}q_\perp\:  e^{\img \vec q_\perp \cdot \vec b_\perp} 
 \,.
\end{align}

The 1-$n$-collinear-loop $2$-Glauber-exchange graph is calculated in the main text in \eq{diagram_planar}. 
Now let's look at the 1-$n$-collinear-loop $i$-Glauber-exchange graph with general $i$,
\begin{align}\label{eq:1niG}
  &-\img\cM_{1ni\text{G}}\nn\\
 &\sim \frac{(-\img)^i}{i!} \int \frac{\dbar^{d'}\! \ell_{1\perp}\cdots \dbar^{d'}\! \ell_{i\perp}
 \dbar^d k \, 
 \left[\abs{k^-}^{-\eta} \, \vec k_\perp\cdot (\vec q_\perp-\vec k_\perp)/k^-\right]
 }{\vec\ell_{1\perp}^{\,2}\cdots \vec\ell_{i\perp}^{\,2}
 \bigl(k^- k^+ -\vec k_\perp^2 +\img0\bigr)
 \bigl[k^-k^+ -(\vec k_\perp-\vec q_\perp)^2 + \img0\bigr]
 \Bigl[p_1^++k^+ - \frac{(\vec p_{1\perp}+\vec k_\perp)^2 - \img0}{p_1^-+k^-}\Bigr] } +\cO(\eta^0) \nn \\
 &\sim \frac{(-\img)^i}{\eta\, i!} \int \frac{\dbar^{d'}\! \ell_{1\perp}\cdots \dbar^{d'}\! \ell_{i\perp}
 }{\vec\ell_{1\perp}^{\,2}\cdots \vec\ell_{i\perp}^{\,2}}
 \int\! \frac{\dbar^{d'}\!k_\perp\, \vec q_\perp^{\,2}
}{\vec k_\perp^2 (\vec k_\perp-\vec q_\perp)^2}+\cO(\eta^0)
\nn\\
& \underset{{\rm F.T.}_\perp}{\implies} \frac{1}{\eta\, i!}[\img\phi(b_\perp)]^i\,.
\end{align}
To get the second line, we have integrated $\ell_1^\pm$, ... $\ell_i^\pm$ using Eq.~(9.16) of Ref.~\cite{Rothstein:2016bsq} with the Glauber $\eta'$ regulator, giving $1/i!$.
We keep only the rapidity divergent term in the numerator, which comes from the second term in \eq{collinear_regulator} for Glauber 1 and $i$, and the first term (the $g_\perp$ term) in \eq{collinear_regulator} for Glauber $2$ to $i-1$; and again, the remaining terms with higher power in $k^-$, are not rapidity divergent in the later $k^-$ integral.
To get the third line, we integrate $k^+$ by contours and then the $k^-$ integral gives an $\eta$ pole.

From this calculation, we note that if the first and the second terms in \eq{collinear_regulator} had different powers of the regulators (say, $\abs{k^-}^{a\eta}$ with some general number of $a$), then the rapidity divergent numerator term in \eq{1niG} would no longer have a regulator as $\abs{k^-}^{-\eta}$ but with some other power. This would then modify the result of \eq{1niG} by some multiplicative factor, which would then spoil the expectation that these graphs exponentiate. In summary, the requirement of Glauber exponentiation dictates the powers of the $\eta$-regulator to be the ones as shown in \eq{collinear_regulator}.

\section{Soft Calculation of Two Glauber Evolution}
\label{app:soft}

In this section, we will explicitly calculate the RRGE for the soft function $S_{(2,2)}$ at LL. To this end, we need to calculate the graphs with one soft loop and two Glauber exchanges.
The sum of these three graphs are convolutions of the one-loop soft function with the two leading order collinear functions,
\begin{align}\label{eq:J0_S1_J0}
  \sum -\img\cM_{(2,2)}^{\text{one $s$ loop}}
  =\,& \frac{-1}{4} \int\!\! \frac{\dbar^{d'}\!\ell_\perp\, \dbar^{d'}\!\ell'_{\perp} }{\vec\ell_\perp^{\,2} \prpsqm{q}{\ell}\, \vec\ell^{\prime2}_\perp \prpsqm{q}{\ell'}} \,
J_{(2)}^{[0]A_1 A_2}(\ell_\perp)  S_{(2,2)}^{[1]A_1 A_2\, A_1' A_2'}(\ell_{\perp}; \ell^{\prime}_{\perp} ) \bar J_{(2)}^{[0] A_1' A_2'}(\ell_\perp^{\prime}) 
\,.\end{align}
Comparing the $\eta$ pole of sum of these one-soft-loop graphs with $-\img\cM_{(2,2)}^{[0]}$ in \eq{Gboxes0}, we can then read off the rapidity RGE for $S_{(2,2)}$ at LL accuracy. When calculating these graphs we use the rapidity regulator discussed in detail in Ref.~\cite{Moult:2022lfy}.

There are three rapidity divergent graphs, which are shown in \fig{diagrams_1soft_loop}.
Graphs involving a quark bubble are rapidity finite and therefore do not contribute to the evolution equation. 
The two graphs $\mathcal{M}_{\rm SE}$ and $\mathcal{M}_{\rm SE}'$ can be calculated by simply iterating the one-loop soft eye graph with an additional rung, while the so-called \textit{H-graph} $\mathcal{M}_{\rm H}$ on the left has additional structure.

Let us first calculate the H-graph.
The momentum labels in the graph are made such that they are parallel to the ones used in \eq{fw_fact}. To calculate the graph, it is easier to parameterize $\ell_{1,2}$ and $\ell_{1,2}'$ as
\begin{align}
  &\ell_1^\mu=\frac{n^\mu}{2}\ell^+ -\frac{\bn^\mu}{2}k^-+\ell_\perp^\mu\,,
  \quad
  \ell_1^{\prime\mu}=\frac{\bn^\mu}{2}\ell^- + \frac{n^\mu}{2}k^+ +(\ell_\perp+k_\perp)^\mu\,,
  \quad
  \ell_2=q_\perp-\ell_1\,,\quad
  \ell_2' =q_\perp-\ell_1'\,,
\end{align}
so that $\ell^\pm\sim\cO(\lambda^2)$. We can think of $\ell$ as a $n-\bn$ Glauber, and $k$ as an independent soft.
We define the prefactor for the H-graph $P_{\rm H}(k,\ell)$, the numerator factor $N_{\rm H}$, and the Glauber denominator $D_{\rm H}^\perp(q_\perp, k_\perp,\ell_\perp)$, as follows\footnote{Similar to the footnote in \sec{BFKL_graphs}, here the $\eta$ regulator in $N_{\rm H}$ (and also in $N_{\rm SE}$ below) can be replaced by other regulators. For example, with an analytic regulator like $(k^z\pm\img0)^{-\eta}$ used in \sec{vanish}, our result does not change at the level of $1/\eta$ poles. One may also consider using the exponential regulator $\exp{\bigl(-\tau e^{-\gamma_E}k^0\bigr)}$~\cite{Li:2016axz}: 
this amounts to changing the $k^z$ integral in \eq{Hgraph} after taking $k^0$ to be $\sqrt{(k^z)^2+\vec k_\perp^2}$ by contours, giving a $\ln\tau$ when expanding $\tau\to0$.
},
\begin{align}
  &P_{\rm H} = \prnth{-\img f^{AA'C}}\prnth{-\img f^{BB'C}} \prnth{T^{A} T^B} \otimes  \prnth{\bT^{A'} \bT^{B'}} \cS^{n\bn}\,,
\nn \\
&N_\text{H}(k,\ell) = 4\pi \img (64 \pi^2 \as^3)\Bigg(k^2 - 2 \vec q_\perp^{\,2} \nn \\
    & \hspace{1.2in} + \frac{2}{k^+ k^-} \left[ \prpsqp{\ell}{k} \prpsqm{q}{\ell} + \vec\ell_\perp^{\,2} \left(\prp{q} - \prp{\ell} - \prp{k} \right)^2 \right] \Bigg) w\reg{k}\,,
\nn \\
&D_\text{H}^\perp(q_\perp, k_\perp,\ell_\perp) = \vec\ell_\perp^{\;2} \prpsqm{q}{\ell} \prpsqp{\ell}{k} \left(\prp{\ell} + \prp{k} - \prp{q}  \right)^2 \, .
\end{align}
The H-graph can then be calculated as
\begin{align}\label{eq:Hgraph}
  -\img\cM_{\rm H} &= 
      \int\!\!\frac{P_{\rm H} \ \dbar^d\ell \; \dbar^dk \;  w'^4 \left| \frac{2 \ell^z}{\nu'} \right|^{-2 \eta'} N_\text{H}(k,\ell)}{
      \prnth{k^2 \!+\! \img0}
      \Bigl[p_1^+\!+\!\ell^+ \!-\! 
      \frac{(\vec p_{1\perp}+\vec\ell_\perp)^2}{p_1^-} \!+\! \img0\Bigr]
      \Bigl[p_2^- \!-\! \ell^- \!-\! 
      \frac{(\vec p_{2\perp}-\vec k_\perp -\vec\ell_\perp)^2}{p_2^+} \!+\! \img0\Bigr]
      D_\text{H}^\perp(q_\perp, k_\perp,\ell_\perp)
      }
  \nn \\[.2cm]
  &= \frac{64 \pi^2 \as^3 P_{\rm H}}{\eta \; \nu^{-\eta}} \int\frac{\dbar^{d'}\!\ell_\perp\; \dbar^{d'}\!k_\perp  }{
  \vec\ell_\perp^{\,2} \prpsqm{q}{\ell} \prpsqp{\ell}{k} \left(\prp{\ell} + \prp{k} - \prp{q}  \right)^2 \bigl(\prpsq{k}\bigr)^{\eta/2} 
  } \nn\\ 
  &\qquad \times \left(\vec q_\perp^{\,2} - \frac1{\prpsq{k}} \left[ \prpsqp{\ell}{k} \prpsqm{q}{\ell} + \vec\ell_\perp^{\,2} \left(\prp{\ell} + \prp{k} - \prp{q}  \right)^2 \right] \right) + \mathcal{O}(\eta^0)
  \nn \\
  &= \frac{64 \pi^2 \as^3 P_{\rm H}}{\eta }\left(\frac{\sqrt{-t}}{\nu}\right)^{-\eta} \int\frac{\dbar^{d'}\!\ell_\perp \; \dbar^{d'}\!\ell_\perp'  }{
    \vec\ell_\perp^{\,2} \prpsqm{q}{\ell}\, \vec\ell^{\prime2}_\perp \prpsqm{q}{\ell'}
    } \nn\\ 
    &\qquad \times \left(\vec q_\perp^{\,2} - \frac1{\prpsqm{\ell}{\ell'}} \left[ \vec\ell^{\,\prime2}_\perp \prpsqm{q}{\ell} + \vec\ell_\perp^{\,2} \prpsqm{q}{\ell^{\,\prime}} \right] \right) + \mathcal{O}(\eta^0)
    \,.
  \end{align}
To get the second line, we first perform the $\ell^0$ and $\ell^z$ integrals using Eq.~(B.5) of Ref.~\cite{Rothstein:2016bsq}, and take $\eta' \to 0$, $w' \to 1$;
then we do the $k^0$ integral by contours, and the remaining $k^z$ integral gives a $\eta$ pole.
In general we must also subtract the zero-bin with the scaling $k^+\to\cO(\lambda^2)$ or/and $k^-\to\cO(\lambda^2)$ to ensure that results are independent of the choice of $\pm i0$ in the soft eikonal propagators. Here we chose the two poles on the same side, e.g. $(k^++\img0)(k^-+\img0)$, so that the zero bin vanishes.
In the final line we swapped $\bigl(\vec k_\perp^2\bigr)^{\eta/2}$ with $\sqrt{-t}^{-\eta}$, and changed variable from $k_\perp$ to $\ell_\perp'$.

We now compute the next two graphs that contribute at this order: the two box graphs with one soft gluon loop (the ``Soft-Eye boxes'').
We define their prefactor, numerator, and Glauber denominator factors as
\begin{align}\label{eq:NSE}
&P_\text{SE} = \prnth{-\img f^{BCD}} \prnth{-\img f^{B'CD}} \delta^{AA'} \prnth{\bT^{A'} \bT^{B'}} \otimes \prnth{T^{A} T^B} \cS^{n\bn}
\nn\\
&\quad~~\,= - N_c \delta^{AA'}\delta^{BB'} \prnth{\bT^{A'} \bT^{B'}} \otimes \prnth{T^{A} T^B} \cS^{n\bn}\,,
\nn
\\
&N_\text{SE}(k,\ell) = 4\pi \img (64 \pi^2 \as^3)\Bigg( d'\,k^+ k^- 
- 2 \Bigl[\prpsq{k} + \prpsqm{k}{\ell}\Bigr] + \frac{\Bigl[ 2\prp{k} \cdot \bigl( \prp{k} - \prp{\ell} \bigr) \Bigr]^2}{k^+k^-}\Biggr) w\reg{k}\,,
\nn \\
&D_\text{SE}^\perp(q_\perp,\ell_\perp) = \bigl(\prp{\ell}^{\;2}\bigr)^2 \, \prpsqm{q}{\ell} \, ,
\end{align}
and calculate the amplitude,
\begin{align}
-\img\cM_{\rm SE} &= \! 
    \int\!\!\frac{P_\text{SE}\ \dbar^dk \; \dbar^d\ell \; w'^4 \left| \frac{2 \ell^z}{\nu'} \right|^{-2 \eta'}N_\text{SE}(k,\ell)}{
    \prnth{k^2 + \img0} [\prnth{\ell-k}^2\!+\!\img0]
    \Bigl[p_1^+ \!+\! \ell^+ \!-\! 
    \frac{(\vec p_{1\perp}\!+\! \vec\ell_\perp)^2}{p_1^-} \!+\! \img0\Bigr]
    \Bigl[p_2^- \!-\! \ell^- \!-\! \frac{(\vec p_{2\perp} \!-\!  \vec\ell_\perp)^2}{p_2^+} \!+\! \img0 \Bigr]
    D_\text{SE}^\perp(q_\perp,\ell_\perp)
    }
\nn \\
&= \frac12 \frac{64 \pi^2 \as^3 P_\text{SE}}{\eta}  \left(\frac{\sqrt{-t}}{\nu}\right)^{-\eta}
\int\!\frac{\dbar^{d'}\!k_\perp \; \dbar^{d'}\!\ell_\perp\left[ 2\prp{k} \cdot \bigl(\prp{\ell}- \prp{k} \bigr) \right]^2 }{
\bigl(\prp{\ell}^{\;2}\bigr)^2\, \prpsqm{q}{\ell} \prpsq{k} \prpsqm{\ell}{k}
} + \mathcal{O}(\eta^0)
\nn \\
&=  \frac12 \frac{64 \pi^2 \as^3 P_\text{SE}}{\eta}  \left(\frac{\sqrt{-t}}{\nu}\right)^{-\eta}
\int\!\!\frac{\dbar^{d'}\!\ell_\perp}{
\vec\ell_\perp^{\,2} \prpsqm{q}{\ell}
}
\int\!\! \frac{\dbar^{d'}\!k_\perp \; \vec\ell_\perp^{\,2}}{\prpsq{k} \prpsqm{\ell}{k}} + \mathcal{O}(\eta^0)
\,.
\label{eq:SoftEye}
\end{align}
Here, to get the second line, we first perform the $\ell^0$ and $\ell^z$ integrals using Eq.~(B.5) of Ref.~\cite{Rothstein:2016bsq}, and take $\eta' \to 0$, $w' \to 1$;
then we do the $k^0$ integral by contours, and the remaining $k^z$ integral gives a $\eta$ pole. Again, we have chosen the poles for $k^+$ and $k^-$ to be on the same side so that the zero bin vanishes.
Notice that we only keep the last term in the big bracket of \eq{NSE}, since it is the only term that is rapidity divergent.
In the last line, the $k_\perp$ integral is recognized to be the Regge trajectory (eg. from Ref.~\cite{Kovchegov:2012mbw}),
\begin{align}
    \omega_G(\ell_\perp) = -\as N_c \int \frac{\dbar^{d'}\!k_\perp \; \vec\ell_\perp^{\,2}}{\prpsq{k} \prpsqm{\ell}{k}} . 
\end{align}
The complementary graph with the soft loop on the right Glauber rung is identical up to the substitution $\ell_\perp \to q_\perp-\ell_\perp$,
\begin{align}
-\img\cM_{\rm SE}' &= \frac12 \frac{64 \pi^2 \as^3 P_\text{SE}}{\eta}  \left(\frac{\sqrt{-t}}{\nu}\right)^{-\eta} \int\frac{\dbar^{d'}\!\ell_\perp}{
\vec\ell_\perp^{\,2} \prpsqm{q}{\ell}
} \int \frac{\dbar^{d'}\!k_\perp \; \prpsqm{q}{\ell}}{\prpsq{k} \bigl(\prp{q}- \prp{\ell}  - \prp{k} \bigr)^2} + \mathcal{O}(\eta^0)
\,.
\label{eq:SoftEyeFlip}
\end{align}
where the $k_\perp$ integral is now $-\frac1{\as N_c} \omega_G(q_\perp-\ell_\perp)$. 

Summing over these three graphs, we get
\begin{align}
  & -\img\cM_{\rm{H}}-\img\cM_{\rm{SE}}-\img\cM_{\rm{SE}}' =\color{black} \frac{64 \pi^2 \as^3 }{\eta }\left(\frac{\sqrt{-t}}{\nu}\right)^{-\eta}  \nn\\ 
    &\qquad \times\Biggl[ P_{\rm H}\int\frac{\dbar^{d'}\!\ell_\perp \; \dbar^{d'}\!\ell_\perp'  }{
      \vec\ell_\perp^{\,2} \prpsqm{q}{\ell}\, \vec\ell^{\prime2}_\perp \prpsqm{q}{\ell'}
      }
      \left(\vec q_\perp^{\,2} - \frac{\vec\ell^{\,\prime2}_\perp \prpsqm{q}{\ell} + \vec\ell_\perp^{\,2} \prpsqm{q}{\ell^{\,\prime}}}{\prpsqm{\ell}{\ell'}}  \right)
    \nn\\
  &\qquad\qquad -\frac12 P_\text{SE}
\int\!\!\frac{\dbar^{d'}\!\ell_\perp}{
\vec\ell_\perp^{\,2} \prpsqm{q}{\ell}
} \bigl(\w_G(\ell_\perp)+\w_G(q_\perp-\ell_\perp)\bigr)\Biggr] + \mathcal{O}(\eta^0)\,.
\end{align}
Using \eq{J0_S1_J0}, we get
\begin{align}
  &S^{[1]A_1 A_2\, A_1' A_2'}_{(2,2)}(\ell_\perp,\ell_\perp',\eta)=\frac{2}{\eta}\Biggl[2 C_{\rm H}^{A_1 A_2\, A_1' A_2'}
  \biggl(\vec q_\perp^{\,2} - \frac1{\prpsqm{\ell}{\ell'}} \left[ \vec\ell^{\,\prime2}_\perp \prpsqm{q}{\ell} + \vec\ell_\perp^{\,2} \prpsqm{q}{\ell^{\,\prime}} \right] \biggr)
  \nn\\
  &\qquad\qquad + C_\delta^{A_1 A_2\, A_1' A_2'}\, \deltaslash^{d'}\!(\ell_\perp-\ell_\perp') \,
  \vec\ell_\perp^{\;2} \prpsqm{q}{\ell}
  \Bigl(\w_G(\ell_\perp)+\w_G(q_\perp-\ell_\perp)\Bigr)\Biggr]
  +\cO(\eta^0)\,.
\end{align}
Comparing with the LO soft function, we get the renormalization equation at LL for the soft function
\begin{align}
  &\quad~ \nu\partial_\nu S^{A_1 A_2\, A_1' A_2'}_{(2,2)}(\ell_\perp,\ell_\perp',\nu)\nn\\
  &= -\int\! \dbar^{d'}\!k_\perp\, 
      \biggl[ \frac2{N_c}\prnth{-\img f^{A_1B_1C}} \prnth{-\img f^{A_2B_2C}} K_\text{NF}(\ell_\perp, k_\perp)
      \nn \\
  & \qquad\qquad\qquad - \delta^{A_1B_1}\delta^{A_2B_2}  \deltaslash^{d'}\!(\ell_\perp- k_\perp) \prnth{\omega_G(\ell_\perp) + \omega_G(\ell_\perp-q_\perp)} \biggr]\, S^{B_1 B_2\, A_1' A_2'}_{(2,2)}(k_\perp,\ell_\perp',\nu)
     \nn\\
  & \qquad  +S^{A_1 A_2\, B_1 B_2}_{(2,2)}(\ell_\perp,k_\perp,\nu)\, \biggl[ \frac2{N_c}\prnth{-\img f^{A'_1B_1C}} \prnth{-\img f^{A'_2B_2C}} K_\text{NF}(\ell'_\perp, k_\perp)
  \nn \\
& \qquad\qquad\qquad - \delta^{A'_1B_1}\delta^{A'_2B_2}  \deltaslash^{d'}\!(\ell_\perp- k_\perp) \prnth{\omega_G(\ell'_\perp) + \omega_G(\ell'_\perp-q_\perp)} \biggr]\\
  &= \frac12\int\! \frac{\dbar^{d'}\!k_\perp}{\vec k_\perp^2 \prpsqm{q}{k}}\, 
  \biggl[
      \gamma_{(2,2)}^{A_1A_2\,B_1B_2}(\ell_\perp,k_\perp)\,
      S^{B_1 B_2\, A_1' A_2'}_{(2,2)}(k_\perp,\ell_\perp',\nu)
      \nn\\
  &\qquad\qquad\qquad\qquad\qquad  +
      S^{A_1 A_2\, B_1 B_2}_{(2,2)}(\ell_\perp,k_\perp,\nu) \,
      \gamma_{(2,2)}^{B_1B_2\,A'_1A'_2}(k_\perp,\ell'_\perp)  \biggr]
\,,\end{align}
where the same $\gamma_{(2,2)}$ as in \eq{gamma22} also appears here, as required by soft-collinear consistency.

Similar to what we have done in \sec{BFKL_color}, we can also decompose the RRGE for the soft function into different color irreps.
Projecting onto different color channel $R$, we obtain the RRGE for $S_{(2,2)}^{R\;A_1A_2\,A'_1A'_2}$ (it is defined in \eq{S_R}),
\begin{align} \label{eq:softRRGE_R}
 &\nu\frac{\partial}{\partial \nu}S_{(2,2)}^{R\; A_1 A_2\,A'_1 A'_2}(\ell_\perp,\ell'_\perp,\nu) \\
 =\,& \int\! \dbar^{d'}\!k_\perp \biggl[ c_R  K_\text{NF}(\ell_\perp, k_\perp)
+ \deltaslash^{d'}\!(\ell_\perp- k_\perp) \prnth{\omega_G(\ell_\perp) + \omega_G(\ell_\perp-q_\perp)} \biggr]
S_{(2,2)}^{R\; A_1 A_2\,A'_1 A'_2}(k_\perp,\ell'_\perp,\nu)
\nn\\
&\quad+\biggl[ c_R  K_\text{NF}(\ell'_\perp, k_\perp)
+ \deltaslash^{d'}\!(\ell'_\perp- k_\perp) \prnth{\omega_G(\ell'_\perp) + \omega_G(\ell'_\perp-q_\perp)} \biggr]
S_{(2,2)}^{R\; A_1 A_2\,A'_1 A'_2}(\ell_\perp,k_\perp,\nu)
\nn\,.
\end{align}
Notice there is no mixing in irrep.~$R$ or operation on the color indices $A_1 A_2\, A'_1 A'_2$, which is special to the two Glauber exchange, as explained in \sec{BFKL_color}. This agrees with the calculation in the main text which was done from the collinear perspective, up to the predicted overall sign.

\bibliographystyle{JHEP}
\bibliography{CollinearOrganization}

\end{document}